\documentclass[twocolumn]{aastex63}

\usepackage{amsmath}
\usepackage{booktabs}
\usepackage{longtable}
\usepackage{xspace}
\usepackage{hyperref}
\usepackage{mathtools}

\usepackage[T1]{fontenc}
\usepackage{lipsum}
\usepackage{xstring} % Needed for IfEqCase
\usepackage{comment}

\newcommand{\Kepler}{\textit{Kepler}\xspace} 
\newcommand{\Gaia}{\textit{Gaia}\xspace}
\newcommand{\ktwo}{\textit{K2}\xspace}

\newcommand{\ntrial}{\ensuremath{n_\mathrm{trial}}\xspace}

\newcommand{\nstar}{\ensuremath{n_\star}\xspace}

\newcommand{\ptr}{\ensuremath{p_{\mathrm{tr}}}}

\newcommand{\pdet}{\ensuremath{p_{\mathrm{det}}}\xspace}

\newcommand{\sigmasb}{\ensuremath{\sigma_\mathrm{sb}}\xspace}

\newcommand{\Mstar}{\ensuremath{M_{\star}}\xspace}
\newcommand{\Mstariso}{\ensuremath{M_{\star,\mathrm{iso}}}\xspace}
\newcommand{\Rstar}{\ensuremath{R_{\star}}\xspace} 
\newcommand{\Rstariso}{\ensuremath{R_{\star,\mathrm{iso}}}\xspace}

\newcommand{\rhostariso}{\ensuremath{\rho_{\star,\mathrm{iso}}}\xspace}
\newcommand{\ageiso}{\ensuremath{\mathrm{age}_{\star,\mathrm{iso}}}\xspace}

\newcommand{\rhostarcirc}{\ensuremath{\rho_\mathrm{\star,circ}}\xspace}

\newcommand{\fe}{\ensuremath{\mathrm{[Fe/H]}}\xspace}

\newcommand{\teff}{\ensuremath{T_{\mathrm{eff}}}\xspace}  
\newcommand{\logg}{\ensuremath{\log g}\xspace} 
\newcommand{\vsini}{\ensuremath{v \sin i}\xspace} 
\newcommand{\kepmag}{\ensuremath{m_\mathrm{Kep}}\xspace}
\newcommand{\kmag}{\ensuremath{m_\mathrm{K}}\xspace}

\newcommand{\Lbol}{\ensuremath{L_{\mathrm{bol}}}\xspace}

\newcommand{\Mbol}{\ensuremath{M_{\mathrm{bol}}}\xspace}

\newcommand{\Mp}{\ensuremath{M_p}\xspace} 
\newcommand{\Mcore}{\ensuremath{M_{c}}\xspace} 
\newcommand{\Menv}{\ensuremath{M_\mathrm{env}}\xspace} 
\newcommand{\Rp}{\ensuremath{R_p}\xspace}
 
\newcommand{\fenv}{\ensuremath{f_\mathrm{env}}\xspace}

\newcommand{\teq}{\ensuremath{T_{\mathrm{eq}}}\xspace}
\newcommand{\Sinc}{\ensuremath{S_\mathrm{inc}}\xspace}
\newcommand{\Sincc}[1]{\ensuremath{S_\mathrm{inc,#1}}\xspace}
\newcommand{\Fxuv}{\ensuremath{\mathcal{F}_\mathrm{xuv}}\xspace}
\newcommand{\Tcirc}{\ensuremath{T_{\mathrm{max,circ}}}\xspace}

\newcommand{\kms}{\ensuremath{\mathrm{km\, s}^{-1}}\xspace}

\newcommand{\Se}{\ensuremath{S_{\oplus}}\xspace}
\newcommand{\Me}{\ensuremath{M_{\oplus}}\xspace} 
\renewcommand{\Re}{\ensuremath{R_{\oplus}}\xspace}

\newcommand{\Rsun}{\ensuremath{R_{\odot}}\xspace }
\newcommand{\Msun}{\ensuremath{M_{\odot}}\xspace}

\newcommand{\dex}{\ensuremath{\mathrm{dex}}\xspace}
\newcommand{\dd}{\ensuremath{\mathrm{d}}\xspace}

\newcommand{\RNum}[1]{\uppercase\expandafter{\romannumeral #1\relax}}

\newcommand{\stat}[1]{%%
\IfEqCase{#1}{%
{n-koi}{6862}%
{n-star}{5869}%
{n-star-field}{177909}%
{n-star-field-pass0}{39667}%
{n-cand}{6862}%
{n-cand-pass0}{2259}%
{n-star-pass0}{1774}%
{n-cand-pass1}{1235}%
{n-star-pass1}{885}%
{n-star-pass1-cks1}{561}%
{n-star-pass1-none}{10}%
{n-star-pass1-smemp}{249}%
{n-star-pass1-smsyn}{65}%
{n-star-pass1-newobs}{314}%
{n-star-pass1-lt4700}{160}%
{sparallax-med}{1.0}%
{sparallax-ferr-med}{2.3}%
{sdistance-med}{965}%
{kmag-med}{12.83}%
{kmag-err-med}{0.03}%
{kepmag-med}{14.6}%
{kepmag-err-med}{14.6}%
{av-med}{0.099}%
{av-min}{0.007}%
{av-max}{0.354}%
{ak-med}{0.007}%
{ak-min}{0.001}%
{ak-max}{0.027}%
{srad-med}{1.0}%
{srad-ferr-med}{3.8}%
}[XX]%
}
\newcommand{\sample}[1]{%%
\IfEqCase{#1}{%
{nstars dr1}{1305}%
{nstars dr2}{411}%
{nstars cxm}{888}%
{nstars dr1 & cxm}{628}%
{nstars dr1 & ~cxm}{677}%
{nstars dr2 & cxm}{260}%
{nstars dr2 & ~cxm}{151}%
{nstars ~dr1 & ~dr2 & cxm}{0}%
{nstars (dr1 | dr2) & cxm}{888}%
{nstars dr1 & dr2}{0}%
{nstars dr1 | dr2}{1716}%
{nstars dr1 | dr2 | cxm}{1716}%
{nstars dr2 & cxm & pre-2018}{81}%
{nstars dr2 & cxm & post-2018}{179}%
{nstars dr2 & post-2018}{330}%
{nstars dr2 & cxm & pre-2018 counts < 1500}{4}%
{kois dr2 & cxm & pre-2018 counts < 1500}{254, 812, 877, 1475}%
{kois ~dr1 & ~dr2 & cxm}{}%
{nplanets planets-cuts1-cut-0-none}{7379}%
{nstars planets-cuts1-cut-0-none}{6319}%
{nplanets planets-cuts1-cut-1-faint}{3307}%
{nstars planets-cuts1-cut-1-faint}{2731}%
{nplanets-rem planets-cuts1-cut-1-faint}{4072}%
{nstars-rem planets-cuts1-cut-1-faint}{3588}%
{nplanets planets-cuts1-cut-2-giantcmd}{2755}%
{nstars planets-cuts1-cut-2-giantcmd}{2203}%
{nplanets-rem planets-cuts1-cut-2-giantcmd}{552}%
{nstars-rem planets-cuts1-cut-2-giantcmd}{528}%
{nplanets planets-cuts1-cut-3-diluted}{2693}%
{nstars planets-cuts1-cut-3-diluted}{2146}%
{nplanets-rem planets-cuts1-cut-3-diluted}{62}%
{nstars-rem planets-cuts1-cut-3-diluted}{57}%
{nplanets planets-cuts1-cut-4-ruwe}{2373}%
{nstars planets-cuts1-cut-4-ruwe}{1872}%
{nplanets-rem planets-cuts1-cut-4-ruwe}{320}%
{nstars-rem planets-cuts1-cut-4-ruwe}{274}%
{nplanets planets-cuts1-cut-5-notreliable}{1425}%
{nstars planets-cuts1-cut-5-notreliable}{1003}%
{nplanets-rem planets-cuts1-cut-5-notreliable}{948}%
{nstars-rem planets-cuts1-cut-5-notreliable}{869}%
{nplanets planets-cuts1-cut-6-lowsnr}{1250}%
{nstars planets-cuts1-cut-6-lowsnr}{891}%
{nplanets-rem planets-cuts1-cut-6-lowsnr}{175}%
{nstars-rem planets-cuts1-cut-6-lowsnr}{112}%
{nplanets planets-cuts1-cut-7-nomcmc}{1246}%
{nstars planets-cuts1-cut-7-nomcmc}{888}%
{nplanets-rem planets-cuts1-cut-7-nomcmc}{4}%
{nstars-rem planets-cuts1-cut-7-nomcmc}{3}%
{nstars planets-cuts1-cut-all}{888}%
{nplanets planets-cuts1-cut-all}{1246}%
{nplanets field-cuts-cut-0-none}{184941}%
{nstars field-cuts-cut-0-none}{184933}%
{nplanets field-cuts-cut-1-faint}{77245}%
{nstars field-cuts-cut-1-faint}{77243}%
{nplanets-rem field-cuts-cut-1-faint}{107696}%
{nstars-rem field-cuts-cut-1-faint}{107690}%
{nplanets field-cuts-cut-2-giantcmd}{47503}%
{nstars field-cuts-cut-2-giantcmd}{47502}%
{nplanets-rem field-cuts-cut-2-giantcmd}{29742}%
{nstars-rem field-cuts-cut-2-giantcmd}{29741}%
{nplanets field-cuts-cut-3-diluted}{46468}%
{nstars field-cuts-cut-3-diluted}{46468}%
{nplanets-rem field-cuts-cut-3-diluted}{1035}%
{nstars-rem field-cuts-cut-3-diluted}{1034}%
{nplanets field-cuts-cut-4-ruwe}{39423}%
{nstars field-cuts-cut-4-ruwe}{39423}%
{nplanets-rem field-cuts-cut-4-ruwe}{7045}%
{nstars-rem field-cuts-cut-4-ruwe}{7045}%
{nstars field-cuts-cut-all}{39423}%
{nplanets field-cuts-cut-all}{39423}%
{nplanets planets-cuts2-cut-0-none}{1246}%
{nstars planets-cuts2-cut-0-none}{888}%
{nplanets planets-cuts2-cut-1-badvsini}{1227}%
{nstars planets-cuts2-cut-1-badvsini}{871}%
{nplanets-rem planets-cuts2-cut-1-badvsini}{19}%
{nstars-rem planets-cuts2-cut-1-badvsini}{17}%
{nplanets planets-cuts2-cut-2-badspecparallax}{1215}%
{nstars planets-cuts2-cut-2-badspecparallax}{862}%
{nplanets-rem planets-cuts2-cut-2-badspecparallax}{12}%
{nstars-rem planets-cuts2-cut-2-badspecparallax}{9}%
{nplanets planets-cuts2-cut-3-sb2}{1214}%
{nstars planets-cuts2-cut-3-sb2}{861}%
{nplanets-rem planets-cuts2-cut-3-sb2}{1}%
{nstars-rem planets-cuts2-cut-3-sb2}{1}%
{nplanets planets-cuts2-cut-4-badimpact}{1193}%
{nstars planets-cuts2-cut-4-badimpact}{843}%
{nplanets-rem planets-cuts2-cut-4-badimpact}{21}%
{nstars-rem planets-cuts2-cut-4-badimpact}{18}%
{nplanets planets-cuts2-cut-5-badimpacttau}{973}%
{nstars planets-cuts2-cut-5-badimpacttau}{704}%
{nplanets-rem planets-cuts2-cut-5-badimpacttau}{220}%
{nstars-rem planets-cuts2-cut-5-badimpacttau}{139}%
{nplanets planets-cuts2-cut-6-badprad}{973}%
{nstars planets-cuts2-cut-6-badprad}{704}%
{nplanets-rem planets-cuts2-cut-6-badprad}{0}%
{nstars-rem planets-cuts2-cut-6-badprad}{0}%
{nplanets planets-cuts2-cut-7-badpradprec}{970}%
{nstars planets-cuts2-cut-7-badpradprec}{703}%
{nplanets-rem planets-cuts2-cut-7-badpradprec}{3}%
{nstars-rem planets-cuts2-cut-7-badpradprec}{1}%
{nstars planets-cuts2-cut-all}{703}%
{nplanets planets-cuts2-cut-all}{970}%
{nstars smemp}{366}%
{nstars smsyn}{1350}%
{kmag-err-med}{0.02}%
{kmag-err-med kepmag<14.2}{0.02}%
{kmag-err-med kepmag>14.2}{0.03}%
{parallax-ferr-med}{1.4}%
{parallax-ferr-med kepmag<14.2}{1.2}%
{parallax-ferr-med kepmag>14.2}{1.8}%
{ak-med kepmag<14.2}{0.01}%
{ak-med kepmag>14.2}{0.01}%
{ak-max kepmag<14.2}{0.07}%
{ak-max kepmag>14.2}{0.06}%
{cks_steff-err-med}{100}%
{cks_smet-err-med}{0.06}%
{gdir_srad-ferr-med}{3.9}%
{gdir_srad-ferr-med kepmag<14.2}{4.0}%
{gdir_srad-ferr-med kepmag>14.2}{4.4}%
{giso_smass-ferr-med}{3.7}%
{giso_smass-ferr-med kepmag<14.2}{3.8}%
{giso_smass-ferr-med kepmag>14.2}{3.7}%
{giso_srad-ferr-med}{2.6}%
{giso_srho-ferr-med}{8.9}%
{giso_srho-ferr-med kepmag<14.2}{8.7}%
{giso_srho-ferr-med kepmag>14.2}{9.7}%
{giso_slogage-err-med}{0.24}%
{giso2_sparallax-ferr-med}{16}%
{dr25_period-ferr-med}{3}%
{dr25_ror-ferr-med}{4.3}%
{dr25_ror-ferr-med kepmag<14.2}{3.9}%
{dr25_ror-ferr-med kepmag>14.2}{4.8}%
{dr25_tau-ferr-med}{2.3}%
{gdir_prad-ferr-med}{5.7}%
{gdir_prad-ferr-med kepmag<14.2}{5.6}%
{gdir_prad-ferr-med kepmag>14.2}{6.3}%
{giso_tau0-ferr-med}{2.9}%
{giso_sma-ferr-med}{1.2}%
{giso_sinc-ferr-med}{9.2}%
{gdir_prad-on_koi-prad_ratio-std}{0.20}%
{gdir_prad-on_koi-prad_ratio-mean}{1.00}%
{occ-nplnt_smass=0.5-1.4}{954}%
{occ-nplnt_smass=0.5-0.7}{185}%
{occ-nplnt_smass=0.7-1.0}{389}%
{occ-nplnt_smass=1.0-1.4}{380}%
}[{{\color{red}XX}}]%
} % numbers
\newcommand{\fit}[1]{%%
\IfEqCase{#1}{%
{sn-m-R0}{2.49 \pm 0.02}%
{sn-m-alpha-mass}{0.25 \pm 0.03}%
{sn-m-alpha-met}{0.00 \pm 0.00}%
{sn-m-alpha-age}{0.00 \pm 0.00}%
{se-m-R0}{1.35 \pm 0.01}%
{se-m-alpha-mass}{0.02 \pm 0.03}%
{se-m-alpha-met}{0.00 \pm 0.00}%
{se-m-alpha-age}{0.00 \pm 0.00}%
{sn-mm-R0}{2.49 \pm 0.02}%
{sn-mm-alpha-mass}{0.25 \pm 0.04}%
{sn-mm-alpha-met}{-0.01 \pm 0.02}%
{sn-mm-alpha-age}{0.00 \pm 0.00}%
{se-mm-R0}{1.35 \pm 0.01}%
{se-mm-alpha-mass}{0.02 \pm 0.03}%
{se-mm-alpha-met}{-0.00 \pm 0.02}%
{se-mm-alpha-age}{0.00 \pm 0.00}%
{sn-mma-R0}{2.51 \pm 0.04}%
{sn-mma-alpha-mass}{0.26 \pm 0.11}%
{sn-mma-alpha-met}{-0.04 \pm 0.03}%
{sn-mma-alpha-age}{-0.01 \pm 0.03}%
{se-mma-R0}{1.35 \pm 0.02}%
{se-mma-alpha-mass}{0.10 \pm 0.11}%
{se-mma-alpha-met}{0.02 \pm 0.03}%
{se-mma-alpha-age}{0.02 \pm 0.02}%
{mps_size-se-alpha}{-0.01 \pm 0.03}%
{mps_size-se-R0}{1.34 \pm 0.01}%
{mps_size-sn-alpha}{0.23 \pm 0.04}%
{mps_size-sn-R0}{2.47 \pm 0.03}%
{fitper_smass=0.5-0.7-prad=1.0-1.7-logf}{-0.37^{ +0.07 }_{ -0.07 }}%
{fitper_smass=0.5-0.7-prad=1.0-1.7-f}{0.42^{ +0.07 }_{ -0.06 }}%
{fitper_smass=0.5-0.7-prad=1.0-1.7-k1}{2.4^{ +1.1 }_{ -1.0 }}%
{fitper_smass=0.5-0.7-prad=1.0-1.7-logx0}{0.4^{ +0.1 }_{ -0.1 }}%
{fitper_smass=0.5-0.7-prad=1.0-1.7-x0}{2.6^{ +0.8 }_{ -0.4 }}%
{fitper_smass=0.5-0.7-prad=1.7-4.0-logf}{0.14^{ +0.06 }_{ -0.06 }}%
{fitper_smass=0.5-0.7-prad=1.7-4.0-f}{1.37^{ +0.19 }_{ -0.18 }}%
{fitper_smass=0.5-0.7-prad=1.7-4.0-k1}{1.7^{ +0.5 }_{ -0.4 }}%
{fitper_smass=0.5-0.7-prad=1.7-4.0-logx0}{0.8^{ +0.1 }_{ -0.1 }}%
{fitper_smass=0.5-0.7-prad=1.7-4.0-x0}{7.0^{ +1.4 }_{ -1.1 }}%
{fitper_smass=0.7-1.0-prad=1.0-1.7-logf}{-0.65^{ +0.05 }_{ -0.05 }}%
{fitper_smass=0.7-1.0-prad=1.0-1.7-f}{0.22^{ +0.03 }_{ -0.02 }}%
{fitper_smass=0.7-1.0-prad=1.0-1.7-k1}{1.5^{ +0.6 }_{ -0.5 }}%
{fitper_smass=0.7-1.0-prad=1.0-1.7-logx0}{0.5^{ +0.1 }_{ -0.1 }}%
{fitper_smass=0.7-1.0-prad=1.0-1.7-x0}{3.3^{ +0.8 }_{ -0.5 }}%
{fitper_smass=0.7-1.0-prad=1.7-4.0-logf}{-0.10^{ +0.04 }_{ -0.04 }}%
{fitper_smass=0.7-1.0-prad=1.7-4.0-f}{0.79^{ +0.08 }_{ -0.08 }}%
{fitper_smass=0.7-1.0-prad=1.7-4.0-k1}{1.5^{ +0.4 }_{ -0.3 }}%
{fitper_smass=0.7-1.0-prad=1.7-4.0-logx0}{1.0^{ +0.1 }_{ -0.1 }}%
{fitper_smass=0.7-1.0-prad=1.7-4.0-x0}{9.7^{ +2.0 }_{ -1.5 }}%
{fitper_smass=1.0-1.4-prad=1.0-1.7-logf}{-0.75^{ +0.05 }_{ -0.05 }}%
{fitper_smass=1.0-1.4-prad=1.0-1.7-f}{0.18^{ +0.02 }_{ -0.02 }}%
{fitper_smass=1.0-1.4-prad=1.0-1.7-k1}{1.9^{ +0.7 }_{ -0.5 }}%
{fitper_smass=1.0-1.4-prad=1.0-1.7-logx0}{0.7^{ +0.1 }_{ -0.1 }}%
{fitper_smass=1.0-1.4-prad=1.0-1.7-x0}{5.2^{ +1.3 }_{ -0.9 }}%
{fitper_smass=1.0-1.4-prad=1.7-4.0-logf}{-0.39^{ +0.04 }_{ -0.04 }}%
{fitper_smass=1.0-1.4-prad=1.7-4.0-f}{0.41^{ +0.04 }_{ -0.04 }}%
{fitper_smass=1.0-1.4-prad=1.7-4.0-k1}{2.2^{ +0.6 }_{ -0.5 }}%
{fitper_smass=1.0-1.4-prad=1.7-4.0-logx0}{0.9^{ +0.1 }_{ -0.1 }}%
{fitper_smass=1.0-1.4-prad=1.7-4.0-x0}{8.6^{ +1.6 }_{ -1.2 }}%
{fitsinc_smass=0.5-0.7-prad=1.0-1.7-logf}{-0.48^{ +0.07 }_{ -0.07 }}%
{fitsinc_smass=0.5-0.7-prad=1.0-1.7-f}{0.33^{ +0.06 }_{ -0.05 }}%
{fitsinc_smass=0.5-0.7-prad=1.0-1.7-k2}{-3.1^{ +0.4 }_{ -0.5 }}%
{fitsinc_smass=0.5-0.7-prad=1.0-1.7-logx0}{2.0^{ +0.1 }_{ -0.2 }}%
{fitsinc_smass=0.5-0.7-prad=1.0-1.7-x0}{89^{ +35 }_{ -31 }}%
{fitsinc_smass=0.5-0.7-prad=1.7-4.0-logf}{0.05^{ +0.06 }_{ -0.06 }}%
{fitsinc_smass=0.5-0.7-prad=1.7-4.0-f}{1.13^{ +0.16 }_{ -0.14 }}%
{fitsinc_smass=0.5-0.7-prad=1.7-4.0-k2}{-3.2^{ +0.3 }_{ -0.3 }}%
{fitsinc_smass=0.5-0.7-prad=1.7-4.0-logx0}{1.4^{ +0.1 }_{ -0.1 }}%
{fitsinc_smass=0.5-0.7-prad=1.7-4.0-x0}{27^{ +6 }_{ -6 }}%
{fitsinc_smass=0.7-1.0-prad=1.0-1.7-logf}{-0.58^{ +0.05 }_{ -0.05 }}%
{fitsinc_smass=0.7-1.0-prad=1.0-1.7-f}{0.26^{ +0.03 }_{ -0.03 }}%
{fitsinc_smass=0.7-1.0-prad=1.0-1.7-k2}{-2.6^{ +0.3 }_{ -0.3 }}%
{fitsinc_smass=0.7-1.0-prad=1.0-1.7-logx0}{2.4^{ +0.1 }_{ -0.1 }}%
{fitsinc_smass=0.7-1.0-prad=1.0-1.7-x0}{265^{ +84 }_{ -76 }}%
{fitsinc_smass=0.7-1.0-prad=1.7-4.0-logf}{-0.12^{ +0.04 }_{ -0.05 }}%
{fitsinc_smass=0.7-1.0-prad=1.7-4.0-f}{0.76^{ +0.08 }_{ -0.08 }}%
{fitsinc_smass=0.7-1.0-prad=1.7-4.0-k2}{-3.1^{ +0.3 }_{ -0.3 }}%
{fitsinc_smass=0.7-1.0-prad=1.7-4.0-logx0}{1.8^{ +0.1 }_{ -0.1 }}%
{fitsinc_smass=0.7-1.0-prad=1.7-4.0-x0}{70^{ +16 }_{ -14 }}%
{fitsinc_smass=1.0-1.4-prad=1.0-1.7-logf}{-0.55^{ +0.05 }_{ -0.05 }}%
{fitsinc_smass=1.0-1.4-prad=1.0-1.7-f}{0.28^{ +0.04 }_{ -0.03 }}%
{fitsinc_smass=1.0-1.4-prad=1.0-1.7-k2}{-3.0^{ +0.4 }_{ -0.5 }}%
{fitsinc_smass=1.0-1.4-prad=1.0-1.7-logx0}{2.8^{ +0.1 }_{ -0.1 }}%
{fitsinc_smass=1.0-1.4-prad=1.0-1.7-x0}{615^{ +159 }_{ -152 }}%
{fitsinc_smass=1.0-1.4-prad=1.7-4.0-logf}{-0.28^{ +0.05 }_{ -0.05 }}%
{fitsinc_smass=1.0-1.4-prad=1.7-4.0-f}{0.52^{ +0.06 }_{ -0.05 }}%
{fitsinc_smass=1.0-1.4-prad=1.7-4.0-k2}{-3.0^{ +0.3 }_{ -0.4 }}%
{fitsinc_smass=1.0-1.4-prad=1.7-4.0-logx0}{2.4^{ +0.1 }_{ -0.1 }}%
{fitsinc_smass=1.0-1.4-prad=1.7-4.0-x0}{244^{ +64 }_{ -55 }}%
}[{{\color{red}XX}}]%
} % numbers
\newcommand{\hand}[1]{%%
\IfEqCase{#1}{%
{n-koi}{6862}%
{gapfit-per-smass=all-m}{$-0.06_{-0.03}^{+0.02}$}% 
{gapfit-per-smass=all-rp}{$1.82^{+0.05}_{-0.04}$}% 
{gapfit-sinc-smass=all-m}{$0.05_{-0.02}^{+0.01}$}% 
{gapfit-sinc-smass=all-rp}{$1.86^{+0.04}_{-0.04}$}% 
{gapfit-per-smass=low-m}{$-0.02 \pm 0.02$}% 
{gapfit-per-smass=low-rp}{$1.70 \pm 0.05$}% 
{gapfit-per-smass=med-m}{$-0.04 \pm 0.03$}% 
{gapfit-per-smass=med-rp}{$1.79 \pm 0.03$}% 
{gapfit-per-smass=hi-m}{$-0.03 \pm 0.02$}% 
{gapfit-per-smass=hi-rp}{$1.95 \pm 0.06$}% 
{gapfit-sinc-smass=low-m}{$0.00 \pm 0.02$}% 
{gapfit-sinc-smass=low-rp}{$0.03^{+0.05}_{-0.01}$}% 
{gapfit-sinc-smass=med-m}{$0.03 \pm 0.02$}% 
{gapfit-sinc-smass=med-rp}{$1.70^{+0.04}_{-0.05}$}% 
{gapfit-sinc-smass=hi-m}{$0.03 \pm 0.02$}%
{gapfit-sinc-smass=hi-rp}{$1.86^{+0.09}_{-0.06}$}% 
}[XX]%
}
\newcommand{\grad}[1]{%%
\IfEqCase{#1}{%
{per-prad-det_smass=0.5-0.7-Rp0}{1.78^{ +0.05 }_{ -0.06 }}%
{per-prad-det_smass=0.5-0.7-m}{-0.12^{ +0.06 }_{ -0.04 }}%
{per-prad-det_smass=0.7-1.0-Rp0}{1.74^{ +0.09 }_{ -0.08 }}%
{per-prad-det_smass=0.7-1.0-m}{-0.13^{ +0.07 }_{ -0.06 }}%
{per-prad-det_smass=1.0-1.4-Rp0}{1.93^{ +0.04 }_{ -0.04 }}%
{per-prad-det_smass=1.0-1.4-m}{-0.06^{ +0.02 }_{ -0.02 }}%
{per-prad-det_smass=0.5-1.4-Rp0}{1.84^{ +0.03 }_{ -0.03 }}%
{per-prad-det_smass=0.5-1.4-m}{-0.11^{ +0.02 }_{ -0.02 }}%
{sinc-prad-det_smass=0.5-0.7-Rp0}{2.00^{ +0.11 }_{ -0.19 }}%
{sinc-prad-det_smass=0.5-0.7-m}{0.07^{ +0.02 }_{ -0.04 }}%
{sinc-prad-det_smass=0.7-1.0-Rp0}{1.86^{ +0.07 }_{ -0.07 }}%
{sinc-prad-det_smass=0.7-1.0-m}{0.10^{ +0.04 }_{ -0.04 }}%
{sinc-prad-det_smass=1.0-1.4-Rp0}{1.85^{ +0.04 }_{ -0.04 }}%
{sinc-prad-det_smass=1.0-1.4-m}{0.05^{ +0.01 }_{ -0.01 }}%
{sinc-prad-det_smass=0.5-1.4-Rp0}{1.86^{ +0.04 }_{ -0.03 }}%
{sinc-prad-det_smass=0.5-1.4-m}{0.06^{ +0.01 }_{ -0.01 }}%
{smass-prad-det_smass=0.5-1.4-Rp0}{1.86^{ +0.03 }_{ -0.03 }}%
{smass-prad-det_smass=0.5-1.4-m}{0.18^{ +0.08 }_{ -0.07 }}%
{smet-prad-det_smass=0.5-1.4-Rp0}{1.82^{ +0.05 }_{ -0.03 }}%
{smet-prad-det_smass=0.5-1.4-m}{0.01^{ +0.05 }_{ -0.06 }}%
{per-prad-occ_smass=0.5-0.7-Rp0}{1.71^{ +0.05 }_{ -0.04 }}%
{per-prad-occ_smass=0.5-0.7-m}{-0.06^{ +0.04 }_{ -0.05 }}%
{per-prad-occ_smass=0.7-1.0-Rp0}{1.70^{ +0.09 }_{ -0.10 }}%
{per-prad-occ_smass=0.7-1.0-m}{-0.10^{ +0.07 }_{ -0.05 }}%
{per-prad-occ_smass=1.0-1.4-Rp0}{1.92^{ +0.06 }_{ -0.05 }}%
{per-prad-occ_smass=1.0-1.4-m}{-0.05^{ +0.03 }_{ -0.03 }}%
{per-prad-occ_smass=0.5-1.4-Rp0}{1.80^{ +0.03 }_{ -0.03 }}%
{per-prad-occ_smass=0.5-1.4-m}{-0.06^{ +0.03 }_{ -0.04 }}%
{sinc-prad-occ_smass=0.5-0.7-Rp0}{1.93^{ +0.11 }_{ -0.13 }}%
{sinc-prad-occ_smass=0.5-0.7-m}{0.06^{ +0.02 }_{ -0.03 }}%
{sinc-prad-occ_smass=0.7-1.0-Rp0}{1.81^{ +0.08 }_{ -0.06 }}%
{sinc-prad-occ_smass=0.7-1.0-m}{0.08^{ +0.04 }_{ -0.04 }}%
{sinc-prad-occ_smass=1.0-1.4-Rp0}{1.84^{ +0.04 }_{ -0.04 }}%
{sinc-prad-occ_smass=1.0-1.4-m}{0.05^{ +0.01 }_{ -0.01 }}%
{sinc-prad-occ_smass=0.5-1.4-Rp0}{1.86^{ +0.04 }_{ -0.04 }}%
{sinc-prad-occ_smass=0.5-1.4-m}{0.06^{ +0.01 }_{ -0.01 }}%
}[{{\color{red}XX}}]%
} % numbers

\begin{document}
\pagenumbering{arabic}
%\thispagestyle{empty}

%\vfill

\title{The California-Kepler Survey. X.\\ The Radius Gap as a Function of Stellar Mass, Metallicity, and Age.}

\author[0000-0003-0967-2893]{Erik A.\ Petigura}
\affiliation{Department of Physics \& Astronomy, University of California Los Angeles, Los Angeles, CA 90095, USA}

\author[0000-0001-7615-6798]{James G. Rogers}
\affiliation{Astrophysics Group, Imperial College London, Blackett Laboratory, Prince Consort Road, London SW7 2AZ, UK}

\author[0000-0002-0531-1073]{Howard Isaacson}
\affiliation{{Department of Astronomy,  University of California Berkeley, Berkeley CA 94720, USA}}
\affiliation{Centre for Astrophysics, University of Southern Queensland, Toowoomba, QLD, Australia}

\author[0000-0002-4856-7837]{James E. Owen}
\affiliation{Astrophysics Group, Imperial College London, Blackett Laboratory, Prince Consort Road, London SW7 2AZ, UK}

\author[0000-0001-9811-568X]{Adam L. Kraus}
\affiliation{Department of Astronomy, The University of Texas at Austin, Austin, TX 78712, USA}

\author[0000-0002-4265-047X]{Joshua N. Winn}
\affiliation{Princeton University, Princeton, NJ 08540, USA}

\author[0000-0003-2562-9043]{Mason G.\ MacDougall}
\affiliation{Department of Physics \& Astronomy, University of California Los Angeles, Los Angeles, CA 90095, USA}

\author[0000-0001-8638-0320]{Andrew W. Howard}
\affiliation{Department of Astronomy, California Institute of Technology, Pasadena, CA 91125, USA}

\author[0000-0003-3504-5316]{Benjamin Fulton}
\affiliation{NASA Exoplanet Science Institute/Caltech-IPAC, MC 314-6, 1200 E. California Blvd., Pasadena, CA 91125, USA}

\author[0000-0002-6115-4359]{Molly R. Kosiarek}
\affiliation{Department of Astronomy and Astrophysics, University of California, Santa Cruz, CA 95064, USA}

\author[0000-0002-3725-3058]{Lauren M. Weiss}
\affiliation{Department of Physics, University of Notre Dame, Notre Dame, IN 46556, USA}

\author[0000-0003-0012-9093]{Aida Behmard}
\affiliation{Division of Geological and Planetary Sciences, California Institute of Technology, Pasadena, CA 91125, USA}

\author[0000-0002-3199-2888]{Sarah Blunt}
\affiliation{Department of Astronomy, California Institute of Technology, Pasadena, CA, USA}

\begin{abstract}
In 2017, the California-\Kepler Survey (CKS) published its first data release (DR1) of high-resolution optical spectra of 1305 planet hosts. Refined CKS planet radii revealed that small planets are bifurcated into two distinct populations: super-Earths (smaller than 1.5 \Re) and sub-Neptunes (between 2.0 and 4.0 \Re), with few planets in between (the ``Radius Gap.'') Several theoretical models of the Radius Gap predict variation with stellar mass, but testing these predictions are challenging with CKS DR1 due to its limited \Mstar range of 0.8--1.4~\Msun. Here, we present CKS DR2 with \sample{nstars dr2} additional spectra and derived properties focusing on stars of 0.5--0.8~\Msun. We found the Radius Gap follows $\Rp \propto P^{m}$ with $m = -0.10 \pm 0.03$, consistent with predictions of XUV- and core-powered mass-loss mechanisms. We found no evidence that $m$ varies with \Mstar. We observed a correlation between the average sub-Neptune size and \Mstar. Over 0.5 to 1.4~\Msun, the average sub-Neptune grows from 2.1 to 2.6~\Re, following $\Rp \propto \Mstar^\alpha$ with $\alpha = \fit{sn-m-alpha-mass}$. In contrast, there is no detectable change for super-Earths. These \Mstar-\Rp trends suggests that protoplanetary disks can efficiently produce cores up to a threshold mass of \Mcore, which grows linearly with stellar mass according to $\Mcore \approx 10~\Me (\Mstar / \Msun)$. There is no significant correlation between sub-Neptune size and stellar metallicity (over $-$0.5 to $+$0.5~dex) suggesting a weak relationship between planet envelope opacity and stellar metallicity. Finally, there is no significant variation in sub-Neptune size with stellar age (over 1 to 10~Gyr), which suggests that the majority of envelope contraction concludes after $\sim$1~Gyr.
\end{abstract}


\section{Introduction}
\label{sec:intro}
The key legacy of NASA's \Kepler mission is its sample of over 4000 extrasolar planets. These planets were discovered from precise, nearly continuous photometry obtained over four years of roughly 200,000 stars. Importantly, the \Kepler planet population covers a wide range of sizes and orbital periods; sizes range from super-Jupiter to sub-Mercury and orbital periods extend from less than a day to more than a year \citep{Thompson18}. The \Kepler exoplanet census has, and will continue to, shed light on the diverse outcomes of planet formation.

The distribution of \Kepler planets encodes key aspects of planet formation physics including the growth of solid cores, the accretion (and loss) of gaseous envelopes, and the prevalence of orbital migration. Gaining insights into these processes requires detailed knowledge of host stars. Until 2017, the properties of the vast majority of \Kepler planet host stars were based on broadband photometry, which limited the accuracy of star and planet properties. Importantly, photometric stellar radii \Rstar were uncertain at the $\sim$40\% level \citep{Brown11}.

To address these limitations, our group conducted the California-\Kepler Survey (CKS; \citealt{Petigura17b}), a spectroscopic survey of 1305 planet hosting stars observed by \Kepler (technically ``\Kepler Objects of Interest'' or ``KOIs''). CKS consisted of several overlapping samples, the largest of which was magnitude-limited and included 960 planet hosts brighter than \Kepler-band magnitude \kepmag = 14.2. Using these spectra, we measured \teff, \logg, and \fe and refined stellar \Mstar, \Rstar, and age along with planet size \Rp and incident stellar flux \Sinc.

CKS stellar properties enabled several insights into the occurrence and properties of planets as a function of stellar metallicity \citep{Petigura18b}, stellar age \citep{Berger18}, planet multiplicity \citep{Weiss18a}, stellar obliquity \citep{Winn17b}, and other properties. One such insight flowed directly from the improvement in \Rstar errors from 40\% to 10\% --- the radius distribution of small planets is bimodal \citep{Fulton17}. This result has been confirmed in samples of planets with even smaller radius uncertainties that have leveraged asteroseismic or parallax constraints \citep{Van-Eylen18,Fulton18b,Berger18b,Petigura20b}.

This Radius Gap was predicted by several groups who considered the effect of photoevaporation on planetary envelopes by X-ray and extreme ultraviolet (XUV) radiation (\citealt{Lopez13,Owen13,Jin14,Chen16}). Other processes that inject energy into planet envelopes may also produce the gap such as the luminosity from cooling cores \citep{Ginzburg18,Gupta19} or the heat from accreting planetesimals \citep{Chatterjee18,Zeng19,Wyatt20}. 

Projecting the planet population along the axis of stellar mass should help illuminate which processes sculpt the distribution of small planets. For example, photoevaporation models predict that small planets will lose their envelopes out to lower bolometric flux levels around low-mass stars due to increased $F_\mathrm{xuv} / F_\mathrm{bol}$.

Such studies are challenging with CKS DR1 because it spans a limited range of $\Mstar \approx0.8$--1.4~\Msun. There are several reasons for this: (1) the \Kepler selection function favored FGK stars \citep{Batalha10}; (2) most of the \Kepler stars below 0.8~\Msun are fainter than \kepmag = 14.2; (3) the spectroscopic techniques used in \cite{Petigura17b} could not return reliable parameters for stars cooler than \teff = 4700~K. Despite the limited \Mstar range, analyses of this dataset indicated an increase in the average size of sub-Neptunes with increasing stellar mass \citep{Fulton18b,Petigura20b}. In addition, \cite{Wu19} and \cite{Berger20a} found a positive \Mstar-\Rp correlation for sub-Neptunes. 

The \teff > 4700~K boundary in CKS poses an additional challenge for interpreting trends with \Mstar. This boundary corresponds to \Mstar > 0.65~\Msun and \Mstar > 0.80~\Msun at $\fe = -0.4$~dex and $+0.4$~dex, respectively, the upper and lower ends of the CKS metallicity distribution. \cite{Owen18b} discussed how this correlation acts as a confounding factor for interpreting both mass and metallicity trends in CKS DR1.

Here, we present a follow-on survey designed to expand the stellar mass range of the CKS sample. We constructed a sample of planet hosts spanning 0.5--1.4~\Msun (\S\ref{sec:sample}) and gathered Keck/HIRES spectra of those absent from CKS DR1. New spectra of \sample{nstars dr2} hosts are presented here as CKS DR2 (\S\ref{sec:hires}). We analyzed the combined DR1/DR2 samples to derive temperatures, metallicities, and limits on stellar companions, which we combined with astrometric and photometric  measurements to derive stellar masses, radii, and ages (\S\ref{sec:host-star}). With updated stellar properties, we refined the precision and purity of the planet catalog (\S\ref{sec:planets}). We analyzed both the distribution of detected planets  (\S\ref{sec:planet-population}) and the underlying occurrence rate (\S\ref{sec:occurrence}). We offer some comparisons to planet formation theory (\S\ref{sec:discussion}) and provide a brief summary and conclusion (\S\ref{sec:conclusion}).

\section{Sample selection}
\label{sec:sample}

\begin{figure*}
\includegraphics[width=0.9\textwidth]{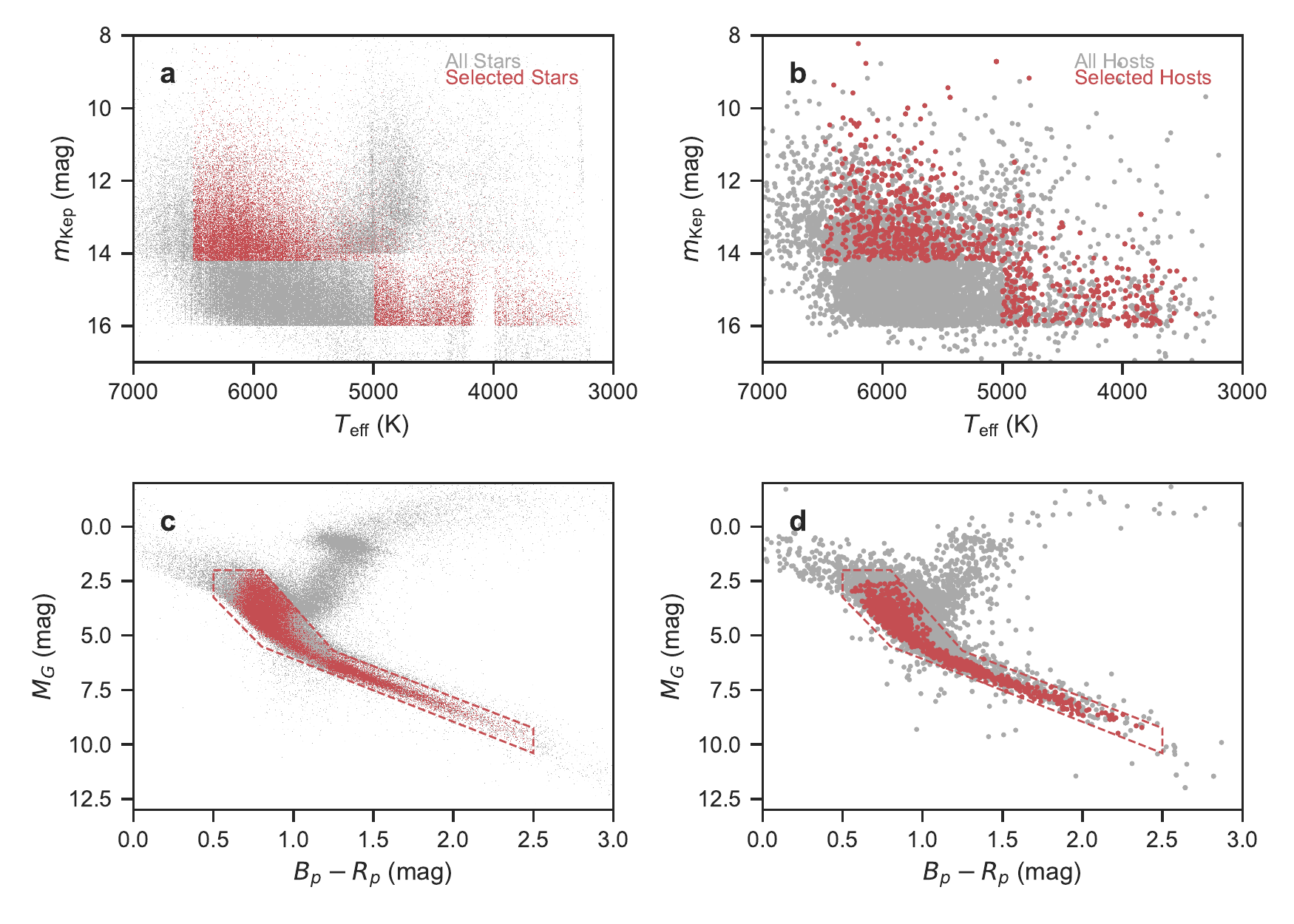}
\centering
\caption{Panel (a): Gray points show effective temperature and apparent magnitude \kepmag of stars observed by \Kepler; red points are stars that pass the selection criteria described in \S\ref{sec:sample}. Panel (b): same as (a), but for planet hosts. Panel (c): same as (a) but showing \Gaia absolute $G$-band magnitude $M_G$ and $B_p - R_p$ color. The boundaries of our main-sequence filter are shown as red-dashed lines. 
%The upper boundary connects the following points $(B_p-R_p, M_G)$ = (0.8,2), (1.25,5.7), and (2.5,9.25); the lower boundary connects (0.5,3.25), (0.8,5.5), and (2.5,10.4). 
%
Panel (d): same as (c) but for planet hosts.} 
\label{fig:cuts-all}
\end{figure*}

Our main goal was to measure planet occurrence as a function of stellar mass. This requires a sample of planets drawn from a well-defined sample of stars spanning a broad range in stellar mass. Before describing our selection function in detail, we first orient the reader with an overview of our survey design and practical considerations. When we designed our survey in 2018, our goal was to increase the number of low-mass stars with CKS-quality parameters given a fixed telescope allocation of 15 Keck/HIRES nights. Simply extending the magnitude-limited survey beyond \kepmag = 14.2 would have included many stars with $\teff \approx 5000$--6500~K that were already well-sampled by DR1 (see Figure~\ref{fig:cuts-all}). Observing all targets down to \kepmag = 16.0 in the manner described in \S\ref{sec:hires} would have required $\sim$75 nights and was impractical. Instead, we adopted a selection function and observing strategy that favored cool stars but mitigated magnitude-dependent effects. The two mitigation strategies are described in detail in \S\ref{sec:sample}--\ref{sec:planets} but briefly involved selecting single isolated stars and collecting spectra of homogeneous quality.

We began with \sample{nstars field-cuts-cut-0-none} stars in the union of the following three catalogs: the \Kepler project Data Release 25 (DR25) stellar properties catalog (\citealt{Mathur17}, M17 hereafter), the \Gaia DR2 catalog \citep{Gaia18}, and the \cite{Berger20a} catalog of stellar properties (B20 hereafter). We then applied the following filters:

\begin{enumerate}
\item {\em Stellar brightness and effective temperature.} Our sample consists of two magnitude-limited components. If \teff = 5000--6500~K (as measured by M17), we required \kepmag < 14.2; if \teff = 3000--5000~K, we required \kepmag < 16.0. A total of \sample{nstars field-cuts-cut-1-faint} stars remained.

\item {\em Main-sequence stars.} We selected main-sequence stars using their position in the \Gaia color-magnitude diagram. Stars were included if they reside within the boundaries shown in Figure~\ref{fig:cuts-all}. A total of \sample{nstars field-cuts-cut-2-giantcmd} stars remained.

\item {\em Single stars: Gaia source catalog.} If a transit is observed within a \Kepler aperture containing more than one star, it is often not possible to identify the star hosting the transiting object. Whichever star is the host, the transit depth is diluted by the neighboring star(s) and the inferred planet radius will be biased to smaller values. To ensure high-precision \Rp, we worked to exclude targets that are diluted by >10\%, which would result in a \Rp error of >5\%. We excluded stars where the cumulative $G$-band flux contribution from all \Gaia-identified neighboring stars within 4~arcsec exceeded 10\%. Companions outside of this limit are resolved as separate sources in the KIC, which had an average FWHM of 2.5~arcsec \citep{Brown11}. The \Kepler project removed the diluting flux from such sources during the photometric extraction (see the discussion of the ``crowding metric'' in \citealt{Stumpe12}). A total of \sample{nstars field-cuts-cut-3-diluted} stars remained.

\item {\em Single stars: Gaia astrometric noise.} Companions within 1~arcsec are often not resolved as separate sources in \Gaia DR2, but can be identified using the residuals to the astrometric fits. \Gaia Re-normalized Unit Weight Error (RUWE) is conceptually similar to the square root of the reduced $\chi^2$ and is close to unity when the astrometric data are consistent with parallax and proper motion only, with no additional variation due to any companions \citep{Lindegren18}. Analyses of stars with high-resolution imaging have demonstrated that requiring RUWE < 1.2 removes companions with $\Delta m_G < 3$~mag and separations of $\rho$ = 0.1--1~arcsec (\citealt{Bryson20}, \citealt{Wood21}, Kraus et al. in preparation). A total of \sample{nstars field-cuts-cut-4-ruwe} stars remained.

\end{enumerate}
We cross-referenced these stars against the \cite{Thompson18} catalog of KOIs, T18 hereafter, which we accessed from the NASA Exoplanet Archive (NEA; \citealt{Akeson13}). This catalog was derived from the final planet search conducted by the \Kepler project (DR25). Among our filtered stellar sample, T18 lists \sample{nstars field-cuts-cut-4-ruwe} stars hosting \sample{nplanets planets-cuts1-cut-4-ruwe} KOIs. We subjected the KOIs to the following additional filters:
\begin{enumerate}
\setcounter{enumi}{4}

\item {\em Candidate reliability.} T18 dispositioned KOIs using a fully automated procedure called the ``Robovetter'' that performed a series of tests on a suite of data quality metrics designed to identify false positives due to data anomalies and eclipsing binaries. When setting the Robovetter thresholds, T18 emulated the human classification of Threshold Crossing Event Review Team (TCERT). Robovetter dispositioned all KOIs as either ``False Positive'' or ``Planet Candidate.'' Some KOIs fall near the boundary between the two dispositions. T18 quantified the proximity of each KOI to this boundary through Monte Carlo simulations. T18 reported a ``disposition score'' which is the fraction of Monte Carlo simulations classified as planet candidates. We required ``Planet Candidate'' status%
\footnote{NEA data column: {\em koi\_pdisposition}}
with a disposition score%
\footnote{NEA data column: {\em koi\_score}}
exceeding 75\% . Finally, we removed a handful of known false positives that cleared the Robovetter that are listed in the DR25 supplemental table.%
\footnote{NEA data column: {\em koi\_pdisposition}}
A total of \sample{nplanets planets-cuts1-cut-5-notreliable} KOIs orbiting \sample{nstars planets-cuts1-cut-5-notreliable} host stars remained. 

\item {\em Candidate signal-to-noise ratio (S/N).} Low S/N candidates are more likely to be false positives (see, e.g., \citealt{Bryson20}) and yield lower overall precision on $\Rp/\Rstar$. The \Kepler project reported S/N as a ``Multiple Event Statistic'' (MES; \citealt{Jenkins02}).%
\footnote{NEA data column: {\em koi\_max\_mult\_ev}}
We required MES > 10. A total of \sample{nplanets planets-cuts1-cut-6-lowsnr} KOIs orbiting \sample{nstars planets-cuts1-cut-6-lowsnr} host stars remained. 

\item {\em Transit Modeling.} To compute planet properties, we used posterior samples of transit parameters derived from Markov Chain Monte Carlo (MCMC). A small fraction of the T18 KOIs were not modeled using MCMC.%
\footnote{NEA data column: {\em koi\_fittype}}
Most of these are dispositioned as false positives. We removed a handful of candidates that passed the previous cuts. A total of \sample{nplanets planets-cuts1-cut-7-nomcmc} KOIs orbiting \sample{nstars planets-cuts1-cut-7-nomcmc} host stars remained. 

\end{enumerate}
Figure~\ref{fig:cuts-all} shows the application of these filters to the field star and planet samples. In this paper, we refer to these \sample{nstars (dr1 | dr2) & cxm} stars as the ``CKS Extended Mass'' (CXM) sample. 

\section{Spectroscopic Observations}
\label{sec:hires}

We worked to compile a homogeneous spectroscopic catalog of all CXM stars from new and archival observations. All spectra were obtained with Keck/HIRES \citep{Vogt94} to ensure common systematics in our analysis. Our goal was to gather spectra with $R \geq 60,000$ and S/N $\geq$ 20~pix$^{-1}$ on blaze at 5500~\AA.

\begin{figure}
\includegraphics[width=0.45\textwidth,trim={6cm 0.5cm 8cm 2cm}]{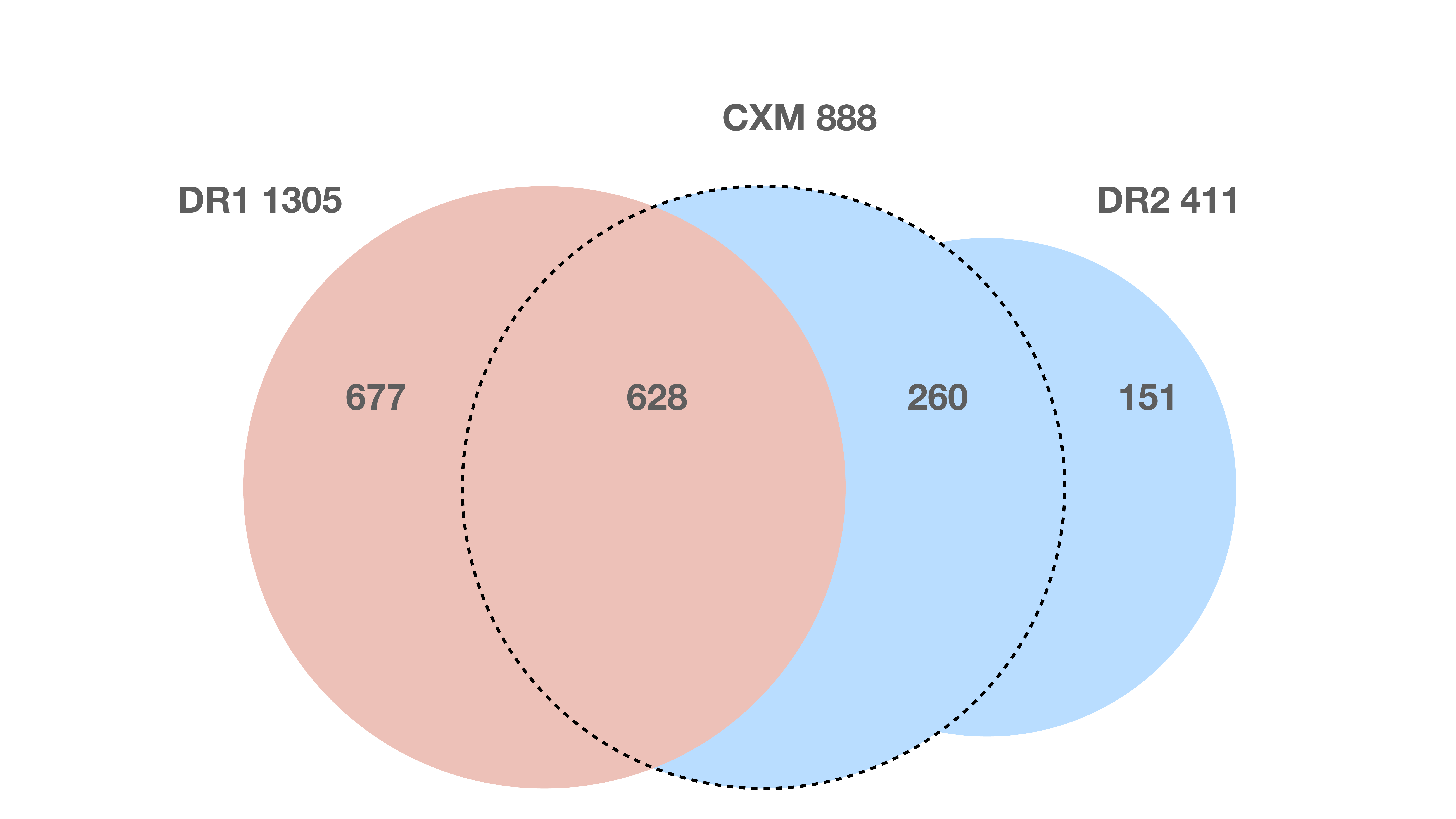}
\centering
\caption{Venn diagram illustrating the number of host stars belonging to various samples described in this paper. CXM---\sample{nstars cxm} stars in extended stellar mass sample described in \S\ref{sec:sample} that forms the basis of the exoplanet demographic work described in \S\ref{sec:planet-population} and \S\ref{sec:occurrence}. DR1---\sample{nstars dr1} stars from \citep{Petigura17b}, DR2---\sample{nstars dr2} stars with new spectra presented in this work.}
\label{fig:venn-diagram}
\end{figure}

The inventory of DR1/DR2 spectra and the CXM host star sample are shown in Figure~\ref{fig:venn-diagram}. By design, there was substantial overlap between the \sample{nstars dr1}-star DR1 sample and the CXM sample (\sample{nstars dr1 & cxm}/\sample{nstars cxm} spectra). The DR1 spectra have $R \geq 60,000$ and S/N $\geq$ 45~pix$^{-1}$ and thus met our spectral quality goals. The \sample{nstars dr2} spectra in DR2 are a mixture of \sample{nstars dr2 & cxm & pre-2018} archival HIRES spectra taken before 2018 and \sample{nstars dr2 & post-2018} new spectra taken  by our team since then. For the archival spectra, all have $R \geq 60,000$ and all but \sample{nstars dr2 & cxm & pre-2018 counts < 1500} had S/N $\geq$ 20~pix$^{-1}$. An additional \sample{nstars dr2 & cxm}/\sample{nstars cxm} CXM stars have spectra in DR2.

Below, we describe the observations of the post-2018 DR2 spectra, which were executed as part of a NASA Key Strategic Mission Support program (PI: Petigura). Our strategy was identical to that of CKS DR1 except we exposed to a lower S/N. We used the standard setup of the California Planet Search (CPS; \citealt{Howard10b}). We observed stars through the ``C2'' slit with sky-projected dimensions of 0.86~arcsec x 14.0~arcsec, which achieves a spectral resolution of $R \approx 60,000$. We used the exposure meter to attain a designed S/N which is measured at the peak of the blaze function at $\lambda = 5500$~\AA.

Given the magnitude-limited nature of DR1, most DR2 stars had \kepmag = 14.2--16.0~mag. It would have been prohibitively expensive to observe these stars to the same S/N level as DR1 (45 pix$^{-1}$), so we used the exposure meter to achieve S/N = 20 pix$^{-1}$ for stars in this magnitude range. In \S\ref{sec:host-star}, we demonstrate that the additional photon-limited uncertainties are smaller than other sources of uncertainty in the extracted parameters.

For faint targets, the sky background is occasionally comparable to the stellar spectrum depending on moon separation, moon phase, and cloud cover. In our spectral reduction, we removed the sky background (see \citealt{Batalha11} for details), but it still contributes to exposure meter counts. Thus, at fixed $m_V$ and exposure meter setting, additional sky background would cause exposures to terminate early. While at the telescope, we used a custom script to compute the sky contribution in realtime and adjusted the exposure meter setting such that we would reach the desired S/N {\em after} sky subtraction. 

\begin{figure*}
\includegraphics[width=1.0\linewidth]{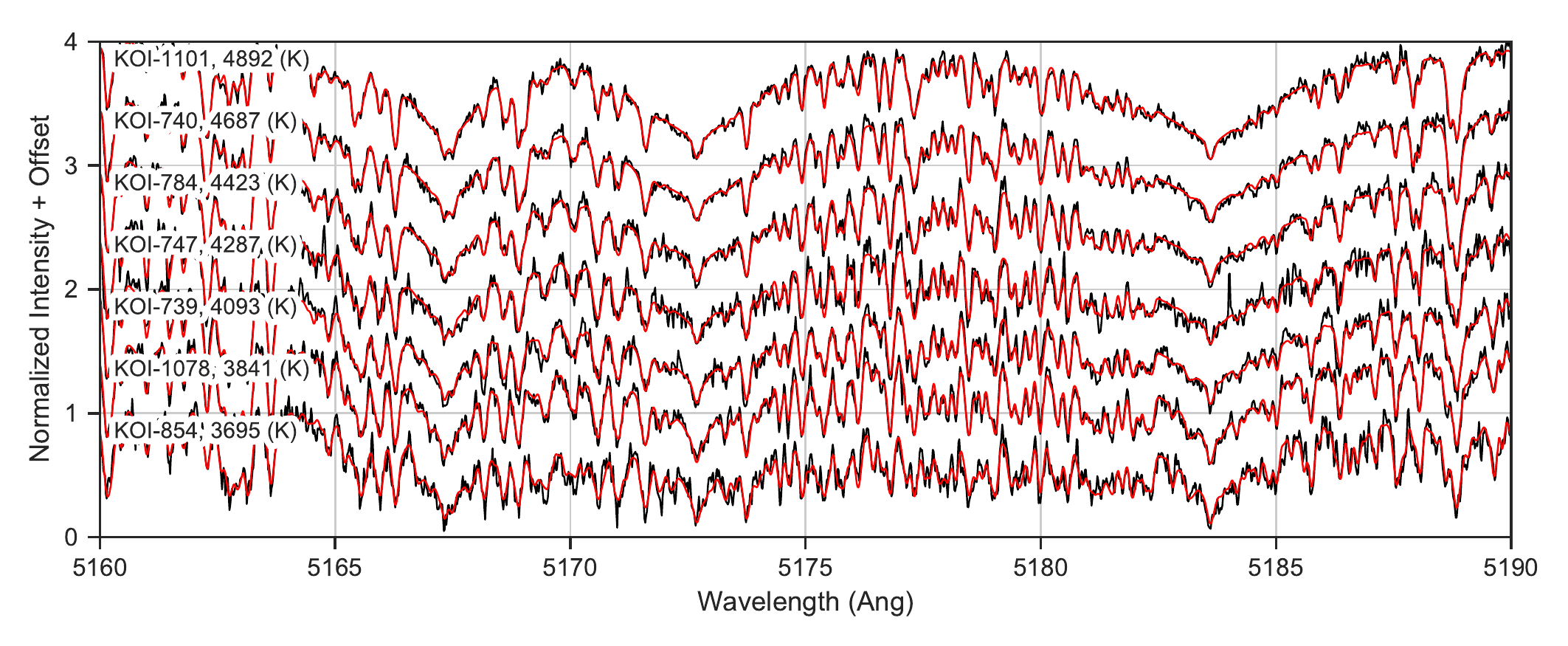}
\centering
\caption{The black lines show a segment of HIRES spectra containing the Mg I b triplet for seven representative spectra observed by our team to illustrate the spectral resolution $R \geq 60,000$ and S/N $\geq 20$~pix$^{-1}$ (see \S\ref{sec:hires}). The red lines are SpecMatch-Empirical fits (see \S\ref{sec:spectroscopic-analysis}).\label{fig:spectra}} 
\end{figure*}

We reduced our spectra according to the standard practices of CPS. We removed the blaze function by dividing our spectra by a composite spectrum of spectrally flat standards compiled by \cite{Clubb18}. One component of our spectroscopic analysis involved registering each spectrum to a common wavelength scale. We used the SpecMatch-Empirical code \citep{Yee17} that includes a ladder of reference spectra that have been registered against the National Solar Observatory solar atlas \citep{Kurucz84}. Figure~\ref{fig:spectra} shows the Mg I b region for  representative spectra over a range of \teff. The deblazed spectra are available via the Keck Observatory Archive%
\footnote{\url{http://www2.keck.hawaii.edu/koa/public/koa.php}}
the CFOP website,%
\footnote{\url{http://cfop.ipac.caltech.edu}}
and a website maintained by our team.%
\footnote{\url{http://astro.caltech.edu/~howard/cks/}}
We have also made available the standard observatory-frame wavelength solution applicable to every spectrum, which is accurate to within one reduced pixel or 1.2 km/s. The registered spectra are also available at the above repositories. 

\section{Host Star Characterization}
\label{sec:host-star} 
\subsection{Spectroscopy}
\label{sec:spectroscopic-analysis}

We derived stellar parameters from our spectra using two related and complementary codes: SpecMatch-Synthetic and SpecMatch-Empirical. Both codes were designed to yield high quality parameters even at moderate or low S/N. A detailed description of SpecMatch-Synthetic is available in \cite{Petigura15thesis} and the code's application to CKS DR1 is given in \cite{Petigura17b}. In brief, the code generates a synthetic spectrum by interpolating within a set of library spectra computed over a grid of \teff, \logg, and \fe. It then applies line broadening kernels that account for the instrumental profile, stellar rotation, and macroturbulence. The library spectra were computed by \cite{Coelho05} under the assumption of local thermodynamic equilibrium (LTE) using the \cite{Castelli04} model atmospheres. The code derives \teff, \logg, \fe, and \vsini using non-linear optimization of an $l_{2}$ metric, which is evaluated over five spectral segments with a combined spectral bandwidth of 380~\AA\ between 5200~\AA\ and 6260~\AA.

The uncertainties of SpecMatch-Synthetic parameters have been extensively vetted for stars having \teff = 4700--6500~K. Errors are dominated by systematic uncertainties which are $\sigma(\teff)$ = 100~K, $\sigma(\logg) = 0.10$~dex, and $\sigma(\fe) =  0.06$~dex. Poisson uncertainties are smaller than systematic uncertainties even at low S/N because the code fits a wide spectral bandpass and combines information from many lines. At our lowest S/N (20 pix$^{-1}$), photon-limited errors are $\sigma(\teff) = 34$~K, $\sigma(\logg) = 0.05$~dex, and $\sigma(\fe) = 0.02$~dex, respectively, below the systematic floor. Above 6500~K, the quality of the derived parameters degrades due to the reduced number of lines and substantial rotational broadening. As \teff drops below 4700~K, the number of lines in SpecMatch-Synthetic wavelength regions increases dramatically due to molecular features (of which MgH is the most prominent). The synthetic spectra do not accurately reproduce these complex spectra and the derived parameters suffer.

By design, the CXM sample included many stars cooler than 4700~K so we needed another method that is better suited to cool stars. In preparation for this project, \cite{Yee17} developed SpecMatch-Empirical, which sidesteps the difficulties of cool star spectral synthesis. In brief, SpecMatch-Empirical compares a target spectrum to a spectral library of 404 stars with precise and accurate \teff, \Rstar, and \fe measured through a combination of the following techniques: interferometry, asteroseismology, line-by-line LTE spectral synthesis, and spectrophotometry. The code identifies the top five matches based on an $l_2$ metric. The target spectrum is then fit using a linear combination of these five spectra over a broad 1000~\AA\ spectral region spanning 5000~\AA\ to 6000~\AA. The extracted parameters are a weighted average of the input parameters. For the representative spectra in Figure~\ref{fig:spectra} we have over-plotted the best-fit spectra.

\cite{Yee17} validated the precision and accuracy of SpecMatch-Empirical across the HR diagram and found that quality of the derived parameters varies in this space due to the relative density of library spectra. For stars having \teff = 3500--5000~K and \fe between $-0.5$ and $+0.5$~dex, the code achieves uncertainties of $\sigma(\teff) = 60$~K and $\sigma(\fe) = 0.12$~dex. Because this code analyzes 1000~\AA\ of spectrum, it is robust to photon-limited uncertainties at low S/N. For our lowest S/N spectra at 20~pix$^{-1}$, Poisson uncertainties amount to $\sigma(\teff) = 6$~K and $\sigma(\fe) = 0.004$~dex and thus contribute negligibly to the overall error budget. 

Table~\ref{tab:star} lists our adopted stellar parameters for the \sample{nstars dr1 | dr2} star union of DR1 and DR2. Given the relative strengths and weakness of the two codes, we split our sample into two groups at \teff = 4800~K, as measured by  SpecMatch-Empirical. We report SpecMatch-Synthetic parameters for the \sample{nstars smsyn} stars above this threshold and SpecMatch-Empirical for the \sample{nstars smemp} below.

In the following section, we combine our spectroscopic \teff and \fe measurements with astrometric and photometric constraints to characterize the stellar hosts. Before proceeding, we note that the input spectra have a bimodal distribution of S/N and consider whether this could bias the DR1 and DR2 stellar parameters relative to one another. As explained above, the photon-limited \teff and \fe errors at S/N = 20~pix$^{-1}$ are less than half the systematic errors. Any S/N dependent offsets are washed out by the larger systematic uncertainties.

\subsection{Stellar Properties}
\label{sec:stellar}
We derived stellar masses, radii, and ages using our spectroscopic measurements of \teff and \fe, along with astrometric and photometric constraints. Our methodology closely follows that of \cite{Fulton18b}. We give a brief summary below, noting differences where relevant. 

We measured \Rstar using the isoclassify package in its ``direct'' mode. This code evaluates \Rstar from the Stefan--Boltzmann law,
\begin{equation}
\label{eqn:rstar}
\Rstar = \left(\frac{\Lbol}{4\pi\sigmasb\teff^4}\right)^{1/2},
\end{equation}
where \Lbol is the bolometric stellar luminosity and \sigmasb is the Stefan-Boltzmann constant.  \Lbol is directly related to bolometric magnitude $\Mbol$, which may be expressed as
\begin{equation}
\label{eqn:mbol}
\Mbol = m - \mu - A + BC,
\end{equation}
where $m$ is the apparent magnitude, $\mu$ is the distance modulus, $A$ is the line-of-sight extinction, and $BC$ is the bolometric correction. We summarize each input below along with their typical uncertainties. Given the discontinuity in our sample selection at \kepmag = 14.2 described in \S\ref{sec:sample}, we report median uncertainties above and below this value. 

\begin{enumerate}
\item {\em Apparent magnitude.} We used 2MASS $K$-band photometric measurements because dust extinction is less severe in the infrared. We used 2MASS \kmag which have a median precision of \sample{kmag-err-med kepmag<14.2}~mag for \kepmag < 14.2 and \sample{kmag-err-med kepmag>14.2}~mag for \kepmag > 14.2.

\item {\em Extinction.} We accounted for $K$-band extinction $A$ using the 3D dust map of \cite{Green19}, an update to the \cite{Green18} map used in \cite{Fulton18b}. Median $A$ is \sample{ak-med kepmag<14.2}~mag for both bright and faint subsets.

\item {\em Distance modulus.} We used \Gaia DR2 parallaxes and applied a correction of +0.053~mas to account for a known systematic offset in the \Kepler field \citep{Zinn18}. The parallaxes have a median precision of \sample{parallax-ferr-med kepmag<14.2}\% for \kepmag < 14.2 and \sample{parallax-ferr-med kepmag>14.2}\% for \kepmag > 14.2. Isoclassify handles the conversion between parallax and distance modulus using a Bayesian framework. 

\item {\em Bolometric correction.} Isoclassify interpolates over a grid of bolometric corrections computed by the MESA Isochrones and Stellar Tracks project (MIST v1.2, \citealt{Choi16,Dotter16,Paxton11,Paxton13,Paxton15}).  Uncertainties in \teff dominate the uncertainty of the $K$-band bolometric correction and a 60~K uncertainty translates to a $\approx$0.03~mag error on $BC$.
\end{enumerate}
Table~\ref{tab:star} lists our measured \Rstar, and Figure~\ref{fig:ferr-hist} shows the distribution of fractional \Rstar uncertainties. The median fractional uncertainty is \sample{gdir_srad-ferr-med kepmag<14.2}\% for \kepmag < 14.2 and \sample{gdir_srad-ferr-med kepmag>14.2}\% for \kepmag > 14.2.

We also ran isoclassify in its ``grid'' mode with the same constraints. In this mode, the code queries the MIST isochrones which provide various stellar properties over a grid of \Mstar, \fe, and age. The grid extends to 20~Gyr to minimize grid edge-effects, which can bias the parameter posteriors. At each grid point, isoclassify computes the likelihood that point is consistent with the input constraints. 

Marginalized values are determined by direct integration. We list \Mstariso, \Rstariso, \rhostariso, and \ageiso in Table~\ref{tab:star}. The ``iso'' subscript indicates that these parameters are forced to be consistent with MIST isochrones. The median fractional uncertainty for \Mstariso is \sample{giso_smass-ferr-med kepmag<14.2}\% for \kepmag < 14.2 and \sample{giso_smass-ferr-med kepmag>14.2}\% for \kepmag > 14.2; for \rhostariso it is \sample{giso_srho-ferr-med kepmag<14.2}\% for \kepmag < 14.2 and \sample{giso_srho-ferr-med kepmag>14.2}\% for \kepmag > 14.2. A handful of stars have \ageiso > 14 Gyr because the grid extends to 20~Gyr for reasons explained above. Age uncertainties vary widely across the HR diagram and are discussed later in \S\ref{sec:planet-population}. 

\begin{deluxetable*}{lRRRRRrRRRRRRRR}
\tablecaption{Stellar Properties\label{tab:star}}
\tabletypesize{\footnotesize}
\tablecolumns{15}
\tablewidth{0pt}
\tablehead{
	\colhead{KOI} & 
	\colhead{\kmag} &
    \colhead{$\pi$} &
	\colhead{\teff} &
	\colhead{\fe} &
	\colhead{\vsini} &
    \colhead{prov} & 
    \colhead{$\Rstar$} & 
	\colhead{\Mstariso} & 
	\colhead{\Rstariso} & 
	\colhead{\rhostariso} & 
    \colhead{\ageiso} &
    \colhead{$\pi_{\mathrm{spec}}$} &
    \colhead{SB2} &
	\colhead{CXM} \\
    \colhead{} & 
    \colhead{mag} & 
    \colhead{mas} &
	\colhead{K} &
	\colhead{dex} & 
	\colhead{\kms} &
    \colhead{} &
    \colhead{\Rsun} & 
	\colhead{\Msun} & 
	\colhead{\Rsun} & 
	\colhead{g/cc} & 
    \colhead{Gyr} & 
    \colhead{mas} &
    \colhead{} &
    \colhead{} 
}
\startdata
%id_koi & m17_kmag & gaia2_sparallax & cks_steff & cks_smet & cks_svsini & cks_sprov & gdir_srad & giso_smass & giso_srad & giso_srho & giso_sage & giso2_sparallax & rm_sb2 & in_cxm
1 & 9.85 & 4.67 & 5857 & 0.02 & 1.3 & syn & 1.04 & 1.02 & 1.03 & 0.91 & 3.7 & 4.76 & 1 & 1 \\ 
2 & 9.33 & 2.96 & 6403 & 0.22 & 5.1 & syn & 2.00 & 1.50 & 1.99 & 0.19 & 1.8 & 3.72 & 1 & 1 \\ 
3 & 7.01 & 26.50 & 4619 & 0.16 & \nodata & emp & 0.78 & 0.76 & 0.75 & 1.79 & 11.2 & 23.80 & 1 & 1 \\ 
4 & 10.20 & 1.29 & 5948 & -0.27 & 38.0 & syn & 3.26 & 1.45 & 3.22 & 0.04 & 2.0 & 0.93 & 3 & 0 \\ 
6 & 10.99 & 2.13 & 6356 & 0.05 & 11.8 & syn & 1.29 & 1.23 & 1.28 & 0.57 & 1.5 & 2.24 & 1 & 0 \\ 
\enddata
\tablecomments{Properties of \sample{nstars dr1 | dr2} planet hosting stars from the union of the DR1 and DR2 datasets. \kmag is 2MASS $K$-band apparent magnitude and $\pi$ is \Gaia DR2 parallax. \teff, \fe, and \vsini were derived by one of two methods listed in the `prov' column: `syn'---SpecMatch-Synthetic and `emp'---SpecMatch-Empirical. \Rstar is our adopted stellar radius from the Stefan-Boltzmann law. Stellar properties with the `iso' subscript incorporate constraints from the MIST isochrones. `SB2' encodes the limits on spectroscopic binaries (SB2s) using the {\em ReaMatch} code \citep{Kolbl15}: 1---No detection of SB2 with $\Delta m_V \lesssim 5$~mag and $\Delta RV \gtrsim 12$~\kms; 2---\teff < 3500~K, star unfit for ReaMatch; 3---\vsini > 10~\kms, star unfit for ReaMatch; 4---Ambiguous detection; 5---Obvious detection. Note: {\em ReaMatch} computes \teff and \vsini independently from SpecMatch-Synthetic and SpecMatch-Empirical. The `CXM' flag is 1 if the star is in the \sample{nstars planets-cuts2-cut-0-none} star CXM subset. Median uncertainties are as follows: \kmag---\sample{kmag-err-med}~mag; $\pi$---\sample{parallax-ferr-med}\%; \teff---\sample{cks_steff-err-med}~K, \fe---\sample{cks_smet-err-med}~dex; \vsini---1~\kms; \Rstar---\sample{gdir_srad-ferr-med}\%; \Mstariso---\sample{giso_smass-ferr-med}\%; \Rstariso---\sample{giso_srad-ferr-med}\%; \rhostariso---\sample{giso_srho-ferr-med}\%; \ageiso---\sample{giso_slogage-err-med}~dex; $\pi_\mathrm{spec}$---\sample{giso2_sparallax-ferr-med}\%. Table \ref{tab:star} is published in its entirety with uncertainties in machine-readable format in the online journal.}
\end{deluxetable*}

\begin{deluxetable*}{lRRRRRRRR}
\tablecaption{Planet Properties\label{tab:planet}}
\tabletypesize{\footnotesize}
\tablecolumns{9}
\tablewidth{0pt}
\tablehead{
	\colhead{Planet} & 
	\colhead{$P$} &
    \colhead{$\Rp/\Rstar$} &
    \colhead{$T$} &
	\colhead{$\Rp$} &
    \colhead{\Tcirc} &
	\colhead{$a$} &
	\colhead{\Sinc} &
	\colhead{samp} \\
    \colhead{} &
    \colhead{d} &
    \colhead{\%} &
    \colhead{hr} &
    \colhead{\Re} &
    \colhead{hr} &
    \colhead{au} &
    \colhead{\Se} &
    \colhead{}
}
\startdata
%id_koicand & koi_period & dr25_ror & dr25_tau & gdir_prad & giso_tau0 & giso_sma & giso_sinc & in_curated
K00001.01 & 2.5 & 12.39 & 1.29 & 14.04 & 2.83 & 0.037 & 814.85 & 0 \\ 
K00002.01 & 2.2 & 7.52 & 3.55 & 16.42 & 4.58 & 0.039 & 3855.68 & 1 \\ 
K00003.01 & 4.9 & 5.80 & 2.23 & 4.92 & 2.84 & 0.053 & 80.94 & 1 \\ 
K00007.01 & 3.2 & 2.45 & 3.84 & 4.07 & 4.29 & 0.046 & 1082.71 & 1 \\ 
K00010.01 & 3.5 & 9.21 & 2.75 & 15.51 & 4.47 & 0.050 & 1262.15 & 1 \\ 
\enddata
\tablecomments{Properties of \sample{nplanets planets-cuts2-cut-0-none} planets orbiting the \sample{nstars planets-cuts2-cut-0-none} star CXM sample. Orbital period $P$, planet-to-star radius ratio $\Rp/\Rstar$, and transit duration $T$ were measured from \Kepler photometry by \cite{Thompson18}. $\Rp$ follows from $\Rp/\Rstar$ and \Rstar. The expected duration of a centrally transiting object on a circular orbit \Tcirc, semi-major axis $a$, and incident bolometric flux $\Sinc$ are determined from Kepler's Third Law and the Stefan-Boltzmann law. The `samp' flag is 1 if planet is in curated sample of \sample{nplanets planets-cuts2-cut-7-badpradprec} planets described at the end of \S\ref{sec:planets}. Median uncertainties are as follows: $P$---\sample{dr25_period-ferr-med} ppm; $\Rp/\Rstar$---\sample{dr25_ror-ferr-med}\%; $T$---\sample{dr25_tau-ferr-med}\%, \Rp---\sample{gdir_prad-ferr-med}\%; \Tcirc---\sample{giso_tau0-ferr-med}\%; $a$---\sample{giso_sma-ferr-med}\%; $\Sinc$---\sample{giso_sinc-ferr-med}\%. Table~\ref{tab:planet} is published in its entirety with uncertainties in machine-readable format in the online journal.}
\end{deluxetable*}

\section{Planet Catalog}
\label{sec:planets}

In the previous section, we computed stellar properties in the combined DR1/DR2 dataset. For the remainder of the paper, we focus on the \sample{nplanets planets-cuts1-cut-7-nomcmc} planets associated with the \sample{nstars cxm} star CXM subsample (see Figure~\ref{fig:venn-diagram}). In this section, we first compute planet properties given our updated stellar properties and then apply an additional set of quality checks to produce a high-reliability sample of planets for demographic work in later sections. 

\subsection{Planet properties} 

\begin{figure}
\centering
\includegraphics[width=0.45\textwidth]{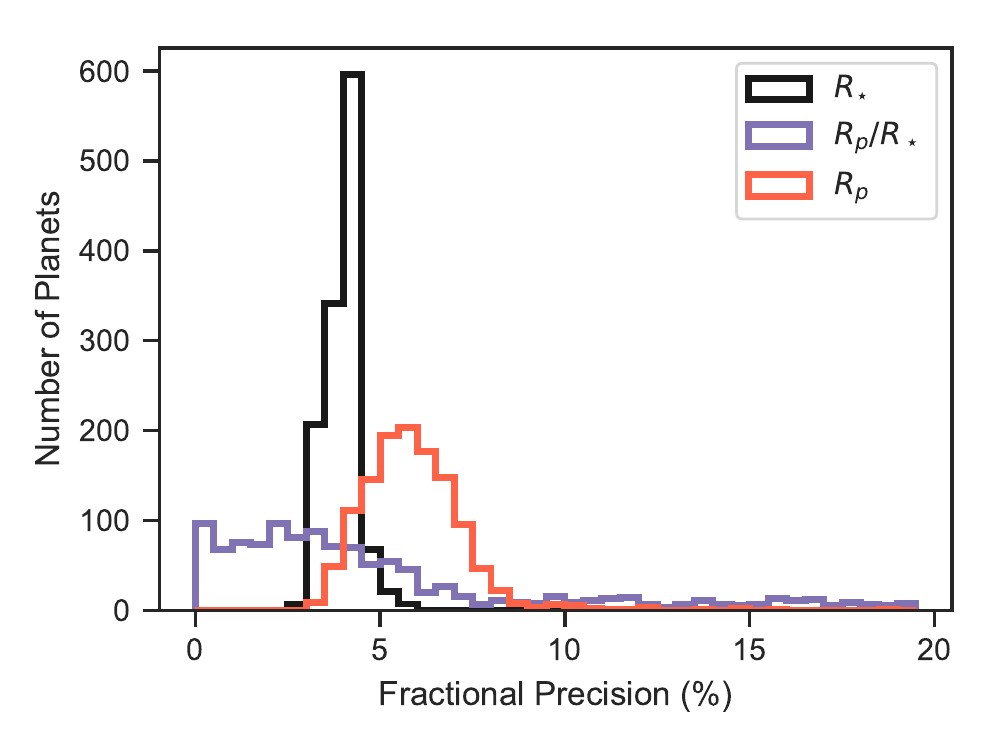}
\caption{Fractional precision of \Rp, \Rstar, and \Rp/\Rstar for the planets in the CXM sample. \label{fig:ferr-hist}}
\end{figure}

We computed \Rp from our \Rstar (listed in Table~\ref{tab:star}) and T18's modeling of \Kepler light curves. We multiplied our posterior samples on \Rstar by the MCMC-derived posterior samples on \Rp/\Rstar (available at the NEA). Median fractional error on \Rp/\Rstar is \sample{dr25_ror-ferr-med kepmag<14.2}\% for \kepmag < 14.2 and \sample{dr25_ror-ferr-med kepmag>14.2}\% for \kepmag > 14.2. Uncertainties in \Rstar and \Rp/\Rstar are independent and propagated into \Rp, and we quote the 16th, 50th, and 84th posterior quantiles in Table~\ref{tab:planet}. The median fractional precision in \Rp is \sample{gdir_prad-ferr-med kepmag<14.2}\% for \kepmag < 14.2 and \sample{gdir_prad-ferr-med kepmag>14.2}\% for \kepmag > 14.2. Figure~\ref{fig:ferr-hist} shows the fractional precision of \Rp, \Rstar, and \Rp/\Rstar. For most of the sample, \Rstar errors are comparable to or larger than \Rp/\Rstar errors. However, for a small subset, \Rp/\Rstar errors dominate. 

A note about the provenance of \Rp/\Rstar: There are a number of published catalogs of \Rp/\Rstar from \Kepler project, of which T18 is the most recent. The NEA table corresponding to the T18 catalog lists the ``best-fit'' \Rp/\Rstar, not the posterior medians. \cite{Petigura20b} found that these best-fit values contribute noise to \Rp/\Rstar from the non-linear optimization and recommended using posterior medians which are a more robust estimator. 

We compared our planet radii to T18, by computing the ratio of the radii $r = R_{p,\mathrm{us}}/R_{p,\mathrm{t18}}$. There was a negligible systematic offset $\mathrm{mean}(r) = 1.00$ but a substantial dispersion $\mathrm{RMS}(r) = 20\%$. The dispersion is dominated by uncertainties in the T18 \Rp. 

Two additional quantities were relevant to our goal of precise planet radii: the observed transit duration $T$ and the expected duration of centrally transiting planet on a circular orbit \Tcirc. Here, $T$ refers to time between the midpoints of first/second contact and third/fourth contact. In the T18 modeling, transit shape was set by the following free parameters: period $P$, transit epoch $T_0$, impact parameter $b$, mean stellar density assuming a circular orbit \rhostarcirc, and $\Rp/\Rstar$. We derived $T$ and its uncertainties according to 
\begin{equation}
T = 2.036~\mathrm{hr} \left(1 - b^{2}\right)^{1/2} \left(\frac{P}{1 \,\mathrm{d}}\right)^{1/3} \left( \frac{\rhostarcirc}{1\, \mathrm{g\, cm}^{-3}}\right) ^{-1/3}
\end{equation}
using the T18 MCMC chains. We computed \Tcirc according to 
\begin{equation}
\Tcirc = 2.036~\mathrm{hr} \left(\frac{P}{1 \,\mathrm{d}}\right)^{1/3} \left( \frac{\rhostariso}{1\, \mathrm{g\, cm}^{-3}}\right) ^{-1/3}
\end{equation}
Both $T$ and \Tcirc are listed in Table~\ref{tab:planet} along with semi-major axis $a$ and incident stellar flux \Sinc.

\subsection{Refining the planet sample} 
\label{sec:refining-the-sample}

\begin{figure*}
\centering
\includegraphics[width=0.9\textwidth]{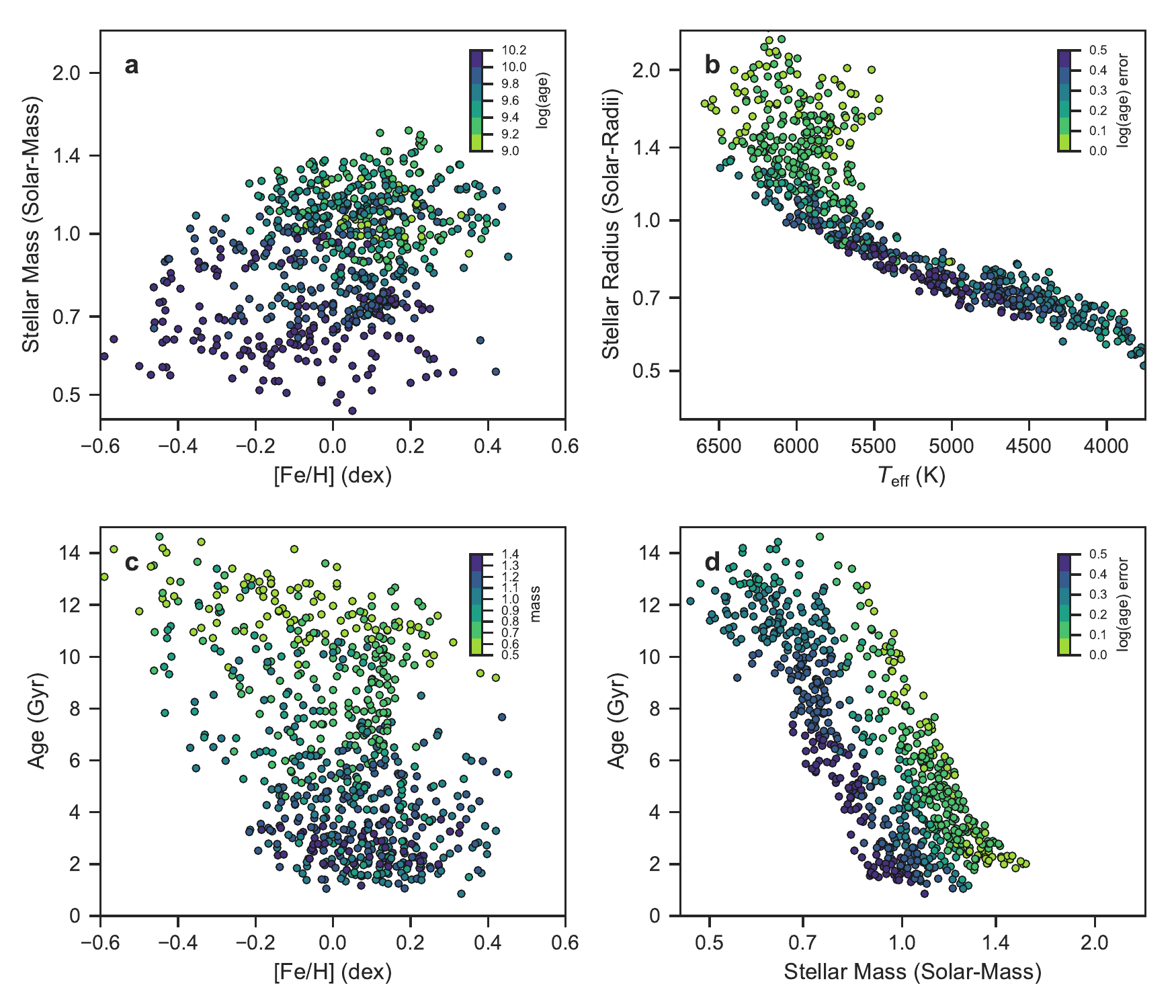}
\caption{Properties of host stars in the filtered CXM sample described in \S\ref{sec:refining-the-sample}. Panel (a): Stellar mass and metallicity. Panel (b): Stellar radius and effective temperature. The color scale conveys the isochrone age uncertainty in dex (half the difference between the 84th and 16th percentiles of the age posterior). Panel (c): Stellar metallicity and age. Panel (d): Stellar mass and age where the color scale is equivalent to panel (b). Several stars have ages exceeding 14 Gyr for reasons we explain in \S\ref{sec:stellar}. We discuss correlations between mass, metallicity, and age in \S\ref{sec:planet-population-mass}.}
\label{fig:hr}
\end{figure*}

While the initial CXM sample was optimized for high-reliability planets (\S\ref{sec:sample}), we incorporated additional diagnostics from our spectra to further increase sample purity. We first filtered on the following stellar properties:

\begin{enumerate}

\item {\em Rotation rate.} Our spectroscopic codes are unreliable at \vsini > 20~\kms \citep{Petigura17b}. We excluded \sample{nstars-rem planets-cuts2-cut-1-badvsini} stars above this threshold which all had \teff > 6500~K as measured by B20. A total of \sample{nplanets planets-cuts2-cut-1-badvsini} KOIs orbiting \sample{nstars planets-cuts2-cut-1-badvsini} hosts remained.

\item {\em Spectroscopic parallax.} Following \cite{Fulton18b}, we used isoclassify to compute a ``spectroscopic parallax'' constrained by our \teff, \fe, \kmag, and the MIST models. If spectroscopic and trigonometric parallaxes differ significantly, we conclude that the input parameters used to compute \Rstar are inconsistent due to unresolved binaries or other effects. We removed \sample{nstars-rem planets-cuts2-cut-2-badspecparallax} stars where these parallaxes differed by $4\sigma$ or more. We chose a $4\sigma$ over a $3\sigma$ threshold because the probability of a $3\sigma$ event is $3\times10^{-3}$ and thus we expect $\approx 3$ such events in a sample of $\approx 10^3$ stars. The probability of a $4\sigma$ event $6\times10^{-5}$ so a single such event is unlikely. A total of \sample{nplanets planets-cuts2-cut-2-badspecparallax} KOIs orbiting \sample{nstars planets-cuts2-cut-2-badspecparallax} hosts remained.

\item {\em Secondary spectra.} We removed \sample{nstars-rem planets-cuts2-cut-3-sb2} star where we identified a secondary set of spectral lines using the methodology of \cite{Kolbl15}, which is sensitive to binaries having $\Delta m_V \lesssim 5$~mag and $\Delta v \gtrsim 12$~\kms. A total of \sample{nplanets planets-cuts2-cut-3-sb2} KOIs orbiting \sample{nstars planets-cuts2-cut-3-sb2} hosts remained.

\end{enumerate}
Next, we filtered based on the following planet properties:
\begin{enumerate}
\setcounter{enumi}{3}

\item {\em Impact parameter.} We removed grazing and high-impact parameter planets by excluding planets where the median of the $b$ posterior exceeded 0.8. \cite{Petigura20b} found, however, that for most \Kepler light curves, this cut is ineffective at excluding high-$b$ transits because $b$ is nearly unconstrained. For such transits, $\Rp/\Rstar$ is biased by 10--20\% due to the strong $b$-$\Rp/\Rstar$ covariance. However, $T/\Tcirc$ is an effective proxy for $b$. In addition to our explicit filter on $b$, we also required $T/\Tcirc > 0.6$ in order to exclude transits with $b > 0.8$. While planets with eccentric orbits that transit near periastron also produce low $T/\Tcirc$, \cite{Petigura20b} showed that fewer than $20\%$ of \Kepler planets with $T/\Tcirc < 0.6$ have $b < 0.8$. A total of \sample{nplanets planets-cuts2-cut-5-badimpacttau} KOIs orbiting \sample{nstars planets-cuts2-cut-5-badimpacttau} hosts remained.

\item {\em Radius precision.} We required fractional errors on \Rp of 20\% or less. A total of \sample{nplanets planets-cuts2-cut-7-badpradprec} KOIs orbiting \sample{nstars planets-cuts2-cut-7-badpradprec} hosts remained.

\end{enumerate}
The properties of the stars that passed these cuts are shown in Figure~\ref{fig:hr}. They span a mass range of $\Mstar \approx 0.5$--1.4~\Msun. 

\section{Distribution of Detected Planets}
\label{sec:planet-population}

From here on, we describe and interpret the features in our curated sample of \sample{nplanets planets-cuts2-cut-all} planets. In this section, we treat the population of detected planets before turning to the occurrence distribution in \S\ref{sec:occurrence}. We explore how the Radius Gap correlates with $P$, \Sinc, \Mstar and age in \S\ref{sec:planet-population-period-radius}. We then inspect how the super-Earth and sub-Neptune populations vary with stellar mass in \S\ref{sec:planet-population-mass}.

\subsection{Dependence of the Radius Gap on period, flux, stellar mass, and stellar metallicity.}
\label{sec:planet-population-period-radius}

\begin{figure*}
\centering
\includegraphics[width=0.9\textwidth]{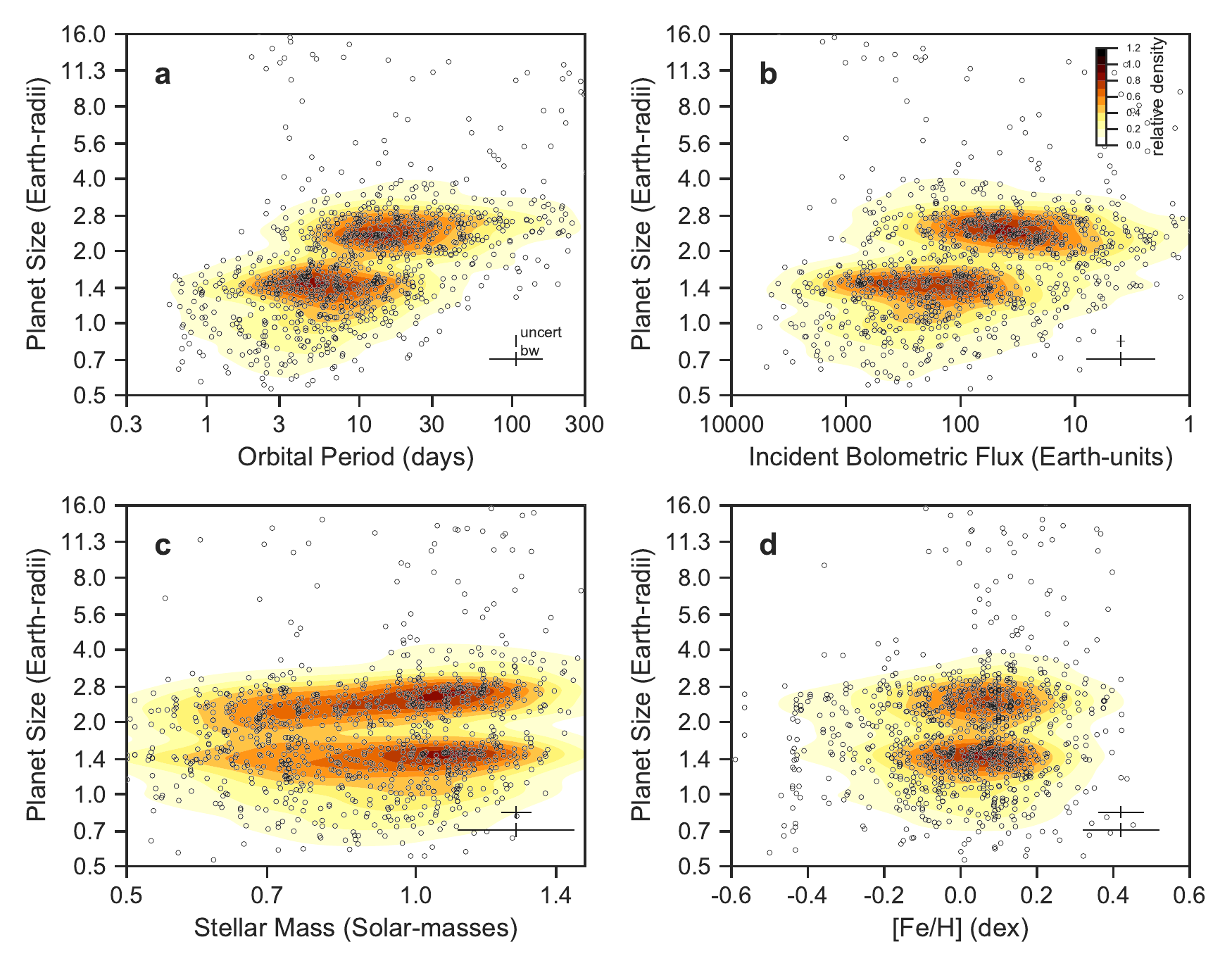}
\caption{Panel (a): sizes and orbital periods of detected planets after applying the filters described in \S\ref{sec:refining-the-sample}. The contours show a Gaussian kernel density estimate (KDE) of the number density of planets in this space. The kernel bandwidth is 0.029~dex in $\Rp$ and 0.18~dex in $P$ as shown by the symbol labelled `bw'. The `uncert' symbol shows the median \Rp uncertainty ($P$ uncertainties are smaller than the line width). Panels (b--d): same as (a), except that x-axis is \Sinc, \Mstar, and \fe and the kernel bandwidths along the horizontal axes are 0.30, 0.061, and 0.1.}
\label{fig:planet-prad}
\end{figure*}

\begin{figure*}
\centering
\includegraphics[width=0.9\textwidth]{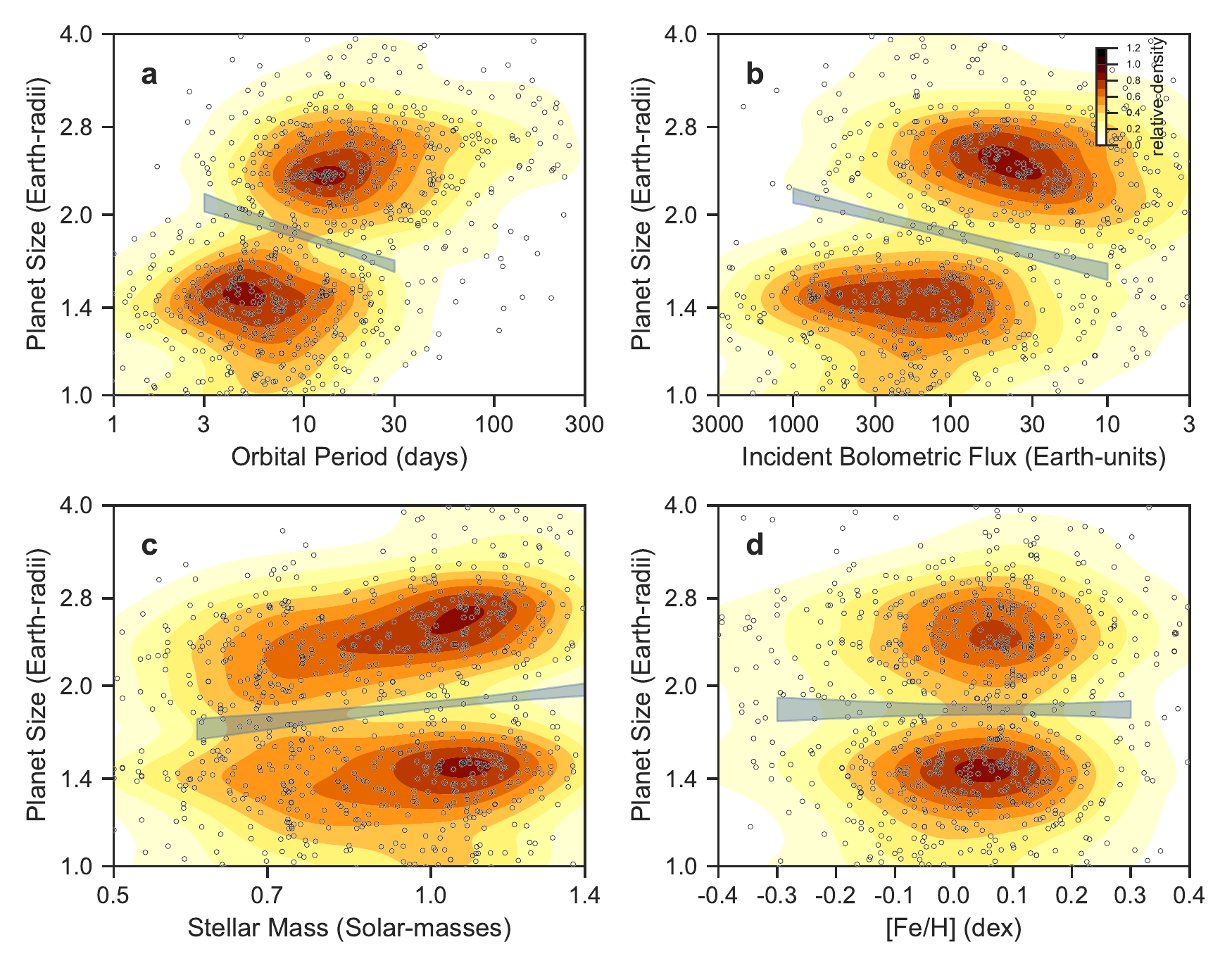}
\caption{A zoomed-in version of Figure~\ref{fig:planet-prad} highlighting the Radius Gap, which is visible in all projections. In the \Mstar projection, the typical sub-Neptune becomes larger with \Mstar, while the super-Earths remain the same size. The bands are power-law fits of Radius Gap (see \S\ref{sec:planet-population-period-radius}).
\label{fig:planet-prad-zoom}}
\end{figure*}

Figure~\ref{fig:planet-prad} shows several views of our planet sample. In the four panels, the y-axes show updated \Rp, and the x-axes show $P$, \Sinc, \Mstar,  and \fe. The contours show the relative density of detected planets in these domains using a Gaussian kernel density estimator (KDE), and the KDE bandwidths are specified in the caption. Figure~\ref{fig:planet-prad-zoom} is analogous to Figure~\ref{fig:planet-prad} but with \Rp restricted to 1--4~\Re for a detailed view of the Radius Gap.

The $P$-\Rp distribution and closely related \Sinc-\Rp distribution of \Kepler planets have been extensively studied in previous works. We begin by remarking on how several known features that appear in our extended stellar mass sample. Most \Kepler planets fall into one of two size categories: ``super-Earths'' and ``sub-Neptunes.'' For definitiveness, we adopt the following size limits for the two planet classes: \Rp = 1.0--1.7~\Re and \Rp = 1.7--4.0~\Re. The Radius Gap separating the populations was observed by \cite{Fulton17} in CKS DR1 (\Mstar $\approx$ 0.8--1.3~\Msun), \cite{Van-Eylen18} in a sample of stars with asteroseismic detections (\Mstar $\approx$ 1.0--1.4~\Msun), and in other subsequent works. We again resolve the Radius Gap, despite the broader \Mstar range of the CXM sample (\Mstar $\approx$ 0.5--1.4~\Msun).

The distribution of super-Earths is centered at shorter periods and higher incident fluxes than the distribution of sub-Neptunes. Compared to super-Earths, there are few sub-Neptunes with $P < 3$~days or $\Sinc > 300$~\Se. This paucity of sub-Neptunes is not a selection effect because larger planets are easier to detect. This zone of low sub-Neptune occurrence is sometimes referred to as the Sub-Neptune Desert (see, e.g., \citealt{Szabo11,Beauge13,Lundkvist16}). In \S\ref{sec:occurrence}, we quantify the sharpness of this boundary and explore its dependence on stellar mass. The relative lack of super-Earths at $P > 30$~days and $\Sinc < 10$~\Se is closely tied survey completeness, also addressed in \S\ref{sec:occurrence}. 

We wish to characterize the slope of the Radius Gap in $P$-\Rp and $\Sinc$-\Rp space because it is a metric by which to test theoretical predictions. However, fitting a parameterized description of the Radius Gap is challenging since it involves characterizing the {\em absence} of planets. \cite{Van-Eylen18} used a support vector classification (SVC) scheme to find the line which maximized the distance between the super-Earth and sub-Neptune populations. The SVC scheme struggles for our sample because the gap is not devoid of planets. (Recently, \citealt{David21} investigated how regularization can assist the SVC identification of the Radius Gap in the CKS sample.)

We adopted a different approach starting with relative number density of detected planets in the $P$-$\Rp$ plane $\dd^2 N / \dd \log \Rp \dd \log P$ as measured by our KDE. This quantity is shown as contours in Figure~\ref{fig:planet-prad-zoom}a. We then found the minimum density along 100 vertical lines spanning $\log \Rp$ = 0.15--0.35 spaced uniformly in $\log P$ over 0.5 to 1.5. We then fit the train of 100 minima with following power-law
\begin{equation}
    \Rp(P) = R_{p,0} \; \bigg( \frac{P}{10~\text{days}} \bigg)^m,
\end{equation}
where the slope $m$ and intercept $R_{p,0}$ are free parameters.  To be clear, $R_{p,0}$ should be interpreted as the midpoint of the Radius Gap at 10-day orbital periods. We chose 10~days as a convenient reference point near the middle of the super-Earth and sub-Neptune populations shown in Figure~\ref{fig:planet-prad-zoom}a. To determine uncertainties we generated 1000 bootstrap resamples with replacement of the planet population shown in Figure~\ref{fig:planet-prad-zoom}a  \citep{Press02}. For each bootstrapped population, we recomputed the planet number density with our KDE, the train of minima, and the best-fitting power-law. We found $m = \dd \log R_p / \dd \log P = \grad{per-prad-det_smass=0.5-1.4-m}$ and $R_{p,0}$ = $\grad{per-prad-det_smass=0.5-1.4-Rp0}$~\Re, which reflect the 16th, 50th, and 84th percentiles. Figure~\ref{fig:planet-prad-zoom} shows the credible gap models. We explored the sensitivity of our method to different period ranges by perturbing the upper and lower boundaries by $\pm0.25$~dex. The derived parameters were consistent to $1\sigma$. We also explored the sensitivity of our method to our adopted KDE period bandwidth. A very wide bandwidth will flatten the Radius Gap (i.e. bias $|m|$ toward smaller values). We repeated our analysis with a much smaller bandwidth of 0.04 dex and found that $m$ and $R_{p,0}$ changed by less than $1\sigma$.

We performed similar analyses to fit the Radius Gap as a function of $\Sinc$, $\Mstar$, and \fe. Table~\ref{tab:gapfits} lists the parametric models used in the fits and the credible range of parameters. Figure~\ref{fig:planet-prad-zoom} shows the range of credible gap models. We observed a positive slope with incident flux ($\dd \log R_p / \dd \log \Sinc = \grad{sinc-prad-det_smass=0.5-1.4-m}$), a positive slope with stellar mass ($\dd \log R_p / \dd \log \Mstar = \grad{smass-prad-det_smass=0.5-1.4-m}$), and no significant correlation with metallicity ($\dd \log R_p / \dd \fe = \grad{smet-prad-det_smass=0.5-1.4-m}$).

\begin{figure*}
\centering
\includegraphics[width=0.9\textwidth]{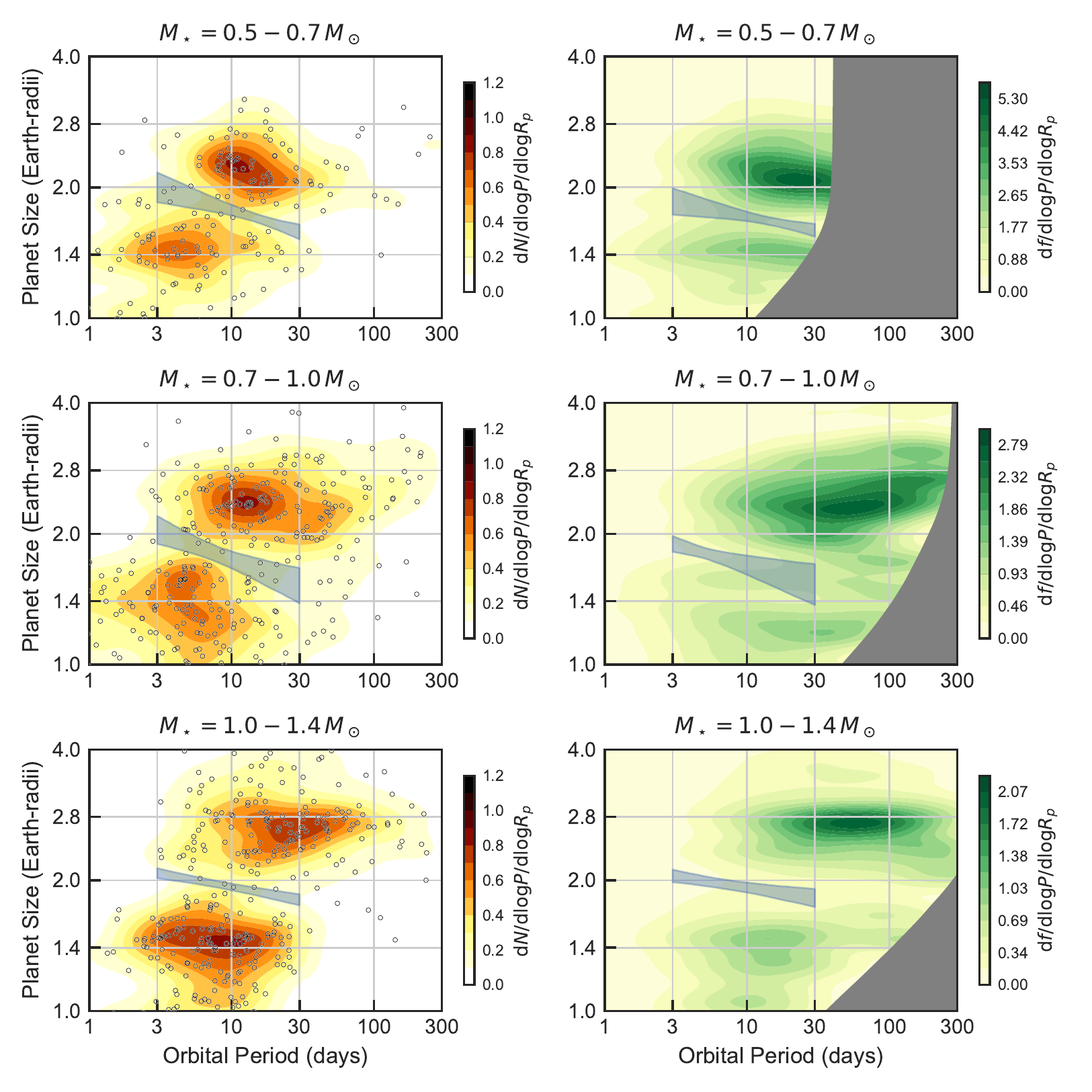}
\caption{Left panels are analogous to Figure~\ref{fig:planet-prad-zoom}a, except we have split the planet sample into three bins of stellar mass. Right panels show the occurrence rate density of planets $\dd f / \dd \log P/ \dd \log \Rp$. The maximum occurrence rate density increases with decreasing \Mstar (note different color scales). Regions of low planet detectability, defined as \ntrial < 50, are gray. The typical sub-Neptune size grows with \Mstar, while super-Earths have a nearly constant size. The bands show fits to the Radius Gap.}
\label{fig:occur-contour-six}
\end{figure*}

We explored how the $P$-\Rp distribution of planets changes with \Mstar. The left panels of Figure~\ref{fig:occur-contour-six} shows this distribution for three different bins of stellar mass with the following boundaries: 0.5, 0.7, 1.0, and 1.4~\Msun. The super-Earth and sub-Neptune populations are distinct in each bin, but their locations vary with \Mstar. As \Mstar increases, the population of super-Earths and sub-Neptunes separates; the super-Earth the population is confined to \Rp < 1.7~\Re while the sub-Neptunes grow in size. We measured the slope and intercept of the Radius Gap in these three bins. They are given in Table~\ref{tab:gapfits} and are consistent to 2$\sigma$. 

\begin{figure*}
\centering
\includegraphics[width=0.9\textwidth]{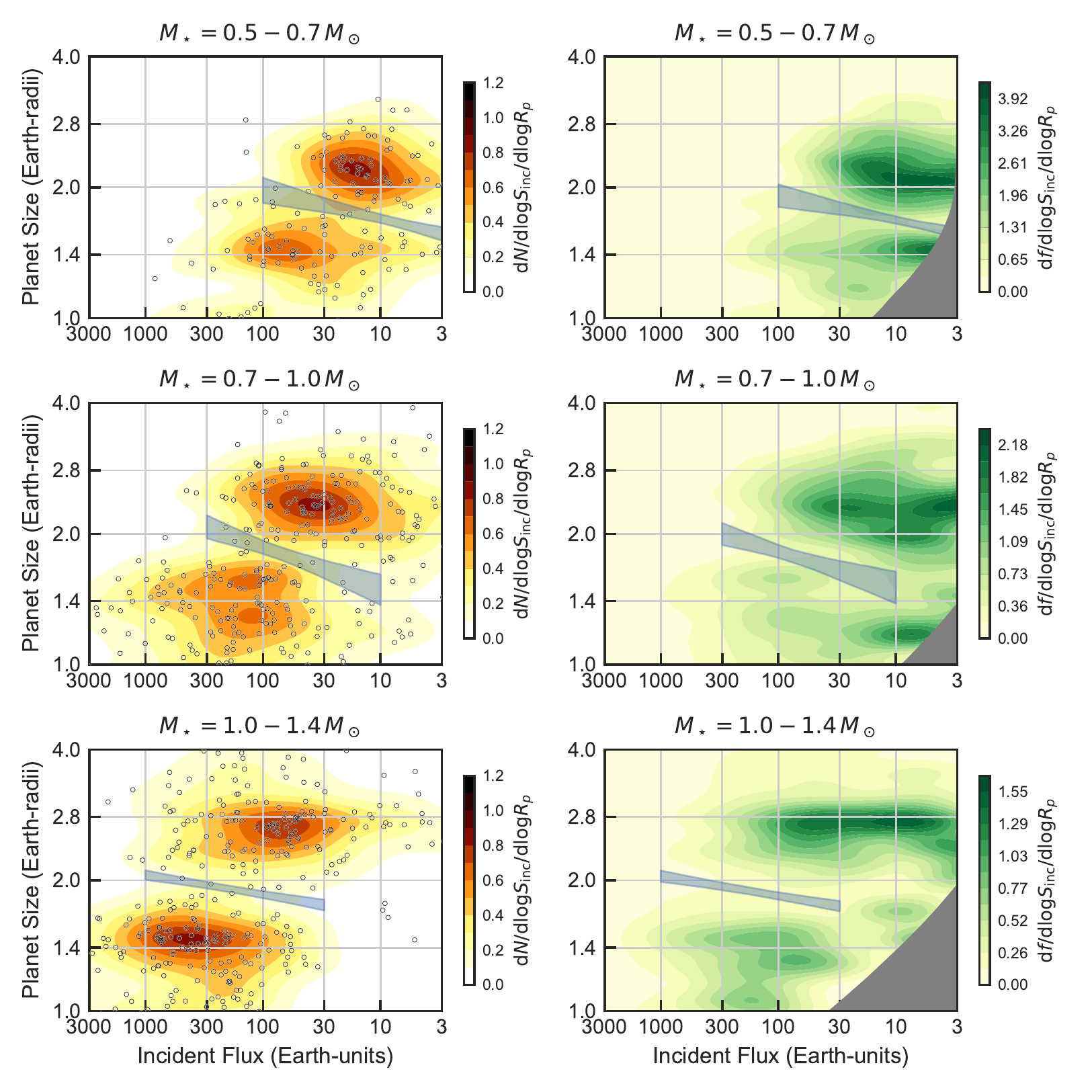}
\caption{Same as Figure~\ref{fig:occur-contour-six} except the x-axis is incident bolometric flux.}
\label{fig:occur-contour-six-sinc}
\end{figure*}

We also explored how the \Sinc-\Rp distribution changes with \Mstar. Figure~\ref{fig:occur-contour-six-sinc} is analogous to Figure~\ref{fig:occur-contour-six} except it shows the \Sinc-\Rp distribution for three bins of stellar mass. The credible models describing the Radius Gap in this space are consistent to 2$\sigma$.

\begin{deluxetable}{LlRRl}
\tablecaption{Power-law fits to the Radius Gap}
\label{tab:gapfits}
\tabletypesize{\footnotesize}
\tablecolumns{5}
\tablehead{
	\colhead{Fit} & 
	\colhead{\Mstar} & 
	\colhead{$m$} & 
	\colhead{$R_{p,0}$} &
	\colhead{Dist} 
}
\startdata
%D
\Rp = R_{p,0} \; \bigg( \frac{P}{10~\text{days}} \bigg)^m   
& 0.5--1.4  & \grad{per-prad-det_smass=0.5-1.4-m} & \grad{per-prad-det_smass=0.5-1.4-Rp0} & D \\ 
& 0.5--0.7  & \grad{per-prad-det_smass=0.5-0.7-m} & \grad{per-prad-det_smass=0.5-0.7-Rp0} & D \\ 
& 0.7--1.0  & \grad{per-prad-det_smass=0.7-1.0-m} & \grad{per-prad-det_smass=0.7-1.0-Rp0} & D \\ 
& 1.0--1.4  & \grad{per-prad-det_smass=1.0-1.4-m} & \grad{per-prad-det_smass=1.0-1.4-Rp0} & D \\ 
% O 
& 0.5--0.7  & \grad{per-prad-occ_smass=0.5-0.7-m} & \grad{per-prad-occ_smass=0.5-0.7-Rp0} & O \\ 
& 0.7--1.0  & \grad{per-prad-occ_smass=0.7-1.0-m} & \grad{per-prad-occ_smass=0.7-1.0-Rp0} & O \\ 
& 1.0--1.4  & \grad{per-prad-occ_smass=1.0-1.4-m} & \grad{per-prad-occ_smass=1.0-1.4-Rp0} & O \\ 
% D
\Rp = R_{p,0} \; \bigg( \frac{\Sinc}{100~\Se} \bigg)^m 
& 0.5--1.4  &\grad{sinc-prad-det_smass=0.5-1.4-m} & \grad{sinc-prad-det_smass=0.5-1.4-Rp0} & D \\ 
& 0.5--0.7  & \grad{sinc-prad-det_smass=0.5-0.7-m} & \grad{sinc-prad-det_smass=0.5-0.7-Rp0} & D \\ 
& 0.7--1.0  & \grad{sinc-prad-det_smass=0.7-1.0-m} & \grad{sinc-prad-det_smass=0.7-1.0-Rp0} & D \\ 
& 1.0--1.4  & \grad{sinc-prad-det_smass=1.0-1.4-m} & \grad{sinc-prad-det_smass=1.0-1.4-Rp0} & D \\ 
% O
& 0.5--0.7  & \grad{sinc-prad-occ_smass=0.5-0.7-m} & \grad{sinc-prad-occ_smass=0.5-0.7-Rp0} & O \\ 
& 0.7--1.0  & \grad{sinc-prad-occ_smass=0.7-1.0-m} & \grad{sinc-prad-occ_smass=0.7-1.0-Rp0} & O \\ 
& 1.0--1.4  & \grad{sinc-prad-occ_smass=1.0-1.4-m} & \grad{sinc-prad-occ_smass=1.0-1.4-Rp0} & O \\ 
\Rp = R_{p,0} \; \bigg( \frac{\Mstar}{1~\Msun} \bigg)^m 
& 0.5--1.4  &\grad{smass-prad-det_smass=0.5-1.4-m} &  \grad{smass-prad-det_smass=0.5-1.4-Rp0} & D\\ 
\Rp = m \fe + R_{p,0} 
& 0.5--1.4  &\grad{smet-prad-det_smass=0.5-1.4-m}  & \grad{smet-prad-det_smass=0.5-1.4-Rp0} & D
\enddata
\tablecomments{We fit the Radius Gap in the {\em detected} planet population in \S\ref{sec:planet-population-period-radius} as function of orbital period, incident stellar flux, stellar mass, and stellar metallicity (see Figure \ref{fig:planet-prad-zoom}). We repeated the orbital period and incident flux fits for narrower ranges of stellar mass (see Figures~\ref{fig:occur-contour-six} and \ref{fig:occur-contour-six-sinc}). We fit the Radius Gap in the de-biased planet {\em occurrence} distribution in \S\ref{sec:occurrence-mass-radius} (see Figures~\ref{fig:occur-contour-six} and \ref{fig:occur-contour-six-sinc}). We elaborate on the following columns: `Fit'---the functional form of the fit; `\Mstar'---range of stellar masses; `Dist'---indicates whether the fit was of the distribution of planet detections `D' or occurrence `O'.}
\end{deluxetable}

\subsection{Dependence of planet size on stellar mass}
\label{sec:planet-population-mass}

In the previous section, we measured the slope of the Radius Gap in the \Rp-\Mstar plane and noted that it increases with \Mstar. Inspecting Figure~\ref{fig:planet-prad-zoom}c, we see qualitatively that the Radius Gap trend driven by the fact that the typical sub-Neptune grows with stellar mass while the typical super-Earth does not. Here, we quantify these trends and explore possible confounding factors.

We first measured the sub-Neptune \Rp-\Mstar correlation by selecting all sub-Neptunes from the population shown in Figure~\ref{fig:planet-prad-zoom} based on their radii (\Rp = 1.7--4.0~\Re). We then fit a power-law
\begin{equation}
\label{eqn:powerlaw-smass}
R_p = R_{p,0} \left(\frac{\Mstar}{\Msun}\right)^{\alpha}.
\end{equation}
We used the Levenberg-Marquardt algorithm as implemented in the {\em lmfit} Python package to find the best-fit parameters and measured uncertainties via 1000 bootstrap resamples. We detected a positive correlation between \Mstar and sub-Neptune size at $8\sigma$ significance  $\alpha = \fit{sn-m-alpha-mass}$. An identical analysis for the super-Earths (\Rp = 1.0--1.7~\Re) revealed no correlation $\alpha = \fit{se-m-alpha-mass}$.

Before interpreting these trends with \Mstar, we must consider confounding effects of other stellar properties that are correlated with \Mstar. (We consider effects from \Mstar-dependent survey completeness in \S\ref{sec:occurrence-mean-size}). Stellar metallicity and age are both concerns because they are astrophysically correlated with mass. This \Mstar-\fe-age correlation is tied to the chemical enrichment of the galaxy and stellar lifespans. A star's main-sequence lifetime is a strong function of its mass; $t_\mathrm{ms} \sim 10 \, \mathrm{Gyr} \, (\Mstar / \Msun)^{-2.5}$. Massive stars are younger on average and formed when the galaxy was more enriched in metals, as shown in Figures~\ref{fig:hr}a and \ref{fig:hr}c. Figure~\ref{fig:hr}b highlights the variable uncertainties of our age measurements which range from $\approx$0.1~dex for stars that have evolved off the main sequence, to $\approx$0.5~\dex for stars with $\teff \lesssim 5500$~K, i.e. indistinguishable from the zero-age main-sequence. 

We quantified possible metallicity and age effects by including them in our regression:
\begin{equation}
\label{eqn:power-law}
R_p = R_{p,0} \left(\frac{\Mstar}{\Msun}\right)^\alpha \left(\frac{N_\mathrm{Fe}}{N_{\mathrm{Fe},\odot}}\right)^{\beta} \left(\frac{\mathrm{age}}{\mathrm{5\, Gyr}}\right)^\gamma.
\end{equation}
When we allowed $\alpha$ and $\beta$ to vary and fixed $\gamma$ to 0, we found $\beta = \fit{sn-mm-alpha-met}$ and $\fit{se-mm-alpha-met}$ for sub-Neptunes and super-Earths, i.e. no significant \Rp-\fe correlation. Finally, we considered the effects of age by allowing $\gamma$ to vary as well. We restricted our analysis to stars with \teff > 5500~K where \ageiso is not prior-dominated.  We found $\gamma = \fit{sn-mma-alpha-age}$ and $\fit{se-mma-alpha-age}$ for the sub-Neptunes and super-Earths, i.e. no \Rp-age correlation.

Based on these fits, we concluded that stellar mass rather than metallicity or age is the property most closely linked to sub-Neptune size variation. We acknowledge two additional caveats to the above analysis which we address in later sections. First, we fit the population of detected planets which includes survey completeness effects that are addressed in \S\ref{sec:occurrence-mean-size}. Second, orbital period is another possible confounding variable if the {\em shape} of the period distribution varies with stellar mass. Fortunately, as we show in \S\ref{sec:occ-period-distribution} the period distribution is nearly independent of stellar mass up to a normalization constant.

\begin{comment}
\begin{deluxetable}{llrrrr}[h!]
\tablecaption{Power-law fits to super-Earth and sub-Neptune sizes}
\label{tab:fits}
\tabletypesize{}\Kepler
\tablecolumns{6}
\tablehead{
	\colhead{Fit} & 
	\colhead{Size} & 
	\colhead{$R_0$} &
	\colhead{$\alpha$} &
	\colhead{$\beta$} & 
	\colhead{$\gamma$} 
}
\startdata
1 & SE & $\fit{se-m-R0}$ & $\fit{se-m-alpha-mass}$ & 0 (fixed) & 0 (fixed)\\ 
  & SN & $\fit{sn-m-R0}$ & $\fit{sn-m-alpha-mass}$ & 0 (fixed) & 0 (fixed)\\ 
2 & SE & $\fit{se-mma-R0}$ & $\fit{se-mma-alpha-mass}$ & $\fit{se-mma-alpha-met}$ & $\fit{se-mma-alpha-age}$ \\ 
  & SN & $\fit{sn-mma-R0}$ & $\fit{sn-mma-alpha-mass}$ & $\fit{sn-mma-alpha-met}$ & $\fit{sn-mma-alpha-age}$ \\ 
\enddata
\tablecomments{Summary of power-law fits (Equation~\ref{eqn:power-law}) to the population of detected super-Earths and sub-Neptunes described in \S\ref{sec:planet-population-mass}}
\end{deluxetable}
\end{comment}

%We note that metallicity appears to correlate more strongly with the dispersion of planet sizes, rather than their mean values. For sub-Neptunes, increasing metallicity correlates a higher dispersion in planet sizes. Planets larger than 4.0~\Re are rare compared to smaller planets at all masses and metallicities. But when they are found, they are almost exclusively associated with stars with super-Solar metallicity. These large planets do not show nearly as strong of a correlation with stellar mass. 

\section{De-Biased Distribution of Planets}
\label{sec:occurrence}

We have characterized the location and slope of the Radius Gap and its dependence on stellar mass. However, one must exercise caution when connecting these trends to planet formation models because the distributions of planets have been filtered through {\em Kepler's} selection function. In this section, we remove selection effects to characterize variation in the planet population with stellar mass. 

In \S\ref{sec:occurrence-mass-radius}, we reexamine the slope of the Radius Gap as a function of period and incident flux (first treated in \S\ref{sec:planet-population-mass}) and find consistent results after accounting for selection effects. In \S\ref{sec:occurrence-mean-size}, we measure the sizes of super-Earths and sub-Neptunes (as in \S\ref{sec:planet-population-mass}) and observe the same qualitative trends after removing selection effects. In $\S\ref{sec:occ-period-distribution}$, we characterize how the period and incident flux distributions of super-Earths and sub-Neptunes vary with stellar mass.

\subsection{Dependence of the Radius Gap on stellar mass}
\label{sec:occurrence-mass-radius}

We return to the $P$-\Rp distribution of planets shown in Figure~\ref{fig:occur-contour-six}. We wish to remove {\em Kepler's} observational biases to measure the planet occurrence rate density (ORD), $\dd^2 f / \dd \log P \dd \log \Rp$. One may integrate the ORD over a specified domain in the $P$-\Rp plane to derive the number of planets per star residing in that domain. 

For every detected planet in a sample of \nstar stars, a large number are missed due to non-transiting geometries or insufficient photometric S/N. We account for both effects here using the inverse detection efficiency method (IDEM, see \citealt{Fulton18b}). Each detected planet $i$ with ($P,\Rp$) represents $w_i = 1 / \left<\ptr\pdet\right>$ total planets where $\left<\ptr\pdet\right>$ is the product of transit and detection probability averaged over the parent stellar population. We computed $\left<\ptr\pdet\right>$ over a grid of $(P,\Rp)$.  The probability that a randomly inclined planet with semi-major axis $a$ will transit with $b < 0.8$ is $\ptr = 0.8 \Rstar / a$, assuming circular orbits. We computed \ptr\ for each star using the B20 \Rstar and deriving $a$ from $P$ and the B20 \Mstar using  Kepler's Third Law. 

\begin{figure}
\centering
\includegraphics[width=0.45\textwidth]{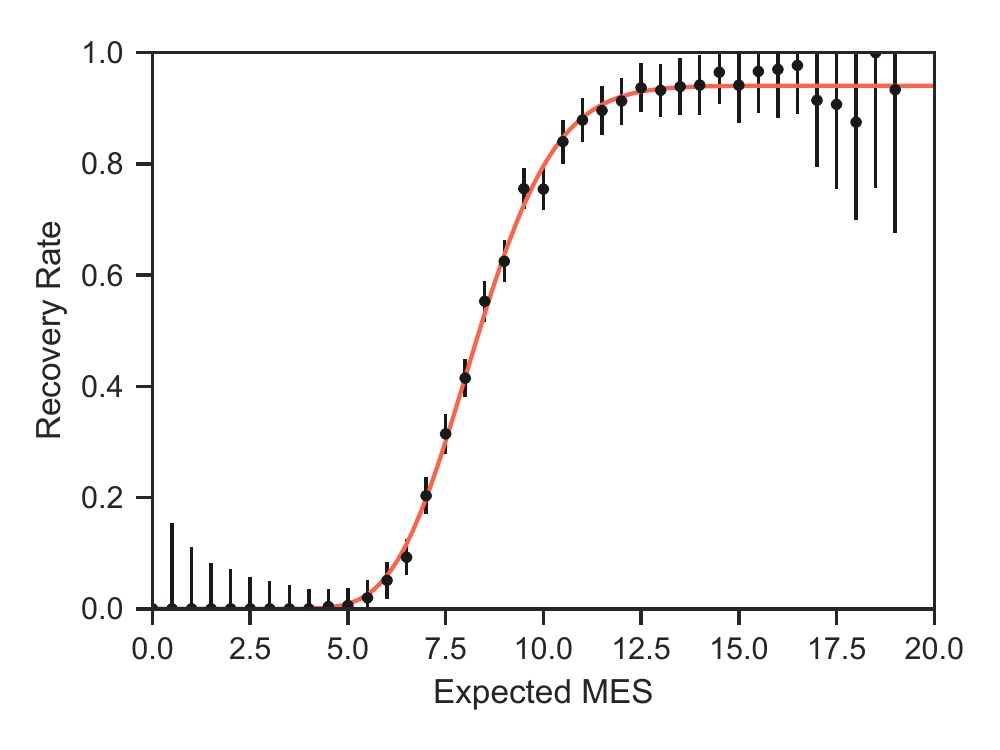}
\caption{Points show the recovery rate of the \Kepler planet detection pipeline of simulated planets injected by \cite{Christiansen20} as a function of multiple event statistic (MES). The curve shows the parameteric model we used to account for pipeline incompleteness in our occurrence calculations described in \S\ref{sec:occurrence}}
\label{fig:completeness}
\end{figure}

We characterized the recovery rate $\pdet$ following the procedure described in \cite{Christiansen20}. These authors injected a suite of 146,295 synthetic transits into the raw \Kepler pixel-level data and searched for transits using the same pipeline used to produce the T18 catalog. The output was a list of injections that were either successfully recovered or not. This list may be used to determine the average recovery rate for any \Kepler sample of interest. We selected the subset of injections with $P = 1-300$~days into our parent stellar population described in \S\ref{sec:sample} and measured the recovery rate as a function of expected MES, a close relative of transit S/N. We modeled the completeness curve as a $\Gamma$ cumulative distribution function
\begin{equation}
\pdet(\mathrm{MES}) = 
\frac{c}{b^a \Gamma(a)} \int_0^\mathrm{MES} t^{a-1} e^{-t/b} \dd t.
\end{equation}
The best-fit model had the following coefficients $(a, b, c) = (26.1, 0.320, 0.941)$ and is shown in Figure~\ref{fig:completeness}. We set the recovery rate to zero when MES < 10 to incorporate our filter described in \S\ref{sec:sample}. We evaluated MES for each star based on the putative planet's transit depth, transit duration, and the star's tabulated photometric noise (Combined Differential Photometric Precision; \citealt{Jenkins10}).

With our calculated weights $w$, we may now compute the total number of planets within a given region of the $P$-\Rp plane by summing the weights of the planets within that region, $\sum_{i} w_i$, or compute the occurrence rate (number of planets per star) via $(1/\nstar) \sum_{i} w_i$.

The IDEM, while mathematically straightforward, has some drawbacks particularly when $\pdet$ is small. In brief, \cite{Hsu18} showed that the IDEM is a biased estimator at low $\pdet$ because the weights $w_i$ are computed assuming no uncertainties in $\Rp$. \cite{Hsu18} proposed a less biased estimator using Approximate Bayesian Computation (ABC). We avoided these complications by restricting our analysis to regions where $\pdet$ > 25\%, where differences between the IDEM and ABC amount to <1$\sigma$ \citep{Fulton18b}. 

We measured the ORD in the three bins of stellar mass introduced in \S\ref{sec:planet-population-mass} via
\begin{equation}
\frac{\dd^2 f}{\dd \log P \dd \log \Rp} =  \frac{1}{\nstar} \sum_{i} w_{i} k(P - P_i, \Rp - R_{p,i}).
\end{equation}
where $k$ is a 2D gaussian with a bandwidth of $\log(2) = 0.3$~dex in $P$ and $\log(1.05)$ = 0.02~dex in \Rp. The results are shown in the right-hand panels of Figure~\ref{fig:occur-contour-six}. The value at each $(P, \Rp)$ is the number of planets per star in a 1~dex $\times$ 1~dex interval centered at $(P, \Rp)$. It is convenient to scale this value to smaller intervals because the ORD varies significantly over 1~dex in $P$ and $\Rp$. For example, to read off the number of planets per star within a box having the dimensions of the kernel FWHM, multiply the ORD by $2.355 \times 0.3 \times 2.355 \times 0.02 = 0.033$.

We do not plot occurrence in regions of low-planet detectability, defined to be where $n_{\star} \left<\ptr\pdet\right> < 50$. The ORD ranges from $\approx$0--4 which corresponds to $\approx$0--0.1 planets per star within the kernel FWHM. For $(P, \Rp)$ values where $n_{\star} \left<\ptr\pdet\right> = 50$, there are effectively 50 stars where such a planet could be detected, and we would expect 0--5 detected planets within the kernel FWHM. At this point Poisson fluctuations become unmanageable. 

We fit a power-law to the Radius Gap for each mass bin. Our methodology, power-law parameterization, fitting domain, and uncertainty analysis are identical to that presented in \S\ref{sec:planet-population}, except that we fit ORD rather than density of detected planets. The credible range of values is given in Table~\ref{tab:gapfits}. The measured slope and intercepts are consistent at 1$\sigma$ to those measured using planet detections only. 
 
\subsection{Dependence of planet size on stellar mass}
\label{sec:occurrence-mean-size}

In \S\ref{sec:planet-population-mass}, we noted a positive \Rp-\Mstar correlation for sub-Neptunes that was not mirrored by super-Earths. However, we must consider the possibility that such correlations (or lack thereof) could stem from the \Kepler selection function. To do so, we first divided the CXM planet hosts and parent stellar population into our three \Mstar bins from \S\ref{sec:planet-population-mass}. In order to treat hosts and non-hosts equally, we used the B20 \Mstar which are available for all stars. 

\begin{figure}
\centering
\includegraphics[width=0.45\textwidth]{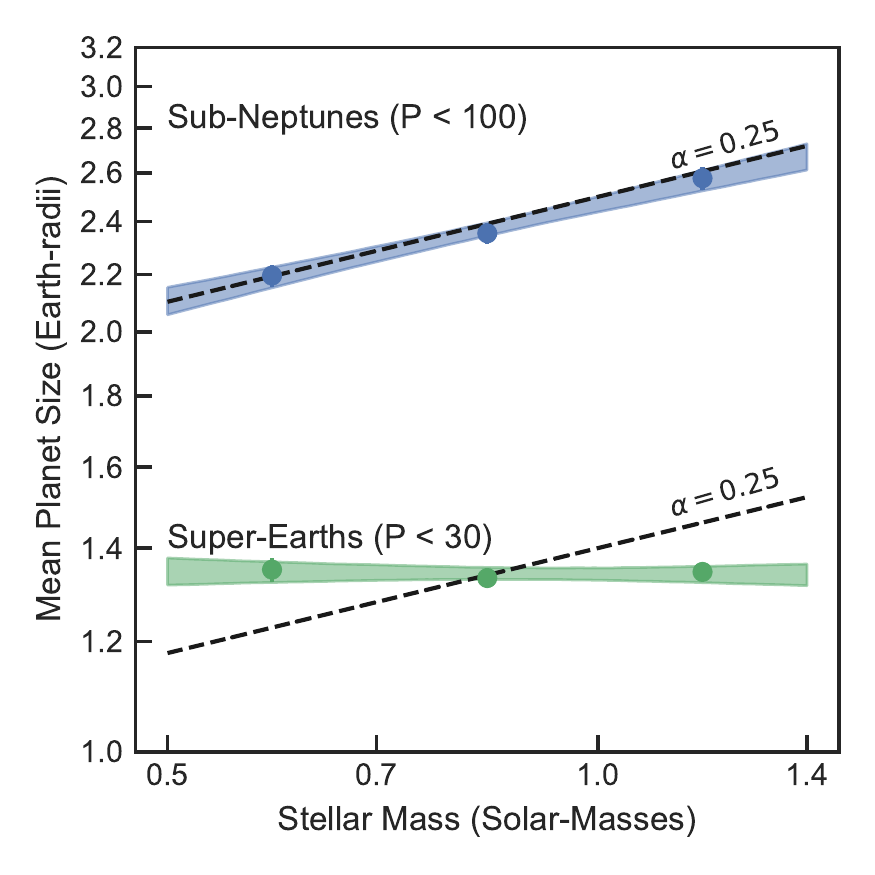}
\caption{Mean size of super-Earths with $P < 30$~days and sub-Neptunes with $P < 100$~days. Over \Mstar = 0.5--1.4~\Msun, the average sub-Neptune grows from 2.1~\Re to 2.6~\Re with a slope of $\alpha = \dd \log \Rp/ \dd \log \Mstar$ = $\fit{mps_size-sn-alpha}$. The mean super-Earth has a nearly constant size of 1.35~\Re with $\alpha$ = $\fit{mps_size-se-alpha}$. \cite{Wu19} reported $\alpha = 0.25$ for both populations and is shown by the dashed lines. Our sub-Neptune results are consistent with this index, but our super-Earth results are inconsistent at 8$\sigma$.}
\label{fig:mean-planet-size}
\end{figure}

Returning to our binned planet population, we computed the occurrence-weighted mean planet size for sub-Neptunes having $P$ < 100~days and display them in Figure~\ref{fig:mean-planet-size}. Fitting the power-law given by Equation~\ref{eqn:powerlaw-smass} to the three binned mean size measurements shown in Figure~\ref{fig:mean-planet-size} yields an index $\alpha$ = $\fit{mps_size-sn-alpha}$ and an intercept of $R_{p,0}$ = $\fit{mps_size-sn-R0}$~\Re (uncertainties from bootstrap resampling). These values are consistent with our fits to the population of detected planets (\S\ref{sec:planet-population-mass}). For the sub-Neptunes, the weights $w$ are dominated by geometrical corrections which are independent of \Rp. Planets with larger $P$ have larger $w$ and are more strongly weighted than in \S\ref{sec:planet-population-mass}. However, given the lack of a significant tilt in the $P$-\Rp distribution, the average value is similar. 

Corrections stemming from low detectability are larger for the super-Earths. While there is no obvious tilt of the super-Earth population in Figure~\ref{fig:planet-prad-zoom}c, completeness effects could conspire to produce a flat dependence when there is actually a positive slope in the underlying population. We computed the occurrence-weighted mean planet size for super-Earths with $P$ < 30~days and found $\alpha$ = $\fit{mps_size-se-alpha}$ and $R_{p,0}$ = $\fit{mps_size-se-R0}$, i.e. no significant correlation. We interpret the different $\Rp$-\Mstar correlations super-Earths and sub-Neptunes and compare our results to previous literature measurements in \S\ref{sec:discussion}.

\subsection{Orbital period and incident flux distributions} 
\label{sec:occ-period-distribution}

One key feature of the exoplanet population is the paucity of sub-Neptunes with $P \lesssim 10$~days, i.e. the ``Sub-Neptune Desert.'' The edge of this desert is significant in both the core and XUV-powered models. It represents a boundary, inside of which core-dominated planets (i.e $\Mcore \gg \Menv$) cannot retain H/He envelopes and are completely stripped. 

Here, we characterize the edge of the Sub-Neptune Desert in terms of period and flux and note how it changes with stellar mass. One can see qualitatively from Figures~\ref{fig:occur-contour-six} and \ref{fig:occur-contour-six-sinc} that the edge of the Sub-Neptune Desert occurs at a similar $P$ for all mass bins but varies in \Sinc by $\sim$10$\times$. We first considered the edge in period space in our three bins of stellar mass introduced in \S\ref{sec:planet-population-mass}. Quantitatively measuring this boundary in $P$ by visually inspecting Figures~\ref{fig:occur-contour-six} is suboptimal for a number of reasons. First, the contours do not convey uncertainties. Second, the Gaussian KDE used to generate the 2D distributions smears out sharp features in the $P$ distributions. Third, the Sub-Neptune Desert is the paucity of short-period Sub-Neptunes {\em relative} to long-period Sub-Neptunes, but the contours show {\em absolute} occurrence which also varies with stellar mass. These shortcomings motivated a different method, which we describe below.

Within each \Mstar bin, modeled the ORD as 
\begin{equation}
\frac{\dd f}{\dd P \dd \Rp}  = f_\star \lambda(P,R_p; \pmb{\theta})
\end{equation}
over a rectangular domain of $P$ and \Rp. Here, $f_\star$ is the mean number of planets per star within the $P$-\Rp domain. In principle, $\lambda$ may be any function that integrates to unity over the same domain. We consider parametric functions described by a vector of parameters $\pmb{\theta}$ that we will constrain through Bayesian inference. In short, the free parameters $f_\star$ and $\pmb{\theta}$ control the normalization and shape of the ORD, respectively. 

Following previous works, we modeled the detected population of planets as a realization of an inhomogeneous Poisson point process (see, e.g., \citealt{Rogers21a}, and references therein). The log-likelihood of observing the full sample is given by 
\begin{equation}
\label{eqn:ppp-likelihood}
\ln \mathcal{L} = -\Lambda +  \sum_{i} \ln \left[ f_{\star} \lambda(P_i,R_{p,i}) \right]
\end{equation}
where
\begin{equation}
\Lambda  = \nstar f_\star \iint \eta(P,R_p) \lambda(P,R_p) \dd P \dd \Rp.
\end{equation}
is the expected number of detected planets given $f_{\star}$ and $\pmb{\theta}$. Here, $\eta = \left<\ptr\pdet\right>$.

For simplicity, we assumed that the period and radius distributions are independent, i.e. $\lambda(P,\Rp) \propto \lambda(P) \lambda(\Rp)$. We also approximated the \Rp distribution log-uniform within the \Rp boundaries, i.e. $\lambda(\Rp) \propto 1 / \Rp$. We precomputed
\begin{equation}
\eta(P) = \frac{1}{\Delta \ln \Rp} \int \eta(P,\Rp) \dd \ln \Rp 
\end{equation}
to reduce $\Lambda$ to a 1D integral

\begin{equation}
\Lambda  = \nstar f_\star \int \eta(P) \lambda(P) \dd P.
\end{equation}

Following \cite{Rogers21a} we described the period distribution with a smooth broken power-law.
\begin{equation}
\label{eqn:sbpl}
\lambda (P)  =  \frac{C}{(P/P_0)^{-k_1} + (P/P_0)^{-k_2}},
\end{equation}
where $C$ is the constant of proportionality that enforces normalization.%
\footnote{
The normalization constant $C$ is 
\begin{eqnarray*}
C & = &  \left[ \frac{P_0 u^{k_1+1}}{k_1+1} {}_2F_1\left(1, \frac{k_1 + 1}{k_1 - k_2}; \frac{k_1+1}{k_1 - k_2} + 1; -u^{k_1 - k_2} \right) \right]_{u = \frac{P_1}{P_0}}^{u = \frac{P_2}{P_0}}
\end{eqnarray*}
where ${}_2F_1$ is the hypergeometric function.
}
In this parameterized description, $P_0$ is the knee of the distribution; for $P \ll P_0$ the distribution approaches a power-law with an index of $k_1$; for $P \gg P_0$ the index is $k_2$.

\begin{figure*}
\centering
\includegraphics[width=0.8\textwidth]{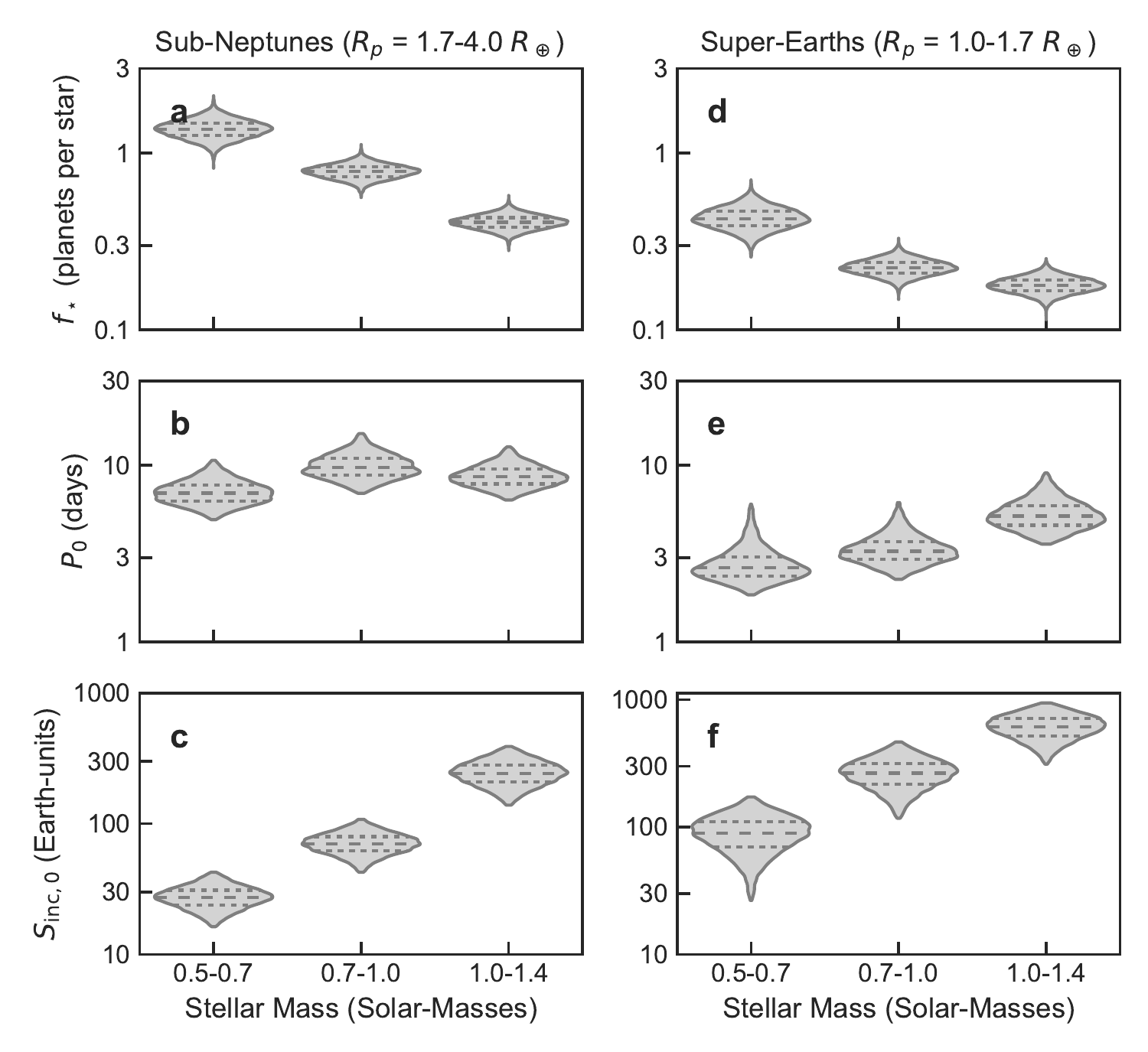}
\caption{Panel (a): posterior probability density of the mean number of sub-Neptunes per star $f_{\star}$ with $P$ = 1--300~days for three stellar mass bins. Dashed lines show the 25\%, 50\%, and 75\% quartiles. Panel (b) shows the period breakpoint $P_0$. Panel (c) shows the flux breakpoint $\Sincc{0}$. Panels (d--f) same as (a--c) but for super-Earths. The mean occurrence in panel (d) corresponds to $P$ = 1--30~days. For both classes of planets, there is a decline in total occurrence with increasing \Mstar For the sub-Neptunes, $P_0$ ranges from $\approx$7.0--9.6~days, consistent to 40\%; for the super-Earths $P_0$ is $\approx$2.7--5.1~days, consistent to a factor of two. For the sub-Neptunes, the breakpoint $\Sincc{0}$ increases by a factor of 9 from $\fit{fitsinc_smass=0.5-0.7-prad=1.7-4.0-x0}$~\Se to $\fit{fitsinc_smass=1.0-1.4-prad=1.7-4.0-x0}$~\Se between the low- and high-mass bin. For the super-Earths, the breakpoint also increases with stellar mass from $\fit{fitsinc_smass=0.5-0.7-prad=1.0-1.7-x0}$~\Se to $\fit{fitsinc_smass=1.0-1.4-prad=1.0-1.7-x0}$~\Se, a factor of 7 increase.}
\label{fig:occur-violin}
\end{figure*}

We first fit the sub-Neptune distribution for our three bins of stellar mass over $P$ = 1--300~days. Our goal was to characterize the knee and slope of the occurrence period distribution. For consistency between the sub-samples, we fixed $k_2 = -1$, i.e. log-uniform occurrence for $P \gg P_{0}$. We maximized the likelihood (Equation~\ref{eqn:ppp-likelihood}) using the L-BFGS-B algorithm and explored the credible range of parameters using the MCMC sampler of \cite{Goodman10}. We imposed log-uniform priors on $f_\star$ and $P_0$ and a linear prior on $k_1$. We sampled the posterior with 8 walkers for 5,000 steps each and discarded the first 1,000 as burn-in. The chains for all parameters were at least 50 times longer than the integrated autocorrelation time, indicating convergence. Table~\ref{tab:fits} lists the 16, 50, 84 percentiles of our posterior samples. Figure~\ref{fig:occur-violin} shows the posterior probability density of $f_\star$ and $P_0$ for different \Mstar bins.

\begin{figure*}
\centering
\includegraphics[width=0.8\textwidth]{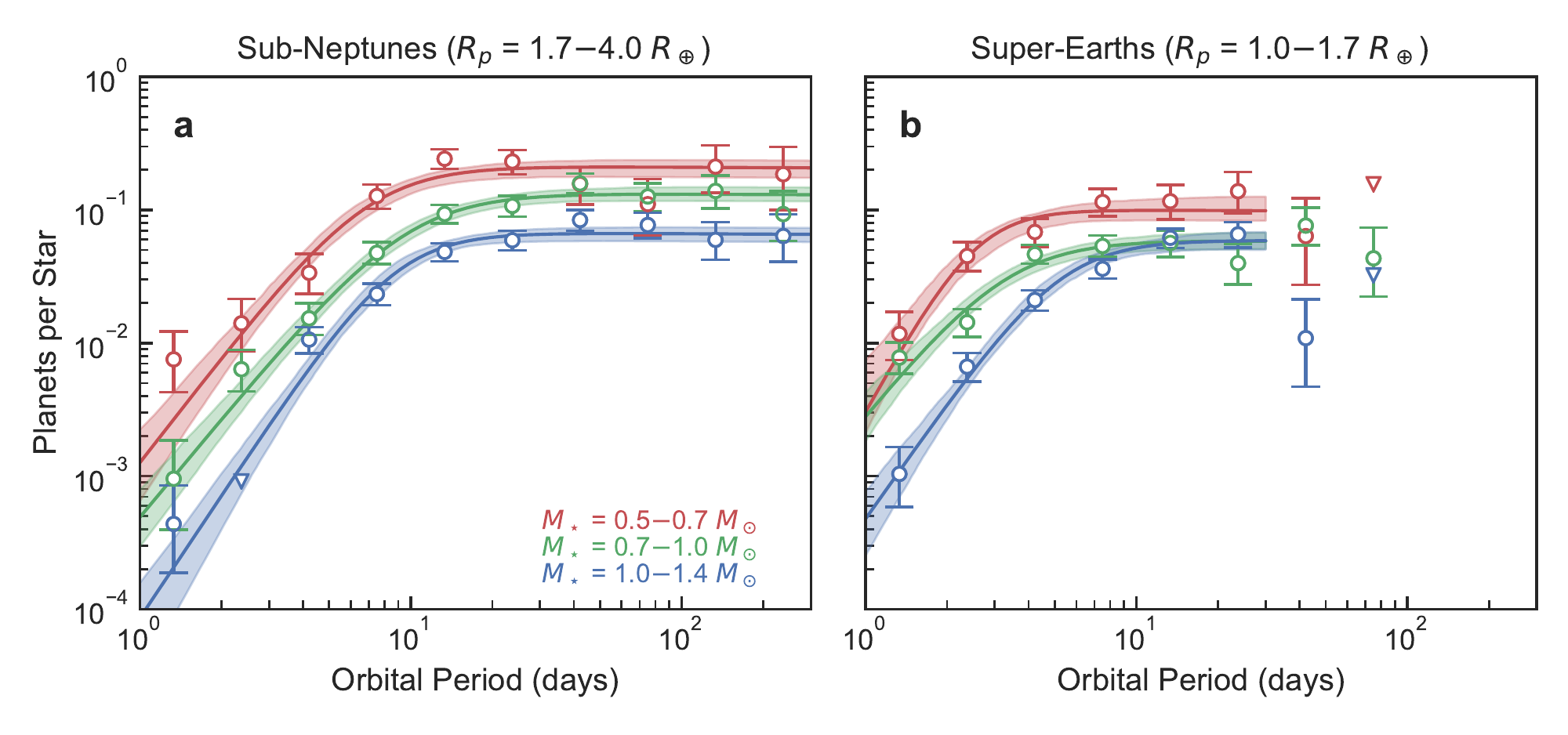}
\includegraphics[width=0.8\textwidth]{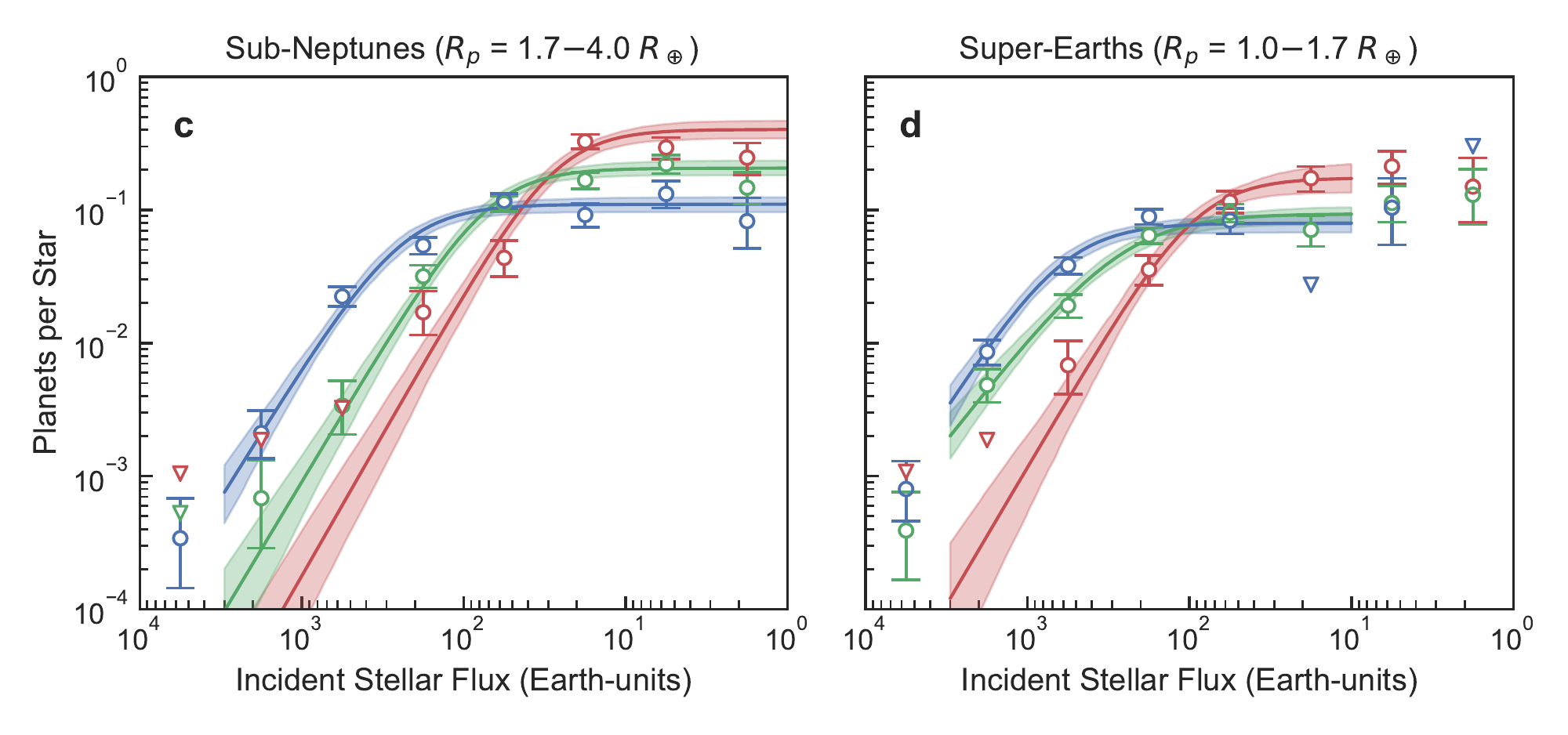}
\caption{Panel (a): period distribution of sub-Neptunes for three bins of stellar mass. The points show the number of planets per star in period bins that are 0.25~dex wide. We modeled the occurrence rate density $\dd f / \dd P$ as a smooth broken power-law which we fit to the unbinned planet population over $P$ = 1--300~days (see \S\ref{sec:occ-period-distribution}). The bands show the credible range of $\dd f / \dd \log P \times 0.25$~dex. The height of each curve at a particular $P$ is the number of planets per star within a 0.25~dex period interval centered at $P$. Panel (b): same as (a) except for super-Earths where the fits extend to 30~days. Panel (c--d): same as (a--b) except showing incident flux distribution. The vertical axis has the following interpretation: number of planets per star within a 0.5~dex flux interval.}
\label{fig:occur-per}
\end{figure*}

Figure~\ref{fig:occur-per} shows the best-fit and credible range of models for our three \Mstar bins. They are similar except for an overall shift in the normalization. For the low-, medium-, and high-mass bins, the breakpoint $P_0$ is at $\fit{fitper_smass=0.5-0.7-prad=1.7-4.0-x0}$, $\fit{fitper_smass=0.7-1.0-prad=1.7-4.0-x0}$, $\fit{fitper_smass=1.0-1.4-prad=1.7-4.0-x0}$~days, respectively, consistent to 2$\sigma$; the power-law indices $k_1$ agree to 1$\sigma$ and are $\fit{fitper_smass=0.5-0.7-prad=1.7-4.0-k1}$, $\fit{fitper_smass=0.7-1.0-prad=1.7-4.0-k1}$, $\fit{fitper_smass=1.0-1.4-prad=1.7-4.0-k1}$; we observed a steady decrease in the absolute number of planets per star from 0.5 to 1.4~\Msun of $\fit{fitper_smass=0.5-0.7-prad=1.7-4.0-f}$, to $\fit{fitper_smass=0.7-1.0-prad=1.7-4.0-f}$, to $\fit{fitper_smass=1.0-1.4-prad=1.7-4.0-f}$. 

The similarity of $P_0$ for the different mass bins is noteworthy because it indicates a key boundary in the planet formation process that is stationary over a wide range of stellar mass. Naturally, stationary $P_0$ implies a variable \Sincc{0} given the relationship between \Mstar and bolometric luminosity. To quantify this boundary in flux space, we again characterized the distribution as a smooth broken power-law, 
\begin{equation}
\label{eqn:sbpl}
\lambda (\Sinc)  =  \frac{C}{(\Sinc/\Sincc{0})^{-k_1} + (\Sinc/\Sincc{0})^{-k_2}},
\end{equation}
but fixed $k_{1}$ to $-$1. Over the three \Mstar bins, we observed factor of 9 increase  in the breakpoint $\Sincc{0}$ from $\fit{fitsinc_smass=0.5-0.7-prad=1.7-4.0-x0}$, to $\fit{fitsinc_smass=0.7-1.0-prad=1.7-4.0-x0}$, to $\fit{fitsinc_smass=1.0-1.4-prad=1.7-4.0-x0}$~\Se.

\begin{figure*}
\centering
\includegraphics[width=0.9\textwidth]{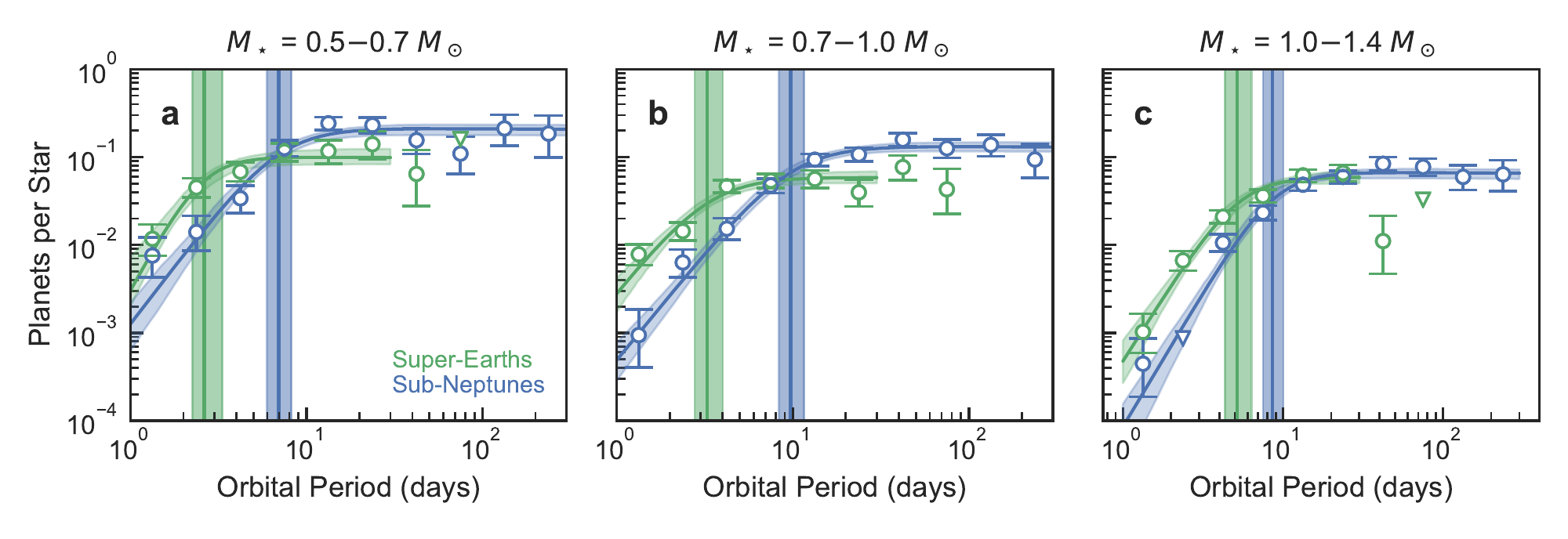}
\includegraphics[width=0.9\textwidth]{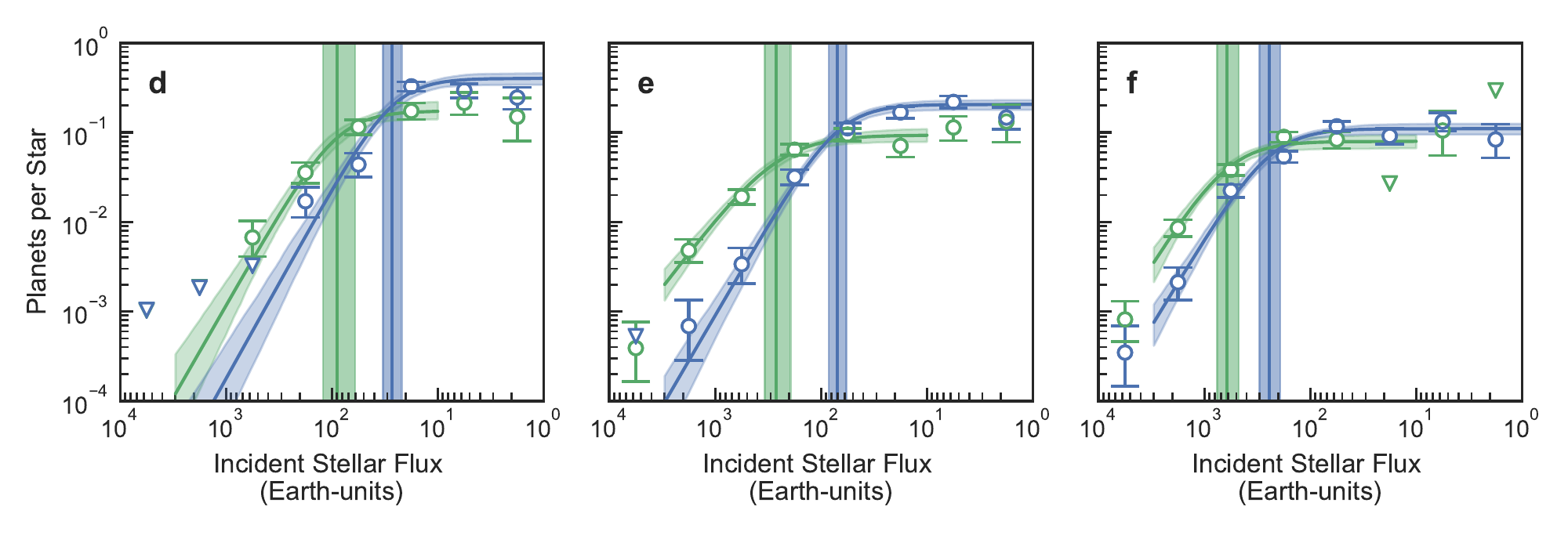}
\caption{Panels (a--c): period distribution of super-Earths (green) and sub-Neptunes (blue) for different bins of stellar mass. Points and curves are the same as shown in  Figure~\ref{fig:occur-per} but grouped according to stellar mass as opposed to planet size. The vertical bands show the range of credible period breakpoints $P_0$. For all stellar mass bins, $P_0$ is smaller for super-Earths than for sub-Neptunes. Panels (d--f): same as (a--c), but x-axis is incident stellar flux. The flux breakpoint $S_\mathrm{inc,0}$ increases with stellar mass, but the relative offset between the two populations is nearly constant.}
\label{fig:occur-per3}
\end{figure*}

We also characterized the period distribution of super-Earths. Our modeling was identical to that of the sub-Neptunes except we only included planets out to 30~days due to low completeness at longer orbital periods. Here, the breakpoint $P_0$ increases with \Mstar from 
$\fit{fitper_smass=0.5-0.7-prad=1.0-1.7-x0}$, to $\fit{fitper_smass=0.7-1.0-prad=1.0-1.7-x0}$, to $\fit{fitper_smass=1.0-1.4-prad=1.0-1.7-x0}$~days.
The breakpoints in flux $\Sincc{0}$ occur at
$\fit{fitsinc_smass=0.5-0.7-prad=1.0-1.7-x0}$, $\fit{fitsinc_smass=0.7-1.0-prad=1.0-1.7-x0}$, and $\fit{fitsinc_smass=1.0-1.4-prad=1.0-1.7-x0}$~\Se. Figure~\ref{fig:occur-per3} contrasts the super-Earth and sub-Neptune breakpoints. The super-Earth breakpoint is shifted to smaller orbital periods and higher incident fluxes for all stellar mass bins. We interpret these shifts in \S\ref{sec:discussion-period-flux}. 

\begin{deluxetable*}{llRRRRRR}
\tablecaption{Fits to the planet period and flux distributions}
\label{tab:fits}
\tabletypesize{\footnotesize}
\tablecolumns{8}
\tablehead{
    \multicolumn{2}{c}{} &
    \multicolumn{3}{c}{Period} & 
    \multicolumn{3}{c}{Flux}\\
	\colhead{Size} & 
	\colhead{\Mstar} & 
	\colhead{$f_\star$} &
	\colhead{$P_0$} &
	\colhead{$k_1$} &
	\colhead{$f_\star$} &
	\colhead{$\Sincc{0}$} &
	\colhead{$k_2$} 
}
\startdata
SE & 0.5--0.7 & \fit{fitper_smass=0.5-0.7-prad=1.0-1.7-f} & \fit{fitper_smass=0.5-0.7-prad=1.0-1.7-x0} & \fit{fitper_smass=0.5-0.7-prad=1.0-1.7-k1} & \fit{fitsinc_smass=0.5-0.7-prad=1.0-1.7-f} & \fit{fitsinc_smass=0.5-0.7-prad=1.0-1.7-x0} & \fit{fitsinc_smass=0.5-0.7-prad=1.0-1.7-k2}\\
   & 0.7--1.0 & \fit{fitper_smass=0.7-1.0-prad=1.0-1.7-f} & \fit{fitper_smass=0.7-1.0-prad=1.0-1.7-x0} & \fit{fitper_smass=0.7-1.0-prad=1.0-1.7-k1} & \fit{fitsinc_smass=0.7-1.0-prad=1.0-1.7-f} & \fit{fitsinc_smass=0.7-1.0-prad=1.0-1.7-x0} & \fit{fitsinc_smass=0.7-1.0-prad=1.0-1.7-k2}\\
   & 1.0--1.4 & \fit{fitper_smass=1.0-1.4-prad=1.0-1.7-f} & \fit{fitper_smass=1.0-1.4-prad=1.0-1.7-x0} & \fit{fitper_smass=1.0-1.4-prad=1.0-1.7-k1} & \fit{fitsinc_smass=1.0-1.4-prad=1.0-1.7-f} & \fit{fitsinc_smass=1.0-1.4-prad=1.0-1.7-x0} & \fit{fitsinc_smass=1.0-1.4-prad=1.0-1.7-k2}\\
SN & 0.5--0.7 & \fit{fitper_smass=0.5-0.7-prad=1.7-4.0-f} & \fit{fitper_smass=0.5-0.7-prad=1.7-4.0-x0} & \fit{fitper_smass=0.5-0.7-prad=1.7-4.0-k1} & \fit{fitsinc_smass=0.5-0.7-prad=1.7-4.0-f} & \fit{fitsinc_smass=0.5-0.7-prad=1.7-4.0-x0} & \fit{fitsinc_smass=0.5-0.7-prad=1.7-4.0-k2}\\
   & 0.7--1.0 & \fit{fitper_smass=0.7-1.0-prad=1.7-4.0-f} & \fit{fitper_smass=0.7-1.0-prad=1.7-4.0-x0} & \fit{fitper_smass=0.7-1.0-prad=1.7-4.0-k1} & \fit{fitsinc_smass=0.7-1.0-prad=1.7-4.0-f} & \fit{fitsinc_smass=0.7-1.0-prad=1.7-4.0-x0} & \fit{fitsinc_smass=0.7-1.0-prad=1.7-4.0-k2}\\
   & 1.0--1.4 & \fit{fitper_smass=1.0-1.4-prad=1.7-4.0-f} & \fit{fitper_smass=1.0-1.4-prad=1.7-4.0-x0} & \fit{fitper_smass=1.0-1.4-prad=1.7-4.0-k1} & \fit{fitsinc_smass=1.0-1.4-prad=1.7-4.0-f} & \fit{fitsinc_smass=1.0-1.4-prad=1.7-4.0-x0} & \fit{fitsinc_smass=1.0-1.4-prad=1.7-4.0-k2}\\
\enddata
\tablecomments{In \S\ref{sec:occ-period-distribution}, we modeled both the period and flux distribution of super-Earths (SE) and sub-Neptunes (SN) for different bins of stellar mass. We used a smooth broken power-law to model the period distribution $\dd f /\dd P \propto \left( (P/P_0)^{-k_1} + (P/P_0)^{-k_2}\right)^{-1}$. Here, $f_\star$ is the mean number of planets per star over the following period ranges: $P$ = 1--30~days (SE); $P$ = 1--300~days (SN). The power-law indices below and above the breakpoint $P_0$ are $k_1$ and $k_2$, respectively. In our modeling, $k_2$ is fixed to $-1$, i.e. $\dd f / \dd \log P \approx$ constant for $P \gg P_0$. Similarly, we also used a smooth broken power-law to model the flux distribution. Here, $k_1$ was fixed to $-1$.}
\end{deluxetable*}

\section{Discussion}
\label{sec:discussion}

\subsection{Positive \Mstar-\Rp correlation suggests massive stars make massive planet cores}
\label{ssec:discussion-mstar-rp}

One of the striking features of Figure~\ref{fig:planet-prad-zoom}c is the divergence of the super-Earth and sub-Neptune populations with stellar mass: the sub-Neptunes grow larger while the super-Earths do not. When we modeled the size dependence of each population as a power-law $\Rp \propto \Mstar^\alpha$, we found $\alpha = \fit{sn-m-alpha-mass}$ and $\alpha = \fit{se-m-alpha-mass}$, respectively. The slope of the Radius Gap in \Mstar-\Rp space (displayed in Figure~\ref{fig:planet-prad-zoom}) lies between those two values $m = \grad{smass-prad-det_smass=0.5-1.4-m}$.

A positive sub-Neptune \Rp-\Mstar correlation has been noted in several previous works. \cite{Fulton18b} observed such a trend over a narrower stellar mass range of $\approx$0.8--1.3~\Msun. Later, \cite{Wu19} stitched together the \cite{Fulton18b} sample with a compilation of $\sim$100 \Kepler and \ktwo planets with hosts of $\Mstar \lesssim 0.7~\Msun$ from \cite{Newton15} and \cite{Dressing17} and reported $\alpha$ = 0.23--0.35, consistent with what we found. 

In contrast to the sub-Neptunes, we observe a flat super-Earth \Rp-\Mstar dependence, $\alpha = \fit{mps_size-se-alpha}$. This result differs from that of \cite{Wu19} who reported $\alpha$ = 0.23--0.35. As a point of comparison, Figure~\ref{fig:mean-planet-size} shows our occurrence-weighted mean super-Earth size with an $\alpha$ = 0.25 dependence overplotted; such a strong relationship is ruled out at $8\sigma$ significance. The differences between the two works may stem from the fact that \cite{Wu19} used an add-mixture of planet samples with heterogeneously derived properties.

\cite{Berger20b} offers a second point of comparison. They derived properties of 2956 KOIs spanning $\Mstar \approx$ 0.6--1.4~\Msun that host 3898 confirmed/candidate planets. They measured the Radius Gap slope in \Mstar-\Rp space and found $m = 0.26^{+0.21}_{-0.16}$, which is consistent with our measured value, albeit with larger uncertainties. 

What causes the super-Earth size to be independent of stellar mass while the sub-Neptune size grows with stellar mass? In our discussion below, we assume planets between 1--4~\Re are Earth-composition bodies that may also include a H/He envelope. This assumption is supported by the growing body of mass measurements of planets in this size range (see, e.g., \citealt{Weiss14,Sinukoff18-thesis}). Most planets smaller than 1.5~\Re have densities consistent with Earth composition bodies; most planets larger than 2.0~\Re are too low-density to be rock/iron bodies and are consistent with having H/He envelopes of $\fenv \gtrsim 1\%$ by mass.

With this two-component model, both the XUV- and core-powered models produce the Radius Gap. In the 1D, energy-limited, XUV-powered models of \cite{Owen17} the mass loss-timescale $\tau_\mathrm{xuv} = M/\dot{M}$ is a strong function of \fenv and the local XUV flux. This timescale is maximized when the size of the core and thickness of the envelope are approximately equal, i.e. $\Rp \approx 2~R_c$. This occurs at $f_\mathrm{env,2} \approx 3\%$. Planets with less envelope experience accelerating mass-loss; those with more experience decelerating mass-loss. Planets are herded toward $f_\mathrm{env}$ = $f_\mathrm{env,2}$ or \fenv = 0 (bare cores).

In the XUV-model, the fate of a planet's envelope depends the time-integrated XUV exposure $\mathcal{F}_\mathrm{xuv} = \int F_\mathrm{xuv} \dd t$. This quantity is strongly correlated with orbital period and weakly correlated with stellar mass \citep{Owen17,Lopez18}. At fixed orbital period, planets around low-mass stars receive less bolometric flux  $F_\mathrm{bol}$, but $F_\mathrm{xuv}/F_{\mathrm{bol}}$ is higher and stars stay XUV-active longer. \cite{McDonald19} found $\Fxuv \propto F_{\mathrm{bol}} \Mstar^{-3}$. Applying following relationships, $P^{2} \propto a^{3} / \Mstar$ (Kepler's Third Law), $F_{\mathrm{bol}} \propto L_{\star,\mathrm{bol}} / a^{2}$, $L_{\star,\mathrm{bol}} \propto \Mstar^{4}$ (\citealt{Kippenhahn90}, for FGK and early M stars), we find $\Fxuv \propto \Mstar^{0.33} P^{-1.33}$. 

If the XUV model is correct and the initial planet population is independent of \Mstar, we expect only minor sub-Neptune size variability with \Mstar. We can get a rough estimate the expected \Rp-\Mstar dependence at fixed $P$ by considering the amount of energy needed to unbind a $f_\mathrm{env} \approx 3\%$ envelope 
\begin{equation}
E_\mathrm{unb} \sim G M_{c} M_\mathrm{env} / R_{c} \propto M_{c}^{1.75}.
\end{equation}
We have approximated the rocky mass-radius relationship as $R_{c} \propto M_c^{0.25}$. The XUV energy received by the planet is approximately
\begin{equation}
E_\mathrm{rec} \sim F_\mathrm{xuv} \pi \Rp^2 \propto F_\mathrm{xuv} M_{c}^{0.5}  \propto M_\star^{0.33} M_c^{0.5}.
\end{equation}
Equating the two energies yields $M_{c}\propto M_\star^{0.26}$ or $R_c \propto \Rp \propto M_\star^{0.06}$, i.e. $\alpha \approx 0.06$.

In the core-powered model of \cite{Gupta20}, planets with $f_\mathrm{env} < 5\%$ have enough thermal energy in their cores to unbind their envelopes and become bare rocky planets. Thus, this process also works to clear the $P$-\Rp plane of planets with \fenv of a few percent or less. The fate of a planet's envelope depends on the mass-flux at the Bondi radius $R_B$, and planets with smaller $R_B$ lose mass faster. Since $R_B \propto M_c / \teq$, planets more readily lose their envelopes when \teq (or equivalently \Sinc) is large, or when $M_c$ is small. If the core-powered model is correct and the initial planet population is independent of \Mstar, we expect significant changes with \Mstar because \Sinc is a strong function of \Mstar. \cite{Gupta20} predicted $\alpha \approx 0.33$.

It is tempting to attribute the larger sub-Neptunes around massive stars as a direct result of higher mass cores. One may, for example, adopt a monotonically increasing mass-radius relationship (e.g. \citealt{Weiss14}), invert it, and infer higher mass sub-Neptunes around more massive stars. However, there is significant astrophysical scatter about the mean mass-radius relationship \citep{Wolfgang16}. Planets of 2--3~\Re have a nearly uniform distribution in mass from 5--20~\Me. In short: radius is not a good predictor of mass. It could also be true that faster or more efficient gas accretion occurs in the protoplanetary disks of massive stars because of differences in gas surface density, ionization structure, or other properties. 

\begin{figure*}
\centering
\includegraphics[width=0.9\textwidth,trim={2cm 3cm 3cm 3cm}]{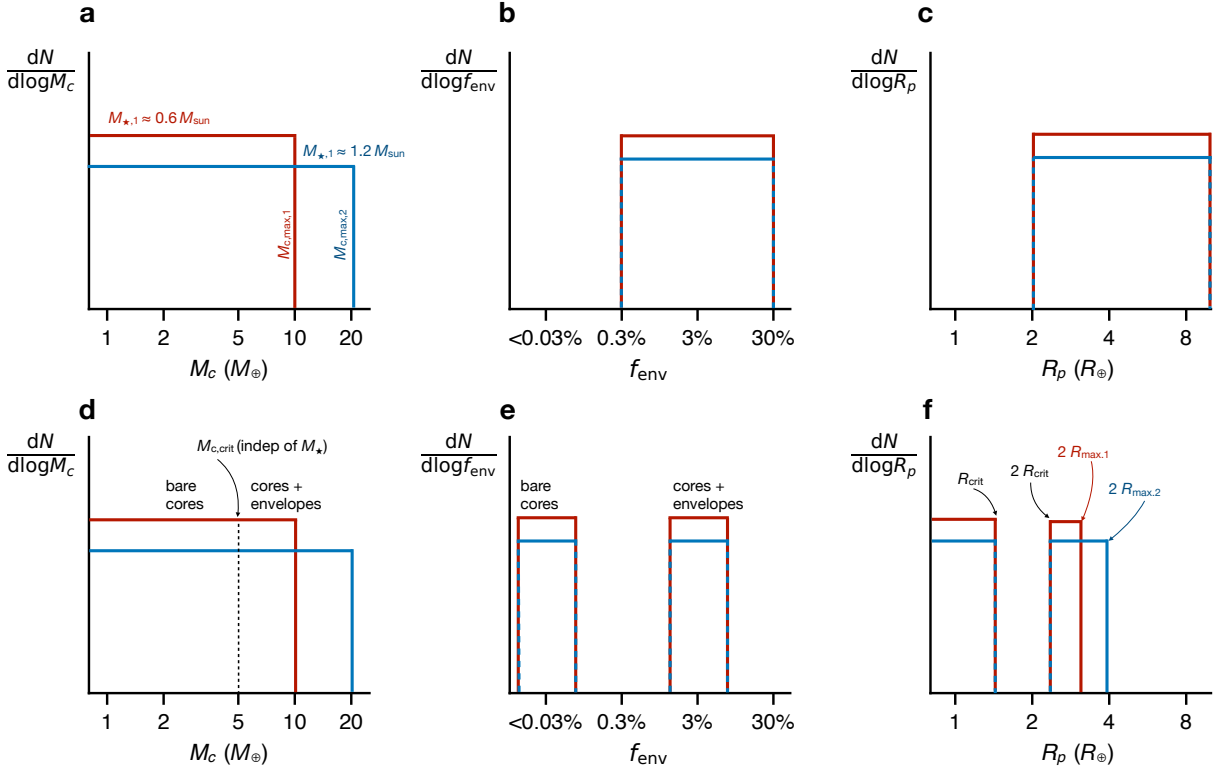}
%\vspace{-3cm}
\caption{A toy model to explore how a correlation between host star mass \Mstar and planet core mass \Mcore impacts super-Earth and sub-Neptune sizes when photoevaporation drives mass-loss. Panels (a--c) show the distributions of \Mcore, envelope fraction $\fenv$, and planet size $\Rp$ at the time of disk dispersal ($t \sim 10$~Myr). Here, low-mass stars have a lower maximum core mass than high-mass stars: for $\Mstar \approx 0.6~\Msun$, $M_\mathrm{c,max,1} \approx 10~\Me$; for $\Mstar \approx 1.2~\Msun$, $M_\mathrm{c,max,2} \approx 20~\Me$. Planets have a broad distribution of envelope fractions $\fenv \approx 0.3$--30\% and a broad distribution of sizes. Panels (d--f) show the same properties after the period of rapid photoevaporation ($t \sim 100$~Myr). Cores up to $M_\mathrm{crit}$ have been stripped bare. We have approximated the time-integrated XUV flux to be independent of \Mstar, so $M_\mathrm{crit}$ is also independent of \Mstar (see \S\ref{ssec:discussion-mstar-rp}). Photoevaporation bifurcates the $\fenv$ distribution into two modes with $\fenv \approx 0\%$ and $\fenv \approx 3\%$. The largest super-Earths have $R_\mathrm{crit}/\Re \approx (M_\mathrm{crit} / \Me)^{0.25}$ so $R_\mathrm{crit}$ is independent of \Mstar. The smallest sub-Neptunes barely retained their envelopes and have $\Rp \approx 2~R_\mathrm{crit}$. The largest sub-Neptune size is set by a $\fenv \approx 3\%$ envelope atop a $M_\mathrm{c,max}$ core or $R_\mathrm{max} = 2 (M_\mathrm{c,max} / \Me)^{0.25}$. Since $M_\mathrm{c,max} \propto \Mstar $, $R_\mathrm{max} \propto \Mstar^{0.25}$. The model is equally applicable to the core-powered framework except mass is lost over $\sim$1 Gyr timescales.}
\label{fig:toy-model}
\end{figure*}

With these caveats in mind, we {\em can} qualitatively explain the increase in sub-Neptune sizes and constant super-Earth sizes in both XUV- and core-powered models as a consequence of more massive cores around more massive stars. We first consider photoevaporation in a toy model shown in Figure~\ref{fig:toy-model}. Here, the cores have a log-uniform mass distribution up to a maximum cutoff mass of $M_\mathrm{c,max}$. The maximum core mass is proportional to stellar mass, $M_\mathrm{c,max} \propto \Mstar$. For simplicity, we show two populations of host stars: (1) where $M_{\star,1} \approx 0.6~\Msun$ and $M_\mathrm{c,max,1} \approx 10~\Me$, and (2) where $M_{\star,2} \approx 1.2~\Msun$ where $M_\mathrm{c,max,2} \approx 20~\Me$. At the time of disk dispersal ($t \sim 10$~Myr), the cores have accreted H/He envelopes with a broad range of $\fenv \approx 0.3$--30\% and have a broad size distribution $\Rp \approx 2$--10~\Re.

After the period of high XUV activity ($t \gtrsim 100$~Myr), cores up to a critical mass $M_\mathrm{c,crit}$ are stripped bare and the largest super-Earth is given by $R_\mathrm{se,max}/\Re \approx (M_\mathrm{c,crit}/\Me)^{0.25}$. Cores above $M_\mathrm{c,crit}$ retain their envelopes and mass-loss concentrates envelope fractions to $\fenv \approx 3\%$. The smallest sub-Neptunes have cores that barely retained their envelopes and obey $R_\mathrm{sn,min} / \Re \approx 2 (M_\mathrm{c,crit}/\Me)^{0.25}$; the largest sub-Neptunes have an $\fenv \approx 3\%$ envelope atop a $M_\mathrm{c,max}$ core and obey $R_\mathrm{sn,max}/\Re \approx 2 (M_\mathrm{c,max}/\Me)^{0.25}$. 

This toy model reproduces some, but not all, of the observed trends. $M_\mathrm{c,max} \propto \Mstar$ so $R_\mathrm{sn,max} \propto \Mstar^{0.25}$, which is consistent with the observed trend. Moreover, approximating $\mathcal{F}_\mathrm{xuv} \propto M^{0.33}$ as constant over our mass range predicts constant  $R_\mathrm{se,max}$ for different \Mstar.

The fact that $R_\mathrm{se,max} \approx 1.7$~\Re implies $M_\mathrm{c,crit} \approx 8$~\Me and $R_\mathrm{sn,min} \approx 3.4$~\Re. Given our assumed $M_\mathrm{c,max}$, we find $R_\mathrm{sn,max} \approx 3.6$~\Re  and $4.2~\Re$. While these have the appropriate dependence on \Mstar, they exceed the upper envelope of the observed sub-Neptunes. Note, however, that these sizes are at $t \sim 100$~Myr and observed planets are $t \approx$~1--10~Gyr and will contract as they radiate away their heat of formation. \cite{Gupta20} found that the radius of the radiative-convective boundary (RCB) shrinks as $\dd \log \Rp / \dd \log t \approx -0.1$. So over $\log t$ = 8 to 9.7 we expect planets to shrink by 0.17~dex or 33\%. The largest sub-Neptunes would shrink to 2.2~\Re and 2.8~\Re for the $M_{\star,1} \approx 0.6~\Msun$ and $M_{\star,2} \approx 1.2~\Msun$ populations, respectively, in rough agreement with observations. However, we note that we observed no age dependence in the mean sub-Neptune size over $\log t$ = 9--10, and comment on this discrepancy later. 

Given the similarities between the core- and XUV-powered models, this toy model may be easily adapted to the core-powered model. The key difference is that the time of rapid mass loss is $t \lesssim 1$~Gyr as opposed to $t \lesssim 100$~Myr. 

To summarize, the \Rp-\Mstar correlation observed for super-Earths and sub-Neptunes is broadly consistent with maximum core-mass that increases with \Mstar. The $\Rp \propto \Mstar^{0.25}$ trend suggests a $M_c \propto \Mstar$ relationship. A linear relationship between core mass and host star mass has also been suggested by \cite{Wu19} and \cite{Berger20b}. 

Why should core mass increase with host star mass? One explanation could be that the core formation process is dust-limited and that the available dust mass $M_\mathrm{dust}$ grows with stellar mass. \cite{Pascucci16} derived $M_\mathrm{dust}$ in protoplanetary disks from ALMA observations at $\sim$1~mm and found a super-linear relationship with stellar mass $M_\mathrm{dust} \propto (\Mstar)^{1.3-1.9}$, with a large 0.8~dex dispersion about this relationship. It is not clear, however, if core-formation is dust-limited due to the difficulties inferring the total inventories of solids from mm emission. \cite{Wu19} offered a different interpretation, positing that planet cores form at a ``thermal mass'' $M_\mathrm{th}$, the mass at which a planet's Hill sphere equals the disk scale height. $M_\mathrm{th} \propto \Mstar^{11/8}$, which is close to the inferred $M_{c} \propto \Mstar$ dependence.

\subsection{Absence of size-metallicity trend suggests weak connection between envelope opacity and stellar metallicity}
\label{sec:discussion-metallicity}

Models of sub-Neptunes depend on the envelope opacity $\kappa$, which sets the cooling rate, location of the RCB, and thus the planet's size. Higher opacity envelopes cool and contract more slowly. 

\begin{figure*}
\centering
\includegraphics[width=0.9\textwidth, trim={0.5cm 10cm 0cm 3.5cm}]{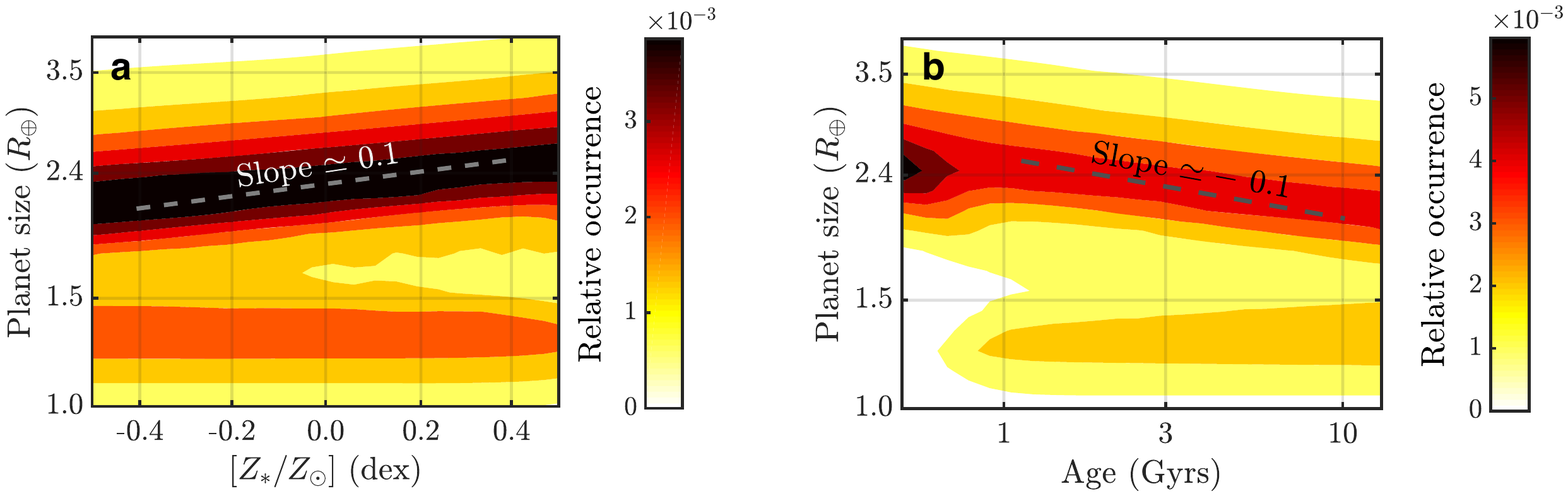}
\includegraphics[width=0.85\textwidth]{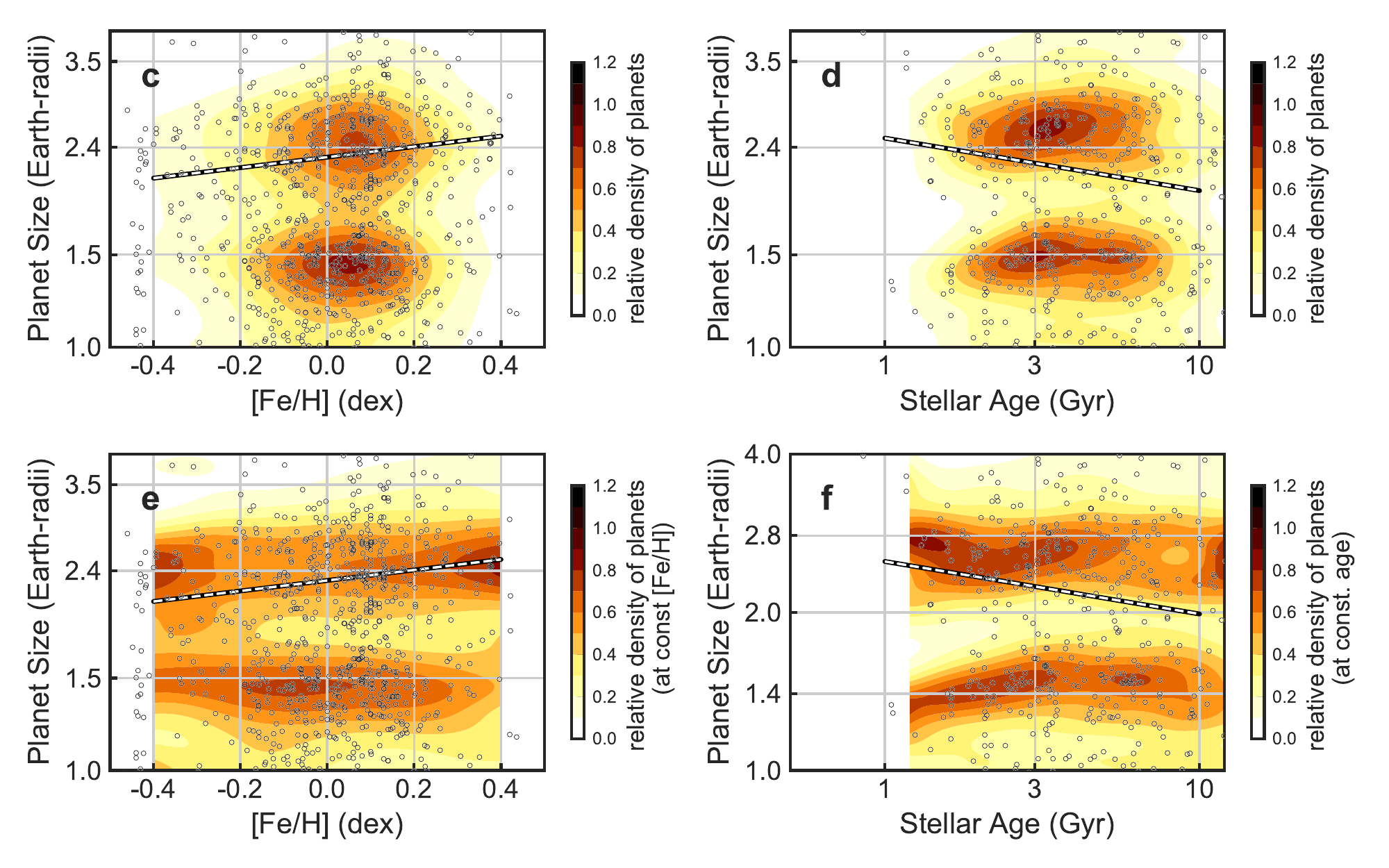}
\caption{Panel (a): predicted planet sizes as a function of stellar metallicity in the core-powered models of \cite{Gupta20}. Here, \Mstar and age were fixed at 1.0~\Msun and 3~Gyr. Panel (c): the relative number density of {\em detected} planets (not occurrence). Panel (e): same as (c), except we have normalized the number density by its integral along columns of constant \fe to accentual features at low/high \fe. In the \cite{Gupta20} models, the typical sub-Neptune is expected to grow with metallicity according to $\Rp \propto Z^{\beta}$, with $\beta = 0.1$. We observed $\beta = \fit{sn-mm-alpha-met}$ and show the predicted $\beta = 0.1$ dependence for comparison in panels (c) and (e). The lack of detectable variation in sub-Neptune size with metallicity when such a trend is predicted in the core-powered models points to missing or incomplete physics, perhaps involving the treatment of envelope opacity (see \S\ref{sec:discussion-metallicity}). 
Panel (b): same as panel (a) except that \Mstar and \fe were fixed to 1.0~\Msun and 0.0~dex and age was allowed to vary. Panel (d) and (f): same as panel (c) and (e) except for age. 
The sub-Neptunes are expected to shrink with time according to $\Rp \propto t^{\gamma}$, with $\gamma = -0.1$. While we observed $\gamma = \fit{sn-mma-alpha-age}$, we could not conclusively rule out a $\gamma = -0.1$ dependence given the age uncertainties (see \S\ref{sec:discussion-age}). 
\label{fig:gupta-comparison}}
\end{figure*}

In their core-powered models, \cite{Gupta20} assumed $\kappa \propto Z$, where $Z$ is the bulk stellar metallicity. To isolate the effects of metallicity, they constructed a population of planets with \Mstar fixed at 1.0~\Msun, \fe ranging from $-0.5$ to $+0.5$, and evolved them for 3~Gyr. The final population is shown in their Figure~9 is reproduced here in Figure~\ref{fig:gupta-comparison}. In the models, sub-Neptune size increases with stellar metallicity as $\Rp \propto Z^{\beta}$, with $\beta = 0.1$. In contrast, in \S\ref{sec:planet-population-mass} we measured $\beta = \fit{sn-mm-alpha-met}$, which rules out the predicted $\beta = 0.1$ at $4\sigma$ significance. 

For visual comparison, we reproduced Figure~\ref{fig:planet-prad-zoom}d in Figure~\ref{fig:gupta-comparison} with a $\beta = 0.1$ relationship over-plotted. The CXM distribution is peaked near \fe = 0.0, but no tilt is observable. To visually enhance the tails of the metallicity distribution, we normalized the density by the integrated density along vertical columns. Both super-Earths and sub-Neptunes show no significant size-metallicity dependence. 

Our results suggests that the dominant opacity at the RCB does not track stellar metallicity. The tension between the observations and the \cite{Gupta20} model does not reflect an insurmountable obstacle for core-powered theory, but does suggest a modification to the opacity treatment. We also note that the XUV-powered models of \cite{Owen17} also assumed a $\kappa \propto Z$ relationship.  

\subsection{Age effects}
\label{sec:discussion-age}

After the period of rapid mass-loss, we expect the sizes of the sub-Neptunes to decrease due to the Kelvin-Helmholtz mechanism. In \S\ref{sec:planet-population-mass}, we found $\Rp \propto (\mathrm{age} / 5\, \mathrm{Gyr})^\gamma$ with $\gamma = \fit{sn-mma-alpha-age}$, consistent with no \Rp-age collocation. This flat age dependence is interesting given that several studies have noted planets younger than 100~Myr seem to have inflated envelopes. In a recent compilation of nine planets with cluster-based ages under 100~Myr by \cite{Bouma20}, all were between 4 and 10~\Re. \cite{Rizzuto18} and \cite{David19} made a similar observation among earlier compilations. 

Planets in this size range are intrinsically rare among field-age stars. As a point of reference, for $P < 100$~days, \cite{Petigura18} measured an occurrence rate of 0.04 planets per star with \Rp = 4.0--11.3~\Re vs. 0.64 planets per star with \Rp = 1--4~\Re among stars with $\Mstar \approx 0.8-1.3$~\Msun. Yet V1298Tau, a 20--30~Myr, solar-mass star hosts {\em four} planets with sizes ranging from 5 to 10~\Re \citep{David19}. Planets in this size range are even rarer among low-mass stars. \cite{Dressing13} measured an occurrence rate of 0.007 planets per star with \Rp = 4.0--11.3~\Re and $P < 50$~days in a sample of stars with $\Mstar \approx 0.4-0.6$~\Msun. Yet K2-33 is a 5--10~Myr, 0.4~\Msun star that hosts a 6~\Re planet \citep{David16}. 

Given the evidence for inflated planets younger than 100~Myr in the literature combined with the flat \Rp-age relationship observed here for ages spanning 1 to 10~Gyr, we conclude that large-scale evolution in the size of sub-Neptune envelopes concludes by $\sim$1~Gyr. 

The envelopes may continue to contract after $\sim$1~Gyr, but the \Rp-age trends may be too subtle to detect given our sample size and age uncertainties. For example, \cite{Gupta20} isolated the age effect in the core-powered context by constructing a planet population with \Mstar and \fe held fixed at 1.0~\Msun and 0.0~dex and evolved them from 1 to 10~Gyr (their Figure~10, reproduced here in Figure~\ref{fig:gupta-comparison}). In this model, the planets obeyed a $\gamma = -0.1$ relationship. However, errors in the measured ages will conspire to flatten such a trend. We explored this effect by injecting a $\gamma = -0.1$ dependence into the observed sub-Neptune population, perturbing their ages by their uncertainties, and refitting our power-law model. We found $\gamma = -0.06$, which is consistent with our measured value of $\gamma = \fit{sn-mma-alpha-age}$ at the 2$\sigma$ level. Thus, we cannot rule out the subtle $\gamma = -0.1$ dependence predicted by \cite{Gupta20} given our current sample size and age uncertainties.

\subsection{Slope of the Radius Gap in period-radius space}

While a number of theoretical models can produce the Radius Gap, they do not produce identical slopes in the $P$-$\Rp$ plane. Thus the slope of the Radius Gap has emerged as a potential discriminant between different models. In the \cite{Owen17} photoevaporation model, the maximum size of a super-Earth decreases with orbital period. The upper envelope is set by the size of a rocky core that is barely stripped by the integrated XUV exposure. Slightly more massive cores retain their envelopes and remain sub-Neptunes. 

At larger orbital periods, $\mathcal{F}_\mathrm{xuv}$ decreases and the cores that can be completely stripped have lower masses and smaller sizes. Assuming a fixed energy-limited efficiency factor $\eta = 10\%$, \cite{Owen17} predicted that the bottom of the Radius Gap will follow $\Rp \propto P^{m}$ with $m = -0.25$. Realistically, $\eta$ should decline as mass increases since escaping gas has more time to cool as it climbs out of a deeper potential well. Adopting $\eta = 10\% (v_{esc} / 15~\kms)^{-2}$ yields a shallower slope $m = -0.16$. 

In the core-powered framework of \cite{Gupta19}, the fate of an envelope around a planet with $P \gtrsim 8$~days depends on the cooling timescale $t_\mathrm{cool} = E / L$ relative to the envelope mass-loss timescale $t_\mathrm{loss} = M_\mathrm{env} / \dot{M}$. Envelopes that escape faster than they cool $t_\mathrm{loss} < t_\mathrm{cool}$ are lost completely. The mass-loss timescale is depends on $\dot{M}$ at the Bondi radius, and $t_\mathrm{loss} \propto  M_\mathrm{env} R_{s}^{2} c_{s} \rho_{rcb} \exp \left(G \Mp / c_{s}^{2} R_{rcb}\right)$ where $R_{s}$ is the sonic point and $c_{s}$ is the sound speed. The cooling timescale may be computed by dividing the combined thermal and gravitational energy of the envelope and core by the luminosity evaluated at the RCB. The exponential behavior of $t_\mathrm{loss}$ results in the dividing line $t_\mathrm{loss} = t_\mathrm{cool}$ being set by $G \Mp / c_{s}^{2} R_\mathrm{rcb} = \mathrm{constant}$, which implies $R_{p} \propto P^{-0.11}$ or $m = -0.11$.

We emphasize that the values of $m$ predicted by the core- and XUV-powered models discussed above have all incorporated assumptions about the properties of planet cores such as the distribution of masses, orbital periods, and bulk compositions. \cite{Rogers21b} showed how changing these input parameters can alter the predicted $m$. {\em However, both the core- and XUV-models categorically predict a negative slope.}

Models that delay the accretion of gas until the nebula is nearly or completely dissipated predict an increasing slope with orbital period \citep{Lee14,Lee16,Lopez18}. Here, the core mass is determined by the solid surface density profile within a feeding zone that is proportional to the Hill radius $R_{H} \propto a (\Mp / \Mstar)^{1/3}$. A \cite{Hayashi81} profile $\Sigma(a) \propto a^{-3/2}$ yields $\Mp \propto a^{0.6} \Mstar^{-0.5}$ or $\Rp \propto P^{0.11} \Mstar^{-0.09}$ or $m = 0.11$.

For the full CXM sample, we measured $m$ = $\grad{per-prad-det_smass=0.5-1.4-m}$. \cite{Van-Eylen18} fit the gap in a sample of 117 planets orbiting 75 stars with asteroseismic detections and found $m$ = $-0.09^{+0.02}_{-0.04}$. Their population of stellar hosts most closely resembles a combination of our middle and high mass bins, for which we measured $m$ = $\grad{per-prad-det_smass=0.7-1.0-m}$ and $\grad{per-prad-det_smass=1.0-1.4-m}$, which straddle the \cite{Van-Eylen18} value. \cite{Martinez19} performed an independent analysis of CKS DR1 spectra including parallax constraints and found $m = -0.11\pm 0.02$. Despite differences in planets used, planet/star parameter provenance, and slope measurement technique, our measurements of slope are consistent with the \cite{Van-Eylen18} and \cite{Martinez19} measurements and with the core- and XUV-powered model predictions. 

Shifting our attention to low-mass stars, we resolved the Radius Gap among the \sample{occ-nplnt_smass=0.5-0.7} planets in our low stellar mass bin \Mstar = 0.5--0.7~\Msun. We measured $m$ = $\grad{per-prad-det_smass=0.5-0.7-m}$, consistent with that of the full sample. As a point of comparison, \cite{Cloutier20} measured the occurrence of planets in a sample of stars having \teff < 4700~K drawn from \Kepler and \ktwo (275 and 53 planets). The host stars were \Mstar = 0.08--0.93~\Msun, but most were between 0.5 and 0.8~\Msun. They resolved the Radius Gap and measured a positive slope in $P$-\Rp space of $m = 0.058\pm0.022$. They interpreted this as a signature of gas-poor formation channels around lower mass stars. However, the Radius Gap is only visible in their full 328 planet sample, but not when they restrict their analysis to the 126 planets around \Mstar = 0.08--0.65~\Msun hosts (their Figure~12). Our measured slope is $\approx 4\sigma$ lower than \cite{Cloutier20}. We found no evidence of a change of slope and thus do not favor an alternate formation pathway for planets around \Mstar = 0.5--0.7~\Msun stars. \cite{Van-Eylen21} arrived at a similar conclusion after analyzing a sample of 27 planets orbiting stars with $\teff < 4000$~K.

Before concluding this section, we wish to remark on some of the challenges associated with using slope of the Radius Gap as a signpost of formation. Note that our uncertainties on $m$ are {\em larger} than \cite{Martinez19} even though we resolved the valley at higher contrast (compare our Figure~\ref{fig:planet-prad-zoom} to their Figure~12). One may understand this through the following limiting case: Consider super-Earth/sup-Neptune populations separated by a gap completely devoid of planets. In such a scenario, selecting the minima used to fit the Radius Gap in \S\ref{sec:planet-population-period-radius} is ill-poised because there is no longer a single minimum \Rp. Thus, a broad range of $R_{p,0}$ and $m$ are allowed. {\em Thus, the statistic for assessing model predictions has the undesirable quality of becoming more uncertain as observational uncertainties improve.}

Moreover, there is no standard approach to fitting the absence of planets. For example, \cite{Van-Eylen18} used a support vector machine scheme, \cite{Berger20b} used the `gapfit' code that subtracts off a trial $\Rp \propto P^m$ relationship and evaluates a 1D KDE of the residuals, and \cite{Martinez19} and this work fit a train of minima computed along 1D projections through the $P$-$\Rp$ plane. All approaches rely on an {\em ad hoc} smoothing parameter which determines the relative influence of the few planets near the edge of the gap relative to the many planets far from the gap on the fit. These differences in fitting method are less troublesome in theoretical studies which can model limitless numbers of planets and do not contend with complications like false positive contamination or mischaracterized measurement uncertainties. 

Recently, \cite{Rogers21a} introduced a different framework for evaluating the agreement between formation theory and the observed planet population that does not rely on measuring the slope of the gap. Here, a model planet population is subjected to a specified set of physical processes, e.g. XUV- or core-powered mass-loss. The synthetic population is transformed into an occurrence rate density distribution. This is then converted into a number rate density distribution by accounting for properties of the parent stellar population, losses due to non-transiting planets, and losses due to pipeline incompleteness. In this framework, one evaluates the likelihood that the observed planets are a realization of an inhomogeneous Poisson point process with the specified number rate density. One may then optimize the model parameters and characterize their uncertainties with using standard techniques. Such a treatment of the CXM sample would be interesting, but is beyond the scope of this paper.

\subsection{Period-flux distribution of planets}
\label{sec:discussion-period-flux}

In \S\ref{sec:occ-period-distribution}, we characterized the period distribution of super-Earths and sub-Neptunes $\dd N / \dd P$ in three bins of stellar mass. We parameterized these distributions as smooth broken power-laws with variable breakpoints $P_0$. From \Mstar = 0.5--1.4~\Msun, the sub-Neptune falloff occurs at a nearly constant $P_0 \approx 10$~days. Similarly, when we modeled the flux distribution, $\dd N / \dd \Sinc$, we found that $\Sincc{0}$ increases by nearly an order of magnitude from $\approx 20$ to 200~\Se. 

Under the assumptions that planet cores are (1) uniformly distributed in log-period and (2) have masses that are uncorrelated with \Mstar, the above observation would be strong evidence for XUV-powered and against core-powered mass-loss. As we explain in \S\ref{ssec:discussion-mstar-rp}, $\Fxuv \propto \Mstar^{0.33}$, and thus only varies by 40\% over our full mass stellar range. Thus the critical \Fxuv that strips the typical sub-Neptune maps to $P_0$ that is constant with \Mstar. Fixed $P_0$ corresponds to $\Sincc{0}$ that grows with \Mstar. Under the same assumptions, $\Sincc{0}$ should be constant with \Mstar in the core-powered model.

Neither of the above assumptions, however, appear to be correct. The steep decline in super-Earths for $P \lesssim P_{0} \approx 5$~days suggests rocky cores struggle to form inside this boundary. Several mechanisms may be responsible, including magnetospheric truncation of the inner disk (e.g., \citealt{Lee17}) or a pressure trap due a silicate sublimation front (e.g., \citealt{Flock19}). Moreover, the \Mstar-\Rp correlation discussed in \S\ref{ssec:discussion-mstar-rp} suggests that massive stars make massive cores.

Therefore, we cannot rule out the core-powered model based on the dependence of $P_0$ and $\Sincc{0}$ on \Mstar. In this framework, stars that are 1.0--1.4~\Msun efficiently produce sub-Neptunes when $\Sinc \lesssim \Sincc{0} \approx 200$~\Se. Stars that are 0.5--0.7~\Msun only efficiently produce sub-Neptunes when $\Sinc \lesssim \Sincc{0} \approx 20$~\Se. The lack of sub-Neptunes in the $\Sinc \approx 20$--$200$~\Se range is not due to envelope stripping but instead due to the absence of suitable cores. Inspecting Figure~\ref{fig:occur-per}, we see that the occurrence rate density of super-Earths declines by a factor of five over $\Sinc = 20$--$200$~\Se.

There is still considerable uncertainty in the distribution of core masses and further uncertainty in how that distribution varies with $P$ and \Mstar. Given these unknowns, both core- and XUV-powered frameworks appear sufficiently flexible to match the observations. Additional mass measurements of sub-Neptunes as a function of $P$-$\Sinc$-$\Mstar$ should help to constrain the models since \Rp and \Mp together constrain both core mass and envelope fraction. 

\section{Summary and Conclusion}
\label{sec:conclusion}

Here, we provide a brief summary of our results and point toward future improvements to this work: 

\begin{itemize}
\item We augmented the 1305-star CKS DR1 spectral library with DR2 containing \sample{nstars dr2} new spectra. 

\item We performed a homogeneous analysis of the combined DR1 and DR2 samples and derived \teff, \fe, \vsini, \Mstar, \Rstar, and age.

\item We constructed a curated sample of \sample{nplanets planets-cuts2-cut-all} planets orbiting \sample{nstars planets-cuts2-cut-all} stars with updated star/planet properties.

\item We resolved the Radius Gap and projected the planet population as a function of $P$, $\Sinc$, $\Mstar$, \fe, and age. 

\item The Radius Gap in the $P$-\Rp plane follows $\Rp \propto P^m$ with $m = -0.10 \pm 0.03$, consistent with both core-powered and XUV-powered models but inconsistent with gas-poor formation models.

\item We observed no significant change in $m$ over 0.5--1.4~\Msun. 

\item Sub-Neptunes tend to be larger around higher mass stars and follow $\Rp \propto (\Mstar/\Msun)^{\alpha}$ with $\alpha = \fit{sn-m-alpha-mass}$. The super-Earths exhibit no measurable \Mstar dependence $\alpha = \fit{se-m-alpha-mass}$. Taken together, these trends are consistent with a core-mass distribution that scales linearly with stellar mass $\Mcore \propto \Mstar$. 

\item The average sub-Neptune is not measurably larger around higher metallicity stars, disfavoring a simple linear relationship between evelope opacity and stellar metallicity $\kappa \propto Z$.

\item The average sub-Neptune does not measurably shrink over 1--10~Gyr. Given the large radii observed among stars younger than 100~Myr, we conclude that the majority of planet radius contraction concludes by $\sim$1~Gyr. 

\item The period distribution of sub-Neptunes has a breakpoint $P_0$ in \Mstar. This is consistent with the predictions of photoevaporation models. However, core-powered models may still be viable given uncertainties in the underlying distribution of planet cores. 
\end{itemize}

The bifurcation of small planets in to two distinct populations was one of the most intriguing results from the \Kepler mission. This feature appears most consistent with models where close-in planets acquire and lose H/He envelopes.  XUV radiation from young stars or heat from cooling cores are leading candidates for the energy source that powers this mass loss. While we worked to identify the dominant mass-loss process, we found both mechanisms are flexible enough to match our observations. 

We look forward to additional observational and theoretical work that could help discriminate between these two theories. With \Gaia, it will soon be possible to extend the analysis presented here to a larger sample of planets (2--3$\times$).  Future \Gaia data releases will include low- and high-resolution spectra for nearly all \Kepler stars. These data will increase the upper \Mstar limit since rapid rotation above 6500~K does not wash out broadband SED information. \Gaia-based refinements in \teff and \fe will be especially useful in determining masses and ages for additional early-type stars. However, increases in sample size from \Gaia may not settle the core-powered vs. XUV-powered question. An analysis of synthetic planet populations by \cite{Rogers21b} suggested that $\gtrsim 5,000$ planets are needed to do so.

We found that heuristic descriptions of the Radius Gap such as its slope in the $P$-$\Rp$ plane were neither straightforward to compute nor straightforward to compare to model predictions, which are based on a previous occurrence analysis. The disconnect between the demographic analysis and the population modeling leads to concerns of self-inconsistency. The modeling approach of \cite{Rogers21a} offers a firmer statistical basis for comparing models and observations. This approach can also naturally accommodate realistic non-independent distributions of host star properties like mass, metallicity, and age. 

Aside from improved host star properties, additional planet observables should shed light on the mass-loss process. Mass measurements of transiting planets spanning $P$, \Sinc, \fe, and age will help constrain the core-mass distribution which contributes to the remaining flexibility of both models since mass and radius together constrain core mass and envelope fraction. Finally, direct measurements of mass-loss as a function of the same properties would be especially valuable. Outflows from a handful of planets have been recently detected using the He-10830~\AA\ line, and \cite{Gupta21} have  highlighted a number of planets that may actively be experiencing mass loss in the core-powered framework. 

\software{The software used in this analysis is available at \url{https://github.com/petigura/CKS-Cool}. We used the following additional packages: {\em Astropy} \citep{Astropy-Collaboration13}, {\em emcee} \citep{Foreman-Mackey13}, {\em isoclassify} \citep{Huber17,Berger20a}, {\em Lmfit} \citep{Newville14}, {\em Matplotlib} \citep{Hunter07}, {\em Numpy} \citep{numpy/scipy}, {\em Scipy} \citep{Jones01}, {\em H5py} (\url{https://doi.org/10.5281/zenodo.5585380}), {\em Pandas} \citep{pandas}, {\em Seaborn} \citep{seaborn}, and {\em Xarray} \citep{xarray}}

\facilities{Keck:I (HIRES), Kepler}

\acknowledgments
We thank Natalie Batalha, Konstantin Batygin, Akash Gupta, Daniel Huber, Eric Lopez, Hilke Schlichting, Joshua Schlieder, Samuel Yee, and Jon Zink for valuable conversations that improved this manuscript. 

Data presented herein were obtained at the W. M. Keck Observatory from telescope time allocated to the National Aeronautics and Space Administration through the agency's scientific partnership with the California Institute of Technology and the University of California. The Observatory was made possible by the generous financial support of the W. M. Keck Foundation

Kepler was competitively selected as the tenth NASA Discovery mission. Funding for this mission is provided by the NASA Science Mission Directorate. We thank the Kepler Science Office, the Science Operations Center, Threshold Crossing Event Review Team (TCERT), and the Follow-up Observations Program (FOP) working group for their work on all steps in the planet discovery process ranging from selecting target stars to curating planet catalogs. 

E.A.P. acknowledges support from the following sources: a Hubble Fellowship grant HST-HF2-51365.001-A awarded by the Space Telescope Science Institute, which is operated by the Association of Universities for Research in Astronomy, Inc. for NASA under contract NAS 5-26555; a NASA Keck PI Data Award (80NSSC20K0308), administered by the NASA Exoplanet Science Institute; a NASA Astrophysics Data Analysis Program (ADAP) grant (80NSSC20K0457); and the Alfred P. Sloan Foundation.

The authors wish to recognize and acknowledge the very significant cultural role and reverence that the summit of Maunakea has long had within the indigenous Hawaiian community. We are most fortunate to have the opportunity to conduct observations from this mountain.

\bibliography{manuscript}

\begin{comment}
\appendix
\startlongtable 
\begin{deluxetable*}{lRRRRRrRRRRRRRR}
\tablecaption{Stellar Properties}
\tabletypesize{\scriptsize}
\tablecolumns{15}
\tablewidth{0pt}
\tablehead{
	\colhead{KOI} & 
	\colhead{\kmag} &
    \colhead{$\pi$} &
	\colhead{\teff} &
	\colhead{\fe} &
	\colhead{\vsini} &
    \colhead{prov} & 
    \colhead{$\Rstar$} & 
	\colhead{\Mstariso} & 
	\colhead{\Rstariso} & 
	\colhead{\rhostariso} & 
    \colhead{\ageiso} &
    \colhead{$\pi_{\mathrm{spec}}$} &
    \colhead{SB2} &
	\colhead{CXM} \\
    \colhead{} & 
    \colhead{mag} & 
    \colhead{mas} &
	\colhead{K} &
	\colhead{dex} & 
	\colhead{\kms} &
    \colhead{} &
    \colhead{\Rsun} & 
	\colhead{\Msun} & 
	\colhead{\Rsun} & 
	\colhead{g/cc} & 
    \colhead{Gyr} & 
    \colhead{mas} &
    \colhead{} &
    \colhead{} 
}
\startdata
\input{tab_star.tex}
\enddata
%\vspace{-2cm}
\tablecomments{}
\end{deluxetable*}

\startlongtable 
\begin{deluxetable*}{lRRRRRRRR}
\tablecaption{Planet Properties}
\tabletypesize{\scriptsize}
\tablecolumns{9}
\tablewidth{0pt}
\tablehead{
	\colhead{Planet} & 
	\colhead{$P$} &
    \colhead{$\Rp/\Rstar$} &
    \colhead{$T$} &
	\colhead{$\Rp$} &
    \colhead{\Tcirc} &
	\colhead{$a$} &
	\colhead{\Sinc} &
	\colhead{samp} \\
    \colhead{} &
    \colhead{d} &
    \colhead{\%} &
    \colhead{hr} &
    \colhead{\Re} &
    \colhead{hr} &
    \colhead{au} &
    \colhead{\Se} &
    \colhead{}
}
\startdata
\input{tab_planet.tex}
\enddata
\tablecomments{}
\end{deluxetable*}
\end{comment}

\end{document}